\documentclass{aa}
 \usepackage{graphicx}
   \begin{document}
 \title{Model simulation of optical light curves for blazar OJ287} 
 \author{S.J.~Qian\inst{1}}
 \institute{National Astronomical Observatories,
    Chinese Academy of Sciences, Beijing 100012, China} 
 \date{Complied by using A\&A latex}
  \abstract{The light curves of optical outbursts observed in blazar OJ287
    during 1983--2015 are analyzed and model-simulated to investigate the
    nature of its optical radiation.}{It is shown that the December/2015
    outburst has its  multi-wavelength variability behavior very similar 
   to that of the synchrotron outburst in March/2016, indicating that the
   2015 outburst may originate from synchrotron process.}{In combination with
   helical motion of superluminal components, the precessing jet nozzle 
   scenario previously proposed is used to model-simulate the light curves
   of all the optical outbursts discussed.}{The optical light-curves for 
    both periodic and non-periodic outbursts observed in blazar OJ287 can
    be well interpreted in terms of lighthouse effect
    due to the helical motion of superluminal optical knots, showing their 
    common origin in synchrotron process.}{A coherent and compatible framework
    is tentatively suggested to understand the entire phenomena in OJ287.
    The double-peak structure of the periodic outbursts might be explained 
    by invoking the cavity-accretion flare models for comparable-mass binary 
    systems in eccentric motion.}
  \maketitle
  \section{Introduction}
   OJ287 (z=0.306) is an optically violent variable BL Lacertae object (BLO)
   and also one of the bright Fermi $\gamma$-ray sources (Ackermann 
   et al. \cite{Ac11}, Hartman et al. \cite{Har99}, Agudo et al. \cite{Ag12}).
     Its strong variability has been observed in all the
    wavebands from radio to $\gamma$-rays with various timescales 
    (hours/days to years).\\
   Its optical variability is particularly exceptional. The optical light curve
   recorded since 1890s reveals quasi-periodic outbursts 
   with a cycle of $\sim$12\,yr (Sillanp\"a\"a et al. \cite{Si88}). 
   Up to now four periodic outbursts with double-peaked
   flares have been observed in 1972--73, 1982--83, 1994--95 and 2005--2007.
   The first flare of the fifth periodic outburst has been 
    observed in December/2015
     and its second flare is predicted to peak on 
    July 31 2019 (Valtonen et al. \cite{Va18}, Dey et al. \cite{De18}
   and references therein).
    The long-lasting quasi-periodicity is believed to be 
   related to the orbital motion of a black hole binary in the center of
    its host galaxy.\\
    In the early works Brown et al. (\cite{Bro89a}, 
   \cite{Bro89b}) showed that the variations  at infrared (IR), optical and 
   ultraviolet wavelengths  are well correlated. Correlation between 
   spectral index and flux density was observed at near-infrared (NIR)
   wavelengths
   (Gear et al. \cite{Ge85}): the source spectrum becomes steeper when it 
   becomes fainter and vice versa. But Sillanp\"a\"a et al. (\cite{Si96a},
   \cite{Si96b}) found that the optical spectral index (or spectral color) was
   very stable during the period 1994--1996 (OJ-94 project, 
   Takalo \cite{Tak96a}). Recently, the multi-wavelength observations 
   performed by Gupta et al. (\cite{Gu16}) during 2015--2017
   also demonstrate the stability of the optical spectral color in OJ287.\\
   Variations observed at centimeter wavelengths usually lag the optical 
   variations (Valtaoja et al. \cite{Val20}, Aller et al. \cite{Al94},
    \cite{Al14}).
   The radio time-delays can be attributed to shock evolution combined with
    opacity effects. But there are some observations revealing 
    simultaneous variations at millimeter and optical wavelengths
   (Sillanp\"a\"a et al. \cite{Si96b}, Valtaoja et al.\cite{Val20}).\\
   OJ287 is a well-known superluminal source on parsec scales and VLBI 
   observations reveal that it has a core-jet structure and superluminal 
   components are steadily ejected from the core (Britzen et al. \cite{Br18},
   Hodgson et al. \cite{Hod17}, Agudo et al. \cite{Ag12},
   Cohen \cite{Co17}, Qian \cite{QiXiv18} and references therein).
   It has been found that there is a close connection between the optical 
   flares and the emergence of superluminal components (Tateyama et al.
   \cite{Ta99}, Qian \cite{QiXiv18}). Recently, based on the analysis of the
   kinematics of superluminal components, Britzen et al. (\cite{Br18}) have
   made an elaborated relativistic jet model (invoking jet precession plus 
   nutation) to explain the radio variability and the kinematics on 
   parsec scales. Based on a potential double-jet scenario, 
   Qian (\cite{QiXiv18}) tentatively 
   derived the total mass of the binary in the range 
   $10^8$--$10^9$$M_{\odot}$, which is consistent with the estimation by
   Gupta et al. (\cite{Gu12}; also see Villforth et al. \cite{Vil10}, 
   Valtaoja et al. \cite{Val20}).\\
    Recently Kushwaha et al. (\cite{Ku18}) and Gupta et al. (\cite{Gu16})
  have monitored the multi-wavelength variations in the NIR-optical-UV bands
   during December/2015\,--\,May/2016, providing 
   new information about the variability behavior in OJ287.
  They showed that the source has a stable color during that period, 
  confirming the finding by
  Sillanp\"a\"a et al. (\cite{Si96a}) and supporting the "single mechanism" for
  the optical flares (periodic major outbursts and non-periodic
   synchrotron bursts) in OJ287.
   Kushwaha et al. (\cite{Ku18}) have found that the December/2015
    optical outburst 
   \footnote{This optical
   outburst has been claimed to be a thermal flare produced by the secondary
   black hole penetrating the disk of the primary hole.} was associated with 
   a simultaneous $\gamma$-ray flare. In addition, another strong synchrotron
   outburst with polarization degree of $\sim$30\%
    was observed in March/2016
    which have its temporal and spectral variations very
   similar to those observed in the December/2015 outburst.\\
   The phenomena in OJ287 are very complex and may involve several different
   mechanisms producing its variations from radio to $\gamma$-rays. Its 
   multi-wavelength variations reveal many prominent features: e.g., (1) 12\,yr
   quasi-periodic optical variability; (2) double-peak structure of the 
   periodic optical outbursts; (3) symmetry of individual optical flare
    profiles; (4) multi-component structure of the major optical outbursts;
   (5) similarity in the variability behavior of individual bursts and major
   periodic outbursts; (6) large range of optical polarization degrees (from 
   $<$2\% to $\sim$40\%); (7) stability of optical spectral index 
   (color stability); (8) connection between radio and optical variations;
   (9) synchronous radio and optical variations 
   (simultaneity and similar  profiles); (10) ejection of superluminal 
    components and jet precession; 
   (11) association of $\gamma$-ray flares with optical flares, etc. \\
       A number of
   models have been proposed (referring to the  discussions in Villforth et 
   al. \cite{Vil10}, Qian \cite{QiXiv18}, \cite{Qi15}). On the whole, 
   these models can be divided into two categories, both involving a black hole
   binary system in the nucleus of OJ287:
   \begin{itemize}
   \item   The precessing binary model (or disk-impact model)
   originally proposed by Lehto \& Valtonen
   (\cite{Le96}; improved versions: Valtonen \cite{Va07}, Valtonen et al.
    \cite{Va06}, Valtonen  et al. \cite{Va18}, Dey et al. \cite{De18})
    has been steadily elaborated to interpret the variability 
   behavior in OJ287, putting the emphasis on  the accurate timing of 
   the first major flares of the periodic outbursts, which were suggested 
   to be bremsstrahlung in origin (unpolarized flares) and produced by 
   the secondary hole penetrating into the accretion disk 
   of the primary hole. In the case of an highly eccentric orbital 
   motion, two impacts would occur near pericenter passages and thus explain
   the double-peak structure of the periodic outbursts. The recent 
   December/2015 optical outburst was studied and interpreted in detail
   by Valtonen et al. (\cite{Va16}, \cite{Va17}). This model requires
   a high inclination angle ($i$\,${\sim}$$50^{\circ}$\,--\,$90^{\circ}$)
   \footnote{In
    this case the jet associated with the secondary hole might not be pointed
    toward us with a small angle, if its spin axis (and jet axis) is 
   approximately parallel to the orbital angular momentum.}
   and a high eccentricity ($e$\,$\sim$\,0.66) and
   a strong constraint on the total mass of the binary, reaching 
   ${\sim}2{\times}{10^{10}}{M_{\odot}}$ with a mass ratio m/M$\sim$0.007. This 
   disk-impact model mainly concentrates on the interpretation of the
   quasi-periodicity of the 12\,yr, double-peak structure and the accurate 
   timing of the periodic outbursts, regarding the periodic outbursts being
   thermal flares due to the impact of the secondary hole penetrating
   the primary disk. This model suggests that the follow-up
    and non-periodic outbursts could be interpreted in terms of the enhanced 
   accretions (disturbances induced by the secondary impacts and tidal 
   effects near pericenter passages). But it can not be used to analyze 
   the complicated phenomena observed in the entire emission (from radio 
   to $\gamma$-rays) and the relationship  between the emission properties 
   and the kinematic  behaviors on parsec
    scales. This model is based on very accurate solution of orbital motion by 
   including the post-Newtonian strong gravitational effects, but invoking 
   a fixed (not variable) disk model. \\
   In contrast, Tanaka (\cite{Tan13}) considered a different mechanism 
  (cavity-accretion flare model) for explaining the
   double-peak structure, assuming the binary having a comparable-mass in a
   coplanar motion. According to the results 
    of hydrodynamic/magnetohydrodynamic (HD/MHD) simulations for binary
   systems surrounded by circumbinary disks, the cavity-accretion
   processes characteristic of comparable-mass binary systems would create 
   two gas streams impacting onto the disks of both the black holes near
    pericenter passages, thus causing the double-peaked outbursts. 
   This model also suggests that the periodic outbursts are bremsstrahlung
   flares caused by the impacts of the gas streams. It is not able to provide
    accurate timing of the periodic outbursts. This cavity-accretion flare 
   model  does not discuss the accretion processes during the intervening
   periods and the interpretation of the follow-up and non-periodic outbursts
    and the related jet behavior.      
  \item Relativistic jet models have been applied  to understand the
   optical and radio variability behavior in OJ287 and discussed by many 
   authors since the earlier years (e.g., Sillanp\"a\"a et al. \cite{Si96a},
   Valtaoja et al. \cite{Val20}), because these models are considered
   to be paradigmatic for explaining the variations  
   (from radio to $\gamma$-rays) observed in blazars. Villata et al. 
   (\cite{Villa98}) considered a precessing double-jet model to explain the 
   periodic double-peak structure. Villforth et al. (\cite{Vil10}) 
   suggested that the periodic outbursts could be interpreted in terms of the
   resonant disk accretion of magnetic field lines.
    Qian (\cite{Qi15}) investigated the possibility that lighthouse effect
    could cause the double-peak structure of the periodic outbursts.
    Recently, based on the analysis of the
    kinematics of the radio superluminal components, Britzen et al.
   (\cite{Br18}) proposed an elaborated model to interpret the radio and
   optical variations, emphasizing the precession and nutation of the 
   relativistic jet being the key ingredients causing the complex phenomena
   in OJ287. In addition, Qian (\cite{QiXiv18}) tentatively suggested that 
   the periodic optical outbursts could be synchrotron flares produced 
   by the superluminal optical knots moving along parabolic trajectories.
    But the explanation of the double-peak structure might have to invoke 
   the cavity-accretion process for comparable-mass binary systems 
   (e.g., Tanaka \cite{Tan13}, Artymowicz \& Lubow \cite{Ar96}, 
   Hayasaki et al. \cite{Ha08}). Relativistic models do not involve 
   disk-impacting events, which cause strong thermal flaring events and 
    the optical/radio phenomena in OJ287 may be explained only by
     invoking the enhanced disk-accretion near pericenter passages and 
    ejection of superluminal optical knots.\\
    It might be worth noticing that, based on the modeling of the kinematics
    of the superluminal components in OJ287 (Qian \cite{QiXiv18}), 
   the total mass of the potential binary has been tentatively determined to be
    $\sim{10^8}$\,--\,${10^9}{M_{\odot}}$
    \footnote{Mass ratio m/M\,$\sim$\,0.3.} which is consistent with the 
    estimations obtained by Gupta et al. (\cite{Gu12}), Villforth et al. 
    (\cite{Vil10}) and Valtaoja et al. (\cite{Val20}).   
   These values seem to favor a binary system  with comparable-mass in 
   a coplanar motion. In this case both the  jets associated
   with the binary holes can point toward us with small angles.
   \end{itemize}
    Unfortunately, all the models currently available can only interpret
    part of the phenomena  observed in  OJ287.
    Some basic issues are  remained to be clarified: mass of th binary,
    double peak structure mechanism, color stability, synchronous 
   radio-optical variations, symmetry in burst
    light curves, similar variability behavior of the flares,
    simultaneity in  optical and $\gamma$-ray flares, etc.
    A comprehensive and coherent framework is imperatively needed to solve 
    all these issues.\\
      In this paper we will apply the precessing jet nozzle model previously
   proposed by Qian et al. (\cite{Qi91a}, \cite{Qi09}, \cite{Qi19}) to make
   simulation of the optical light curves for the six periodic major 
   outbursts (in 1983.00, 1984.10, 1994.75, 2005.76, 2007.70 and 2015.87)
   by using lighthouse effect due to the helical motion of superluminal
    optical knots.
   In particular, the multi-wavelength light curves of both the outbursts in
    December/2015 and in March/2016 will be simulated and compared,
    demonstrating the distinct similarity in their temporal
    and spectral variations (multi-wavelength light curves with similar rising
   and decaying time scales and similar broken power-law spectra). Since the
    outburst in March/2016 is a highly polarized synchrotron flare, the similar 
   variability behaviors of the December/2015 outburst (peaking at 2457360) 
   and the March/2016 outburst (peaking at 2457450) may imply that they have 
    a common emission mechanism and  the December/2015 outburst may  be 
   synchrotron in origin. 
      \begin{figure*}
    \centering
    \includegraphics[width=6cm,angle=-90]{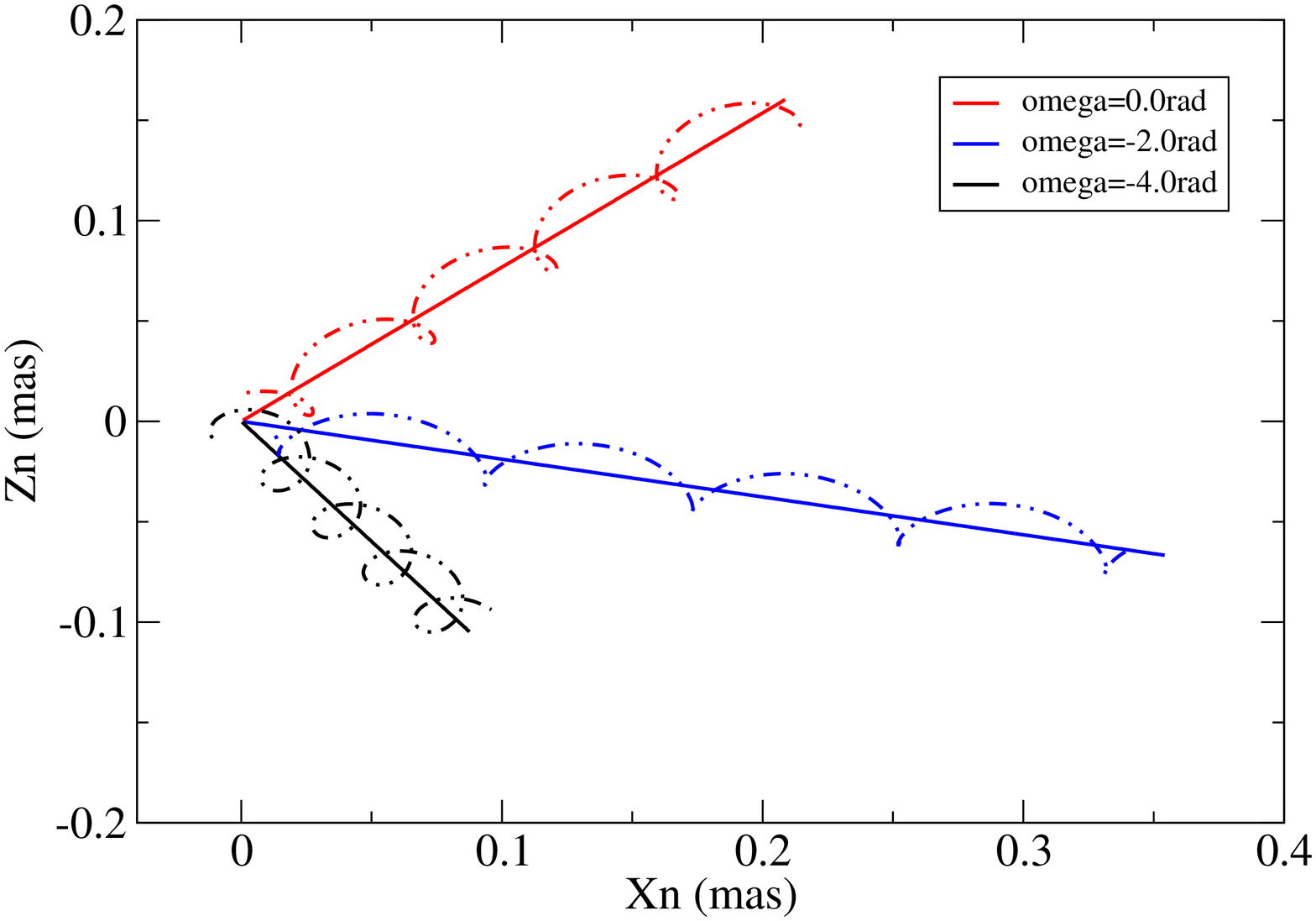}
    \includegraphics[width=6cm,angle=-90]{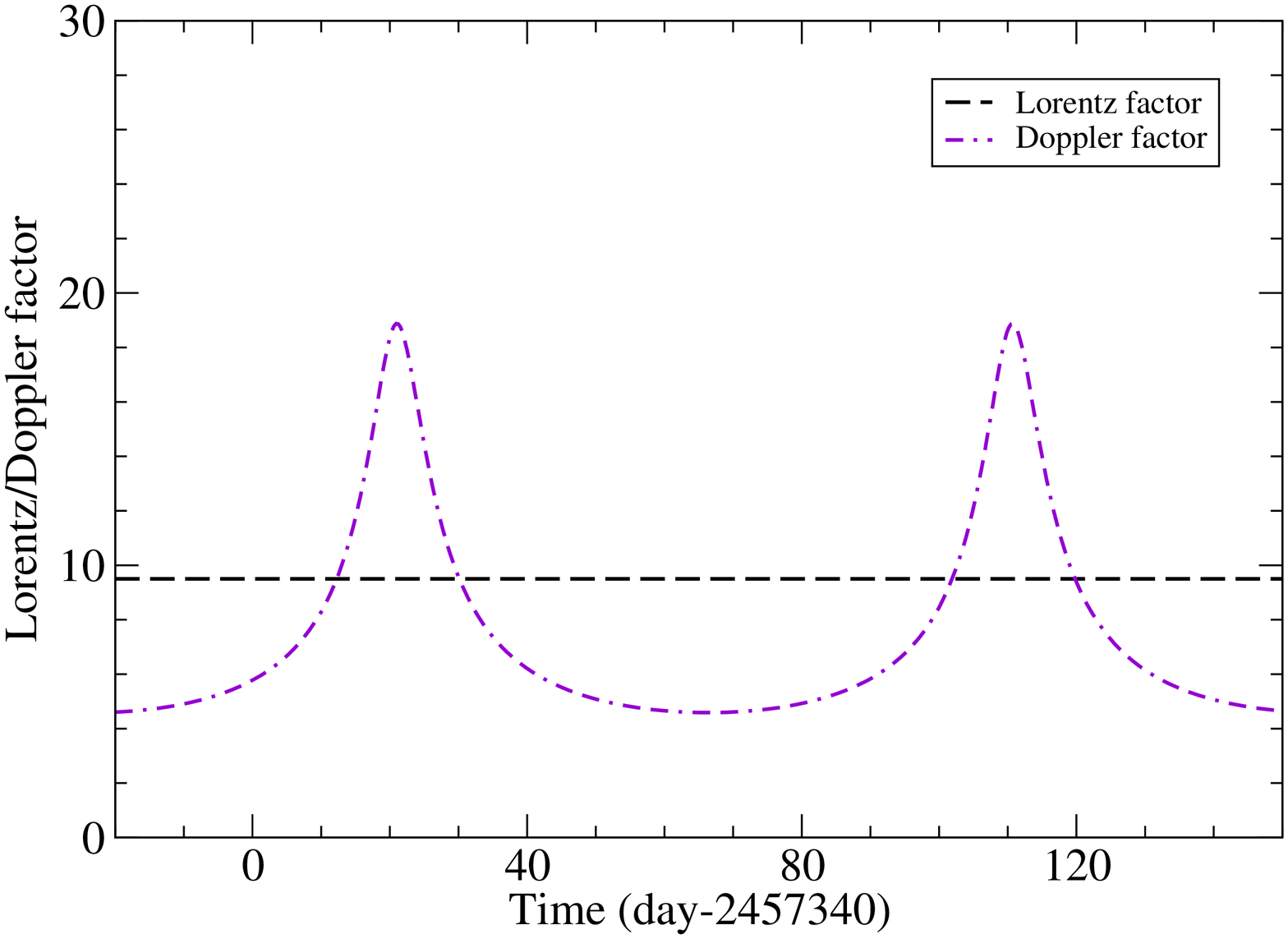}
    \caption{Left panel: A sketch of the precessing jet nozzle scenario with
    helical motion. The straight lines denote the precessing 
    jet axis (projected on the plane of the sky) which is described by the
    precession phases of $\omega$=0.0\,rad, -2.0\,rad and -4.0\,rad,
    respectively. The helices indicate the trajectories of the optical knots 
    moving along the jet axis in perfect collimation zones.
    The helical trajectory defined by $\omega$=-2.0\,rad is used to simulate
    the light curves  of the optical outbursts in December/2015 and March/2016.
    The corresponding Lorentz and Doppler factors are shown in the right panel.
    The helical motion in the perfect collimation zone is assumed to start
     at z=0.}
    \end{figure*}
    \begin{table}
    \caption{Model parameters for the helical trajectories of optical knots.}
    \begin{flushleft}
    \centering
    \begin{tabular}{ll}
    \hline
    Parameter & fixed value \\
    \hline
    $\epsilon$ & $3^{\circ}$ \\
    $\psi$  & 0.0\,rad \\
    $\omega$ & -2.0\,rad \\
    $a$   & 0.0402 \\
    $x$   &  1.0 \\
    $A_0$ & 0.0138\,mas \\
    d$\phi$/d$z_0$ & -7.04\,rad/mas \\
    \hline
   \end{tabular}
   \end{flushleft}
   \end{table}
   \begin{table}
   \caption{Base-level (underlying jet) spectrum for OJ287.}
   \begin{flushleft}
   \centering
   \begin{tabular}{ll}
   \hline
   Waveband & Flux (mJy) \\
   \hline
   K & 8.0 \\
   J & 5.0 \\
   I & 6.2 \\
   R & 3.5 \\
   V & 3.0 \\
   B & 2.0 \\
   U & 1.5 \\
   UVW1 & 0.8 \\
   UVM2 & 0.8 \\
   UVW2 & 0.8 \\
   \hline
   \end{tabular}
   \end{flushleft}
   \end{table}
    \begin{figure*}
    \centering
    \includegraphics[width=4.5cm,angle=-90]{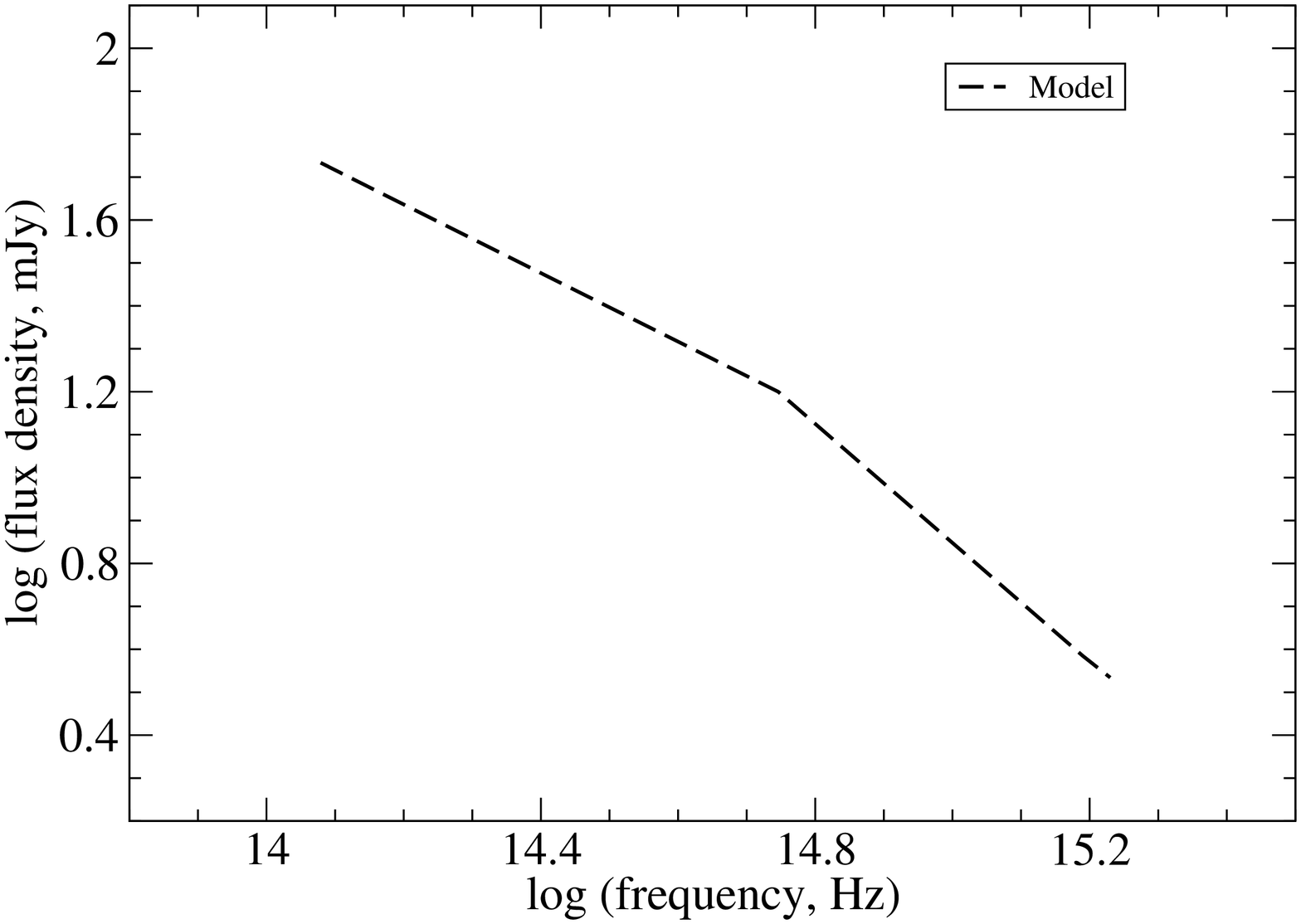}
    \includegraphics[width=4.5cm,angle=-90]{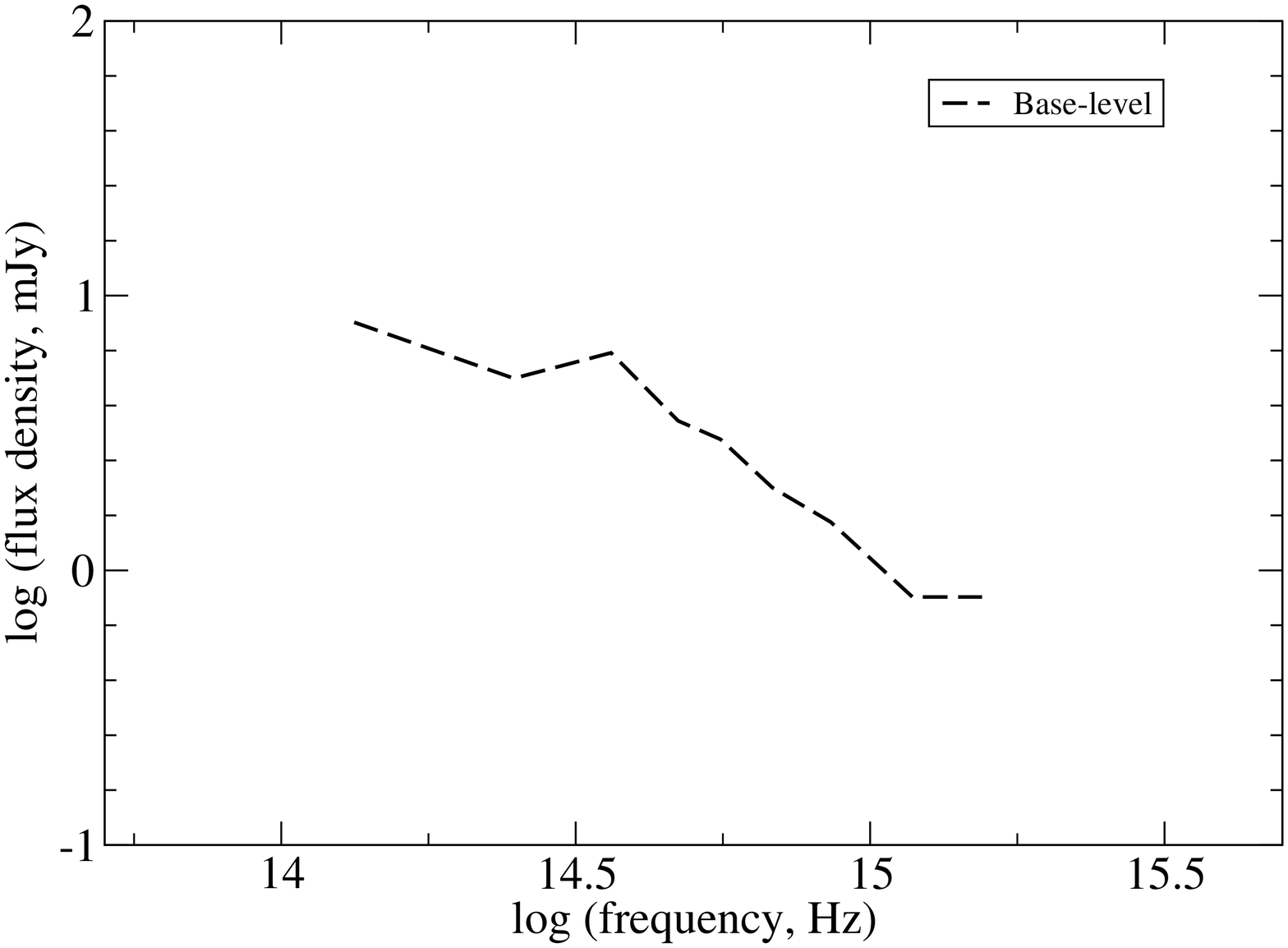}
    \includegraphics[width=4.5cm,angle=-90]{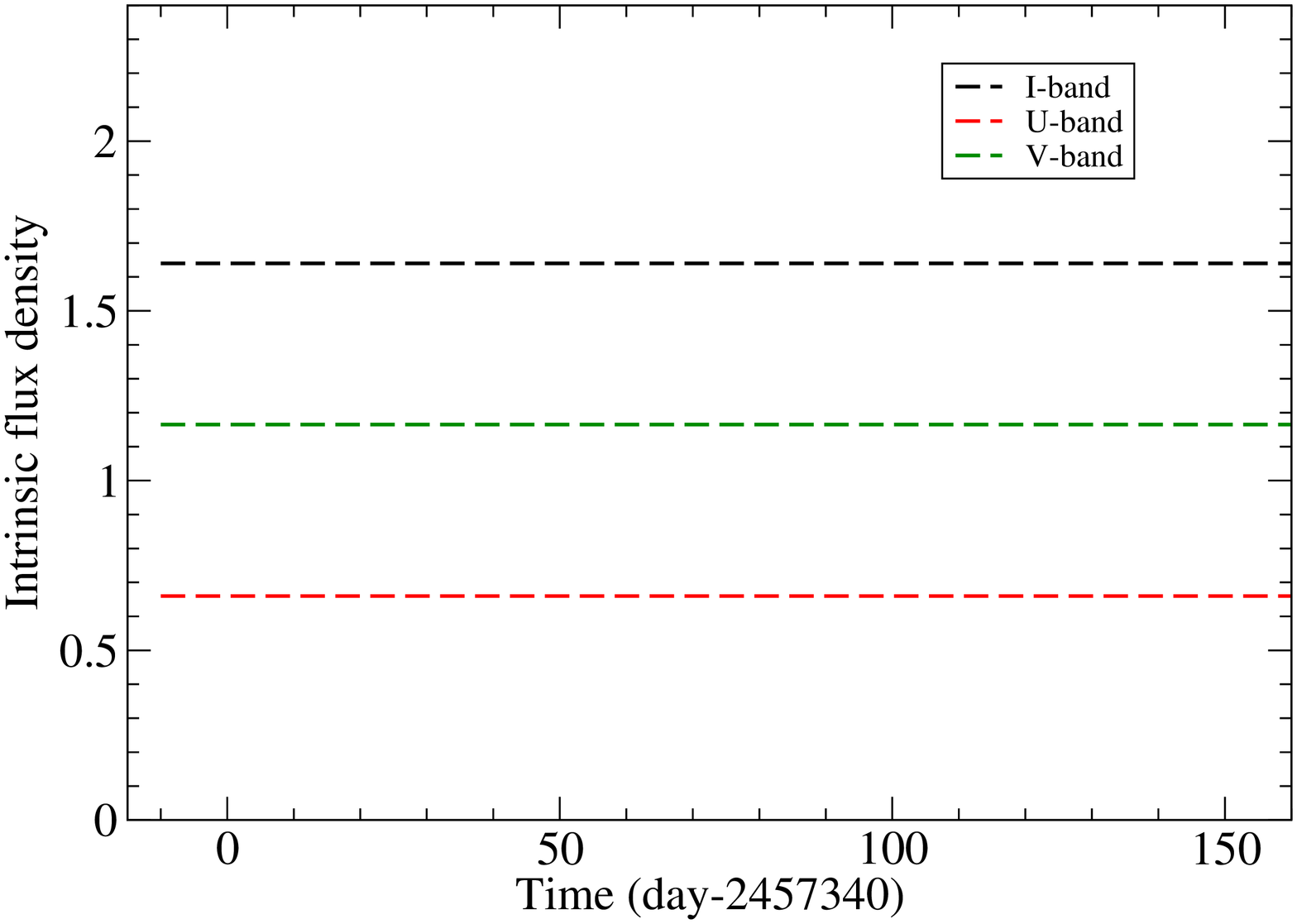}
    \caption{Left panel: the  modeled broken power-law synchrotron 
    spectrum with a spectral break at V-band of $\Delta{\alpha}$=0.5
     (from $\alpha$=0.8 to $\alpha$=1.3, 
   $S_{\nu}$\,$\propto$\,${\nu}^{-\alpha}$).
    The base-level spectrum is shown in the middle panel.
    The modeled intrinsic flux densities (in the comoving frame of the
     optical knot) at I-, V- and U-bands are shown in the right panel
    (unit=$10^{-4}$\,mJy).}
    \end{figure*}
   \begin{table}
    \caption{Parameters for model simulation of the periodic outburst in 
   December/2015 (peaking at JD2457360) and the March/2016 outburst (peaking
    at JD2457450): $\Gamma$--Lorentz factor 
   of the optical knot, maximum Doppler
   factor $\rm{{\delta}_{max}}$, 
   ratio $\rm{{\delta}_{max}}$/$\rm{{\delta}_{min}}$,
   $\rm{S_{int}}$(mJy)--intrinsic flux density of the optical knot, base-level 
   (underlying jet) flux density $\rm{S_b}$=3.5\,mJy at R-band, FWHM (full
   width at half maximum of the modeled light curve; day). $t$\,=\,flare 
   time\,=\,day--2457000.}
   \begin{flushleft}
   \centering
   \begin{tabular}{llllll}
   \hline
   $t$ & $\Gamma$ & $\rm{{\delta}_{max}}$ & ratio & $\rm{S_{int}}$ & FWHM  \\
   \hline
   360 & 9.5 & 18.88 & 4.11 & 1.16$\times{10^{-4}}$ & 5.9 \\
   450 & 9.5 & 18.88 & 4.11 & 1.16$\times{10^{-4}}$ & 5.9 \\
   \hline
   \end{tabular}
   \end{flushleft}
    \end{table}
      \section{Assumptions and approaches}
   In order to better understand the entire phenomena observed in the blazar 
   OJ287, we  would perform detailed simulation of the light curves of its
   optical outbursts, which include not only the periodic major outbursts
    (or the "impact outbursts" claimed to be bremsstrahlung flares caused by
   the  evolving gas-bubbles torn out from the primary disk by the
    secondary-hole penetrations),   but also
   the non-periodic outbursts (usually recognized as synchrotron flares with
   high polarization). The multi-wavelength light curves of the December/2015 
   outburst (peaking at 2457360) and the March/2016 outburst (peaking at 
   2457450) are analyzed and compared for finding their common properties:
   the symmetry of their light-curves and  similarity in their temporary and
   spectral variations.\\
   We will apply the precessing jet-nozzle model previously proposed for 
   several blazars (3C345: Qian et al. \cite{Qi91a}, \cite{Qi09}; 3C279: 
   Qian \cite{Qi11}, \cite{Qi12},  \cite{Qi13}, Qian et al. \cite{Qi19};
   3C454.3: Qian et al. \cite{Qi14}; 
   NRAO 150: Qian \cite{Qi16}; B 1308-326: 
   Qian et al. \cite{Qi17}; PG 1302-102: Qian et al. \cite{Qi18}; 
   OJ287: Qian \cite{QiXiv18})
   to investigate the kinematics of the optical superluminal
   components in OJ287 and propose an interpretation for the multi-wavelength 
   light curves of the optical outbursts in December/2015 and March/2016 
   obtained by Kushwaha et al. (\cite{Ku18}). We will also perform model
   simulation of the V-band light-curves of the six periodic major outbursts in
   1983.00, 1984.10, 1994.75, 2005.76, 2007.70 and 2015.87, suggesting a 
   coherent scenario 
   to understand the entire phenomena in OJ287.\\
   We describe the main assumptions and relevant approaches
   for the model simulation of the outburst light curves as follows.\\
    \subsection{Parameters of precessing nozzle model}
    In order to make model simulation of the light curves of the optical
    outbursts in OJ287, we would need to use an appropriate and specific
    scheme of the precessing nozzle model. Here we we adopt the geometric 
    parameters applied in the previous precessing nozzle models 
    (details referring to Qian \cite{QiXiv18}) and introduce the parameters 
    of helical motion.\\
    We assume that the superluminal optical knots move along helical 
    trajectories around the jet axis which precesses around the precession 
    axis, as shown in Figure 1 (left panel). The precession axis is defined by
    parameter $\epsilon$\,=\,$3^{\circ}$ and $\psi$\,=\,0.0\,rad. The jet axis
    ia assumed to be a straight line  with parameters $a$\,=\,0.0402 and 
    $x$\,=\,1.0 (details referring to Qian \cite{QiXiv18}), which precesses 
    around the precession axis with a period of 12\,yr. Optical superluminal
    knots are assumed to be ejected from the jet nozzle, moving outward along
    helical trajectories. The helical motion of the optical knots are skechily
    shown in Figure 1 (left panel). Using the the precessing nozzle model we 
    can study the helical motion of superluminal optical knots ejected at 
    different precession phases. For concrete model simulations, we will
    assume that the superluminal optical knots are ejected along the jet axis
    defined by the precession phase $\omega$\,=\,--2.0\,rad, moving along the 
    helical trajectories which are defined by parameters $A_0$ and 
    d$\phi(z)$/dz: $A_0$ represents the amplitude of the helical trajectories
    and d$\phi(z)$/dz represents the rotation rate of the helical motion.
    We will take $A_0$\,=\,0.0138\,$\rm{mas}$ and 
    d$\phi$/dz\,=\,--7.04\,$\rm{rad/mas}$. The model parameters are summarized
     in Table 1. In order to demonstrate the common features of  helical
    motion of the superluminal optical components, parameters $A_0$ and
    d$\phi$/dz are assumed to be constant for all the optical flares and only
    their bulk Lorentz ($\Gamma$) and intrinsic flux density ($S_{int}$)
     are chosen to
    model-fit their light curves. Constant $A_0$ and d$\phi$/dz describe 
    uniform helical motion in a perfect collimation zone.\footnote{Introducing
     $A_0$ and d$\phi$/dz as functions of distance $z$, one can study various
    patterns of helical motion.}
    \subsection{March/2016 outburst: a synchrotron flare}
    As shown by Kushwaha et al. (\cite{Ku18}), the strong optical outburst
   observed in March/2016 (peaking at 2457450) is a synchrotron flare with a
   high polarization degree of $\sim$30\% (at R-band). This is consistent with
   the observations made by Valtonen et al. (\cite{Va17}). Both the optical
   polarization observations rule out the possibility that the March/2016 
   outburst is a bremsstrahlung-dominated flare. The reason is:  if this 
   outburst comprised of two components (one thermal and one synchrotron) and 
   was thermal-dominated, then in order to
   explain its 30\% polarization, the synchrotron component would be required
   to have a polarization degree of at least $>$60\%. Such high polarization
   degrees have never been observed in OJ287 (e.g., see the photopolarimetric
   light curves (R-band) during 2005--2008 in Villforth et al. \cite{Vil10}).
   Thus we suggest that the March/2016 outburst (peaking at 2457450)  
   originates from synchrotron process, definitely a non-thermal flare.\\
   This argument is a simple and natural one in blazar physics, but seems quite
   important for understanding the physical processes in OJ287. For example,
   the spectral energy distribution (SED) and its variation of the March/2016 
   outburst are very similar to those of the December/2015 outburst 
   (Fig.4 in Kushwaha et al. \cite{Ku18}; 
   also see Figures 4\,--\,9 displayed below). The SED of both the 
   outbursts reveals two prominent 
   features: (1) an offset between the visible spectrum and the near-infrared
   (NIR) spectrum; (2) a 
   transition from a rather steep visible spectrum to a flatter UV spectrum.
   Superficially, these features look like those observed in other blazars
   (e.g., 3C345, 3C454.3, BL Lac and AO 0235+106; Bregman et al. \cite{Bre86},
   Raiteri et al. \cite{Ra07}, Villata et al. \cite{Villa04},\cite{Villa02}, 
   Raiteri et al. \cite{Ra05}), which have been claimed to 
   be constructed from the "big blue bump" and "little blue  bump"
   produced by the accretion disks and emission lines in the broad-line regions
   (BLR). However, in the case of OJ287, the variations in the UV-band are 
   simultaneous with the NIR-optical variations and  no color changes occur
    during the March/2016 synchrotron outburst (as well during the 
   December/2015 outburst).
   This variability behavior seems indicating that the emission from the
   NIR-optical to the UV-bands are all produced in the jet and the emitting 
   source might have a peculiar inhomogeneous structure (Raiteri et al.
   \cite{Ra07}, \cite{Ra06}, Ostorero et al. \cite{Os04}). We will continue
   to argue for this possibility below, especially based on the connection
    between the radio and optical outbursts.
     \subsection{Similarity between 2015 and 2016 outbursts}
   In the following sections, we will perform model simulation of the 
   multi-wavelength (NIR-optical-UV) light curves of the December/2015 and 
   March/2016 outbursts
   (Kushwaha et al. \cite{Ku18}, Gupta et al. \cite{Gu16}) and show that
    the temporary and spectral variations of the December/2015 outburst 
   are very similar to those of the March/2016 outburst, although the 
   March/2016 outburst occurred $\sim$90\,days later, but was as strong as 
   the December/2015 outburst and had high polarization degrees. They have 
   similar multi-wavelength light curves in the
   NIR-optical-UV bands having symmetric profiles with very similar rising 
   and declining timescales.   While the December/2015 outburst 
   has been claimed to be an ``impact thermal outburst'', 
   occurring at a location $\sim$18,000\,AU away from the primary black hole 
   and originating from an evolving gas-bubble torn out from the accretion disk
   of the primary hole (Lehto \& Valtonen \cite{Le96}, Valtonen et al. 
   \cite{Va16}), the March/2016 outburst is definitely a non-thermal flare,
   originating from synchrotron process in the jet. It is rather difficult
   to understand why the December/2015 thermal outburst could have 
   its temporary and spectral variability behavior so resembling to that
    of the March/2016 synchrotron outburst. Our model simulation
   of their light-curves indicates that the March/2016 and December/2015
   outbursts could be interpreted in terms of the lighthouse effect due to 
   helical motion of one superluminal optical knot in a perfect collimation
    zone of the jet via two helical revolutions. Thus 
   the resemblance in the temporary and spectral variations observed in
   December/2015 and March/2016 outbursts may imply that both December/2015 
   and March/2016 outbursts originate from a common radiation mechanism and 
   they are non-thermal (synchrotron) flares produced in the jet. 
    The two outbursts may be combined into  "one flaring event"
   \footnote{The data-points of the March/2016 outbursts need to be shifted
   backward in time by 89.4\,days.} and their observational data-points 
   are superposed to analyze the common properties of their temporary and 
   spectral variations. For the R-band light curves, the data-points measured
   by Valtonen et al. (\cite{Va16}) are also incorporated in the analysis, 
   providing sufficiently complete temporal coverage for the simulation of
   the multi-wavelength light curves.
   \subsection{Nature of 2015 outburst: $\gamma$-ray observations}
   The nature of the radiation of the outburst in December/2015 (peaking at 
   2457360) is still a debatable issue: whether  it is a 
   synchrotron flare or a bremsstrahlung-dominated one. We argue that the
   December/2015 optical outburst may be a synchrotron flare. \\
   According to Valtonen et al. (\cite{Va16}), the December/2015 outburst
   is composed of two components: one bremsstrahlung component and 
   one synchrotron component, and it is bremsstrahlung-dominated.
    In order to  explain its observed polarization
   degree of $\sim$6\%,  the non-thermal component is assumed to be highly
   polarized with a polarization degree of 40\%.\footnote{Here
    we do not consider the case that the base-level makes a 10\% contribution
    to the polarization degree.}
   In this case the thermal component is much stronger than the non-thermal 
   component, making
   $\sim$68\% and $\sim$17\% contributions to the total flux of the outburst,
   respectively.\footnote{The base-level emission makes a steady $\sim$15\%
    contribution to the total flux density.} However, the 
   relationship between the thermal component and the non-thermal component
   was not clarified: (1) where this highly-polarized component is produced:
   in the jet of the primary hole or in the jet of the secondary hole ? 
  (2) How could the flux variation of the non-thermal component be 
   simultaneous with that of the thermal component, because the two emission 
  components appeared at different locations
   (not co-spatial): the thermal flare occurred at $\sim$0.1\,pc away from the 
   primary hole and its jet; (3) How could  the variable 
   non-thermal component (with a constant polarization degree) be possible to
    closely match the behavior of the thermal component, because they originate
   from different emission mechanisms: bremsstrahlung from an evolving
    gas-bubble and  synchrotron emission from a shock in the jet, having 
   different evolution behaviors with different timescales. \\
   It is worth noticing that the December/2015 outburst emits $\gamma$-rays
    and the  variations in the $\gamma$-ray bands  are simultaneous with  the
    variations in the NIR-optical-UV bands without  time lags 
   (Kushwaha et al. \cite{Ku18}), having similar variability time scales.
     Obviously, this $\gamma$-ray flare should be associated with
   the synchrotron flare component and both emitting regions must be co-spatial
    within the relativistic jet. Thus the  simultaneity in the variations and 
    the similarity in the variability behavior between the $\gamma$-ray
    flare and the bremsstrahlung-dominated outburst seems unlikely,
    because the bremsstrahlung flare is believed to be produced by 
    an evolving gas-bubble torn off the
    primary hole accretion disk by the second black hole penetrating into the
    primary disk (Lehto \& Valtonen \cite{Le96}), occurring at a distance of 
    $\sim$18,000\,AU away from  the primary black hole and its jet
    (Valtonen et al. \cite{Va17}), while the synchrotron component flare and 
   its associated $\gamma$-ray flare are produced in the jet of the primary 
   black hole: they occur at different locations (not co-spatial) through
    different mechanisms. The only plausible interpretation for the 
    simultaneous variations in the $\gamma$-ray and optical bands 
   may be that the December/2015 optical outburst is originated from the jet
   through synchrotron process. This may be the most persuasive argument
   for the  December/2015 outburst being a non-thermal flare.
     \subsection{Spectral energy distribution of December/2015 outburst}
    As typically observed in generic blazars, the spectral energy distribution
    of the December/2015 outburst consists of two bumps: one in the 
    NIR-optical-UV bands and the other one in the $\gamma$-ray bands (Kushwaha 
    et al. \cite{Ku18}). These two bumps are normally interpreted  in terms of
    synchrotron and inverse-Compton processes, respectively. In the one-zone
    scenario (e.g., Qian et al. \cite{Qi98a}, \cite{Qi98b}, Ghisellini et al.
    \cite{Gh07}, Vercellone et al. \cite{Ve10}, \cite{Ve12}), 
    the simultaneity of
    the NIR-optical-UV and $\gamma$-ray variations and their similar 
    variability time scales (rising and declining timescales) would suggest
    that the NIR-optical-UV emitting region and the $\gamma$-ray  emitting
    region are  co-spatial in the jet of the primary hole. It seems difficult to
    understand that the $\gamma$-ray flare could be simultaneous with the
    NIR-optical-UV variations in a bremsstrahlung-dominated outburst. Moreover, 
    under the  bremsstrahlung-dominated assumption for the December/2015 
   outburst, the low-frequency bump of its SED has to be decomposed into 
    two parts (Kushwaha et al. \cite{Ku18}): one thermal and one non-thermal.
     The peak frequency of the non-thermal part has to be shifted to the 
   far-infrared  regime and its optical-UV power has to be lowered 
    to only a half of the observed optical-UV power, reaching the power levels
     during the quiescent states in OJ287
    (e.g., Seta et al. \cite{Set09}, Kushwaha et al. \cite{Ku13}, 
    Kushwaha et al. \cite{Ku18}). This seems inconsistent with the
     normal behavior 
    observed in $\gamma$-ray blazars: the synchrotron bump moves to higher 
     frequency with higher peak power (${\nu}{F_{\nu}}$) 
    during $\gamma$-ray flaring states, 
    when the high-energy bump shifts to higher energy $\gamma$-ray bands. 
    Therefore, it seems more likely that the low-frequency bump of the
    December/2015 outburst may be entirely synchrotron in origin.
    \footnote{Note that, the peak frequency of the low-frequency bump
    usually observed in OJ287 does not show significant 
    difference between the quiescent and flaring states, but the peak power 
    increases during the flaring states (Seta et al. \cite{Set09}).}\\
     \subsection{Broken power-law spectrum}
    The spectral break detected in the optical--UV wavebands for the 
    December/2015 outburst (Kushwaha et al. \cite{Ku18}) has been 
    interpreted as due to
    the thermal emission of the accretion disk surrounding the primary
    black hole with a mass of $1.8{\times}{10^{10}}{M_{\odot}}$. It is noted
     that the March/2016 outburst has a similar spectral break, which may not
    originate from the primary disk and due to the superposition of the 
    synchrotron emission from different jet regions. 
     Moreover, both outbursts has no color 
    variations, as Gupta et al. (\cite{Gu16}) observed. Thus in the following
   simulations of the multi-wavelength light curves of the two outbursts,
    a common broken power-law spectrum will be assumed with a spectral break
    of $\Delta{\alpha}$=0.5: in the K- to V-bands 
    $S_{\nu}$\,${\propto}$\,${{\nu}^{-\alpha}}$ with $\alpha$=0.8, 
    and in the V- to UV-bands $S_{\nu}$\,${\propto}$\,${{\nu}^{\-\alpha}}$ with
    $\alpha$=1.3. The spectrum is sketchily shown in Figure 2 (left panel).
    This kind of broken power-law spectrum can be resulted from local
    continuous injection or re-acceleration of relativistic electrons in the 
    superluminal optical knots  under synchrotron/IC radiative
    losses (Kardashev \cite{Ka62}, Pacholczyk \cite{Pa70},
     Qian \cite{Qi78}, \cite{Qi96a},
    \cite{Qi96b}, \cite{Qi97}, Sahayanathan et al.
    \cite{Sa03}). The synchrotron  spectrum assumed here is 
    essentially different from the bubble-emitting bremsstrahlung spectrum 
    produced by the disk-impacting process. It is noted that the thermal 
    spectrum predicted by the disk-impact model was much flatter at optical-UV
    wavelengths. According to Valtonen \& Ciprini (\cite{Va12})
    and Valtonen et al. (\cite{Val12}), the thermal spectrum of the 2005 
    outburst was derived from the observed spectrum by correcting the host
    galaxy extinction with a hydrogen column density of
     6.3$\times{10}^{20}$/${\rm{cm}}^2$. However,
    if correction of extinction in the host galaxy were needed, the synchrotron
    spectrum of the March/2016 outburst would also be converted into a thermal
    spectrum. This seems unlikely. Thus the host galaxy extinction will not be
    included here and we suggest that both the December/2015 and March/2016 have
    similar non-thermal spectra.   
    \subsection{Spectral variability}
     As explained in Sect.2.1, the spectral energy distribution of the 
   December/2015 outburst (peaking at 2457360) in the NIR-optical-UV bands 
   exhibits two distinct
    features: the offset between the NIR and optical portions and  the 
    transition from the rather steep optical spectrum to a flatter 
     spectrum in the UV portion. If this NIR-optical-UV spectrum could be 
      interpreted as composed of two constituents: a thermal spectrum 
     emitted from the accretion disk of the primary black hole (with a mass of
      $\sim{2\times}{10^{10}}{\rm{M_{\odot}}}$) dominating the optical-UV
     emission and a synchrotron spectrum emitted from a shock in the jet
     dominating the IR-radio emission, one would have to explain why the two
     emitting sources could vary simultaneously. Moreover, the March/2016
      outburst (peaking at 2457450) has its spectrum and spectral variations
     very similar to those of the December/2015 outburst  and both outbursts 
     exhibit no color changes (Gupta et al. \cite{Gu16}). As the March/2016 
     outburst is a highly-polarized synchrotron one, its spectral 
     variations should not be related to the thermal emission from the
      primary disk and its color stability must be a characteristic feature 
    of the  synchrotron source itself. The color stability commonly observed 
     in both December/2015 and March/2016 outbursts may be a significant
     signature demonstrating the nature of their emission. Thus the 
    similarity in the spectral variations between the December/2015 and 
    March/2016 outbursts may imply that the December/2015 outburst also
     originate from  the jet. In fact, according to the disk-impact model,
     the December/2015
     outburst is caused by the evolving gas-bubble torn off the primary
     disk by the secondary hole impacting. The thermal emission from an 
     evolving gas-bubble during its adiabatic expansion and cooling 
     (from ${\sim}10^5$K to lower temperatures) would be color-changeable,
      inconsistent with  the observations. 
    \subsection{Symmetry in light curve profiles}
      As Sillanp\"a\"a et al. (\cite{Si96a}) pointed out that during
    the OJ-94 project period (1993.8--1996.1) the two major outbursts had a 
    strong symmetry. Through detailed inspection, we recognized that 
    the two major outbursts could be decomposed into a number of subbursts, 
    each having symmetric light curves with similar rising and declining
    timescales. A few isolated  moderate outbursts also have symmetric 
    light curves. Most interestingly, the December/2015 outburst peaking 
    at JD2457360 (Valtonen et al. \cite{Va16}) does not exhibit the 
    "standard light curve" expected for ``impact outbursts'' 
     (Valtonen et al.  \cite{Va11}),  but showing a symmetric profile.
     During the period of September/2015\,--\,May/2017, a number of  rather 
    isolated moderate bursts (e.g., peaking at JD-2457379, -2457450, -2457759
     and -2457885; Valtonen et al. \cite{Va17}) were observed to exhibit 
     symmetric profiles.\\
      As discussed in Sect.4 below, the five periodic outbursts
    (in 1983.00, 1984.10, 1994.75, 2005.76 and 2007.70) could be decomposed 
    into a number of subflares with each subflare (or "elementary flare")
    having a symmetric light 
    curve. This can be clearly seen in Figures 10\,--\,12 and 14\,--\,15.\\
    Since symmetry in the light curves is a common characteristic feature,
   the  periodic and non-periodic outbursts in OJ287 should originate 
   from similar mechanisms. However, both the  
    evolving gas-bubble emitting mechanism (Lehto \& Valtonen \cite{Le96}) 
   and the shock-in-jet models (e.g., Marscher \& Gear \cite{Ma85}, 
    Qian \cite{Qi10}), could not produce outbursts with symmetric 
    light-curves. In this work, we would suggest that the symmetric
    light-curves observed in OJ287  are produced by lighthouse effect
    due to the helical motion of superluminal optical knots in
    the jet (Qian \cite{Qi15}). 
      \subsection{Connection between radio and optical flares}
     Investigations of the connection between the radio and optical  variations
     may provide important clues for the entire phenomena in OJ287 and help to 
    understand the nature of its multi-waveband emissions. Centimeter
    radio bursts (e.g., at 8\,GHz) are typically observed to be delayed with 
    respect to the optical outbursts by a month or so.  The bump-like
    structure in the radio light curves are connected with  the spike-like 
    structure in the optical light-curves (Britzen et al. \cite{Br18},
     Tateyama et al. \cite{Ta99}, Qian \cite{QiXiv18}). 
    This kind of radio-optical connection can be 
   understood as a result of the evolution of the superluminal optical knots
   (shocks or blobs) combined with the opacity effects at radio 
   frequencies (Qian \cite{Qi10}). However, simultaneous flares have been
   observed  in OJ287 at millimeter and optical wavelengths. For example, 
   Sillanp\"a\"a (\cite{Si96b}) observed a simultaneous  behavior at the
   beginning of the year 1992 (JD2448610\,--\,JD2448670): the variations at
   optical V-band and at 37\,GHz  were not only simultaneous but also had 
   very similar profiles. According to Valtaoja et al. (\cite{Val20}), during 
   the period 1990.5--1994.0, the 37\,GHz variations  were mostly simultaneous
   with the optical variations with no measurable time delays. In addition, the
   major 37\,GHz outburst (peaking in 1996.61) has an approximately symmetric
    profile with its declining phase closely tracking the optical flare.\\
    Most interestingly, the two major optical flares of the double-peaked 
   outbursts during 1994.7--1996.1 had different connections between the
   millimeter flares and the optical flares. For the first optical flare 
  (1994.8--1995.3) there was no  simultaneous 37\,GHz counterpart observed
  (Sillanp\"a\"a et al. \cite{Si96b}). But for the second optical flare
  (1995.90--1996.10) simultaneous 37\,GHz and optical variations 
   were observed. According to Valtaoja et al. (\cite{Val20}, Fig.8 therein),
    both the optical and millimeter flares have complex multi-component 
   structures.  The millimeter
    and the optical variations are not only simultaneous but also have very
   similar envelopes. Thus both the optical and radio/mm light curves
    could be decomposed into a number of subflares (or elementary flares)
   with symmetric profiles and 
   interpreted in terms of the helical motion model. \\
     \begin{figure*}
     \centering
     \includegraphics[width=6cm,angle=-90]{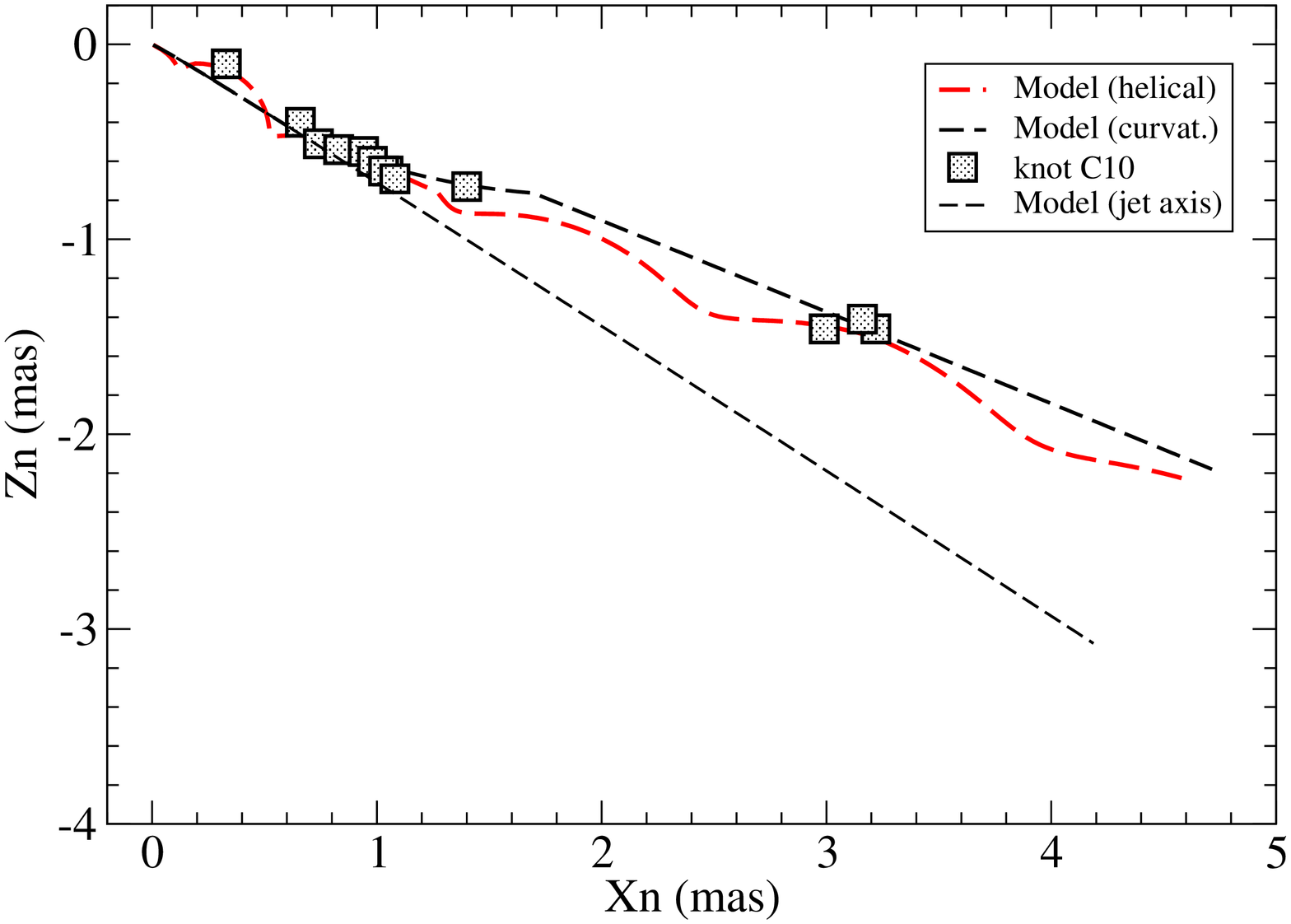}
     \includegraphics[width=6cm,angle=-90]{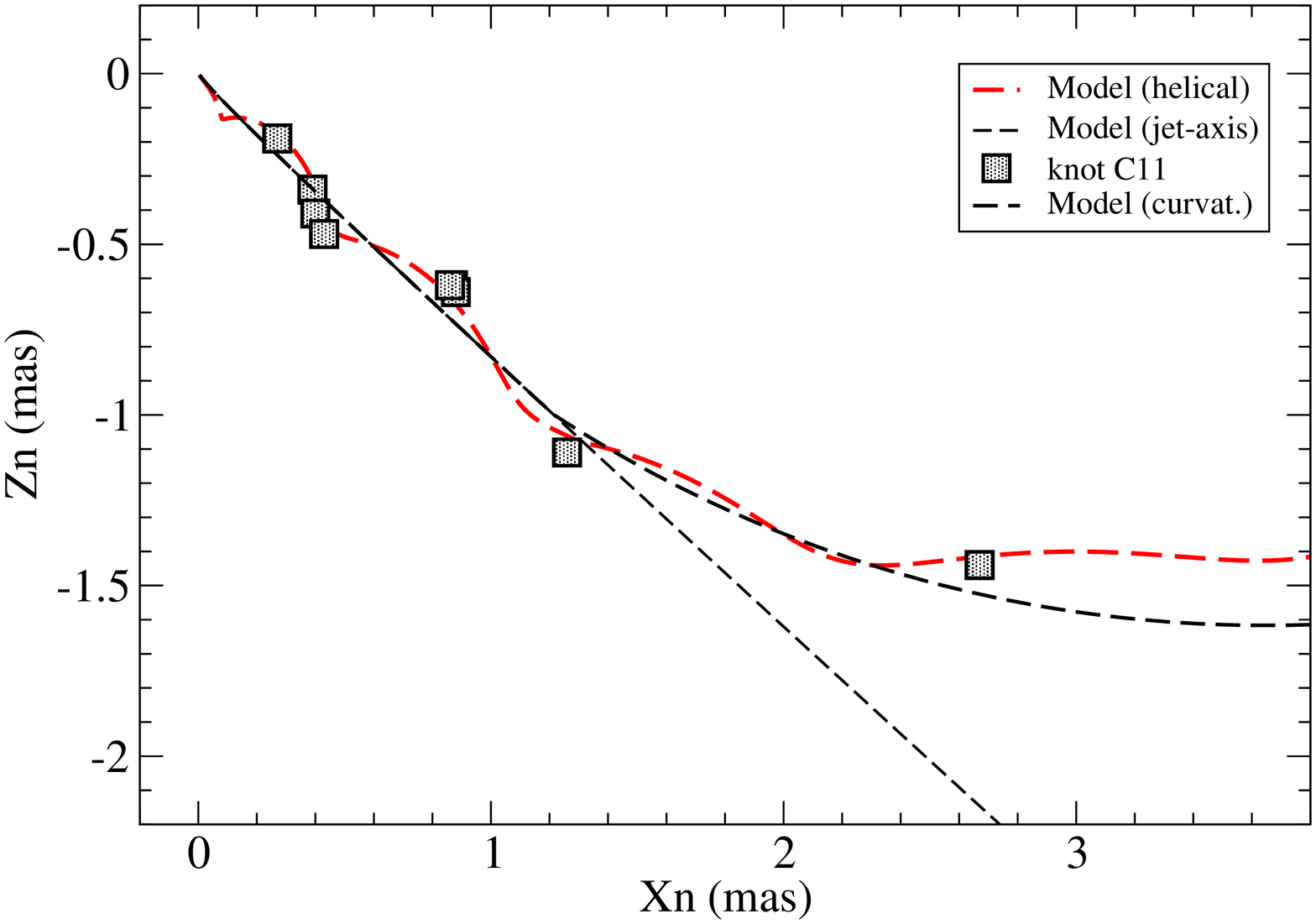}
     \caption{Model simulation of the superluminal motion of knot C10 (left
    panel) and knot C11 (right panel). The black  thin dashed lines denote the 
    precessing common trajectories. The black thick dashed lines represent
    the model trajectories with outer curvatures. the red dashed lines indicate
   the helical trajectory fits. The helical motions of both knots have very
   small pitch angles on parsec scales (Qian \cite{QiXiv18}).}
     \end{figure*}
   Obviously, the close connection between the millimeter and optical flares 
   observed in the 1995.9 periodic outburst seems important, implying that:
   (1) The optical and radio flares should be produced in 
   co-spatial emitting regions\footnote{Here ``co-spatial'' would mean some
    special structure (or emission distribution) of the optical-millimeter 
   source in the direction perpendicular to its motion.}and originate from
    a common synchrotron process in the relativistic jet; (2) The 
   simultaneity in the millimeter and optical variations may disfavor 
   shock-in-jet models, because optical shocks typically evolve through 
   three stages (Compton\,-\,synchrotron\,-\,adiabatic stages), resulting
    in different optical-radio relationships (Qian et al. \cite{Qi10},
    Qian \cite{Qi96a}, \cite{Qi96b}, \cite{Qi97},
     Marscher \& Gear \cite{Ma85}, Valtaoja et al. \cite{Val88}, 
   \cite{Val92},  Litchfield et al. \cite{Li95}).\footnote{This does not 
    exclude the possibility that the superluminal components are steady shocks
    on time-scales of ten days or so.}
    Thus shock-in-jet models seem not able to produce simultaneous mm-radio and
   optical variations with very similar symmetric light curves.
    We would suggest that lighthouse effect due to helical motion of
    superluminal optical knots (shocks or blobs) may be the most plausible
     mechanism to interpret the simultaneity and symmetry observed in the
   mm-radio and optical variations observed in OJ287. The proposed helical
    motion model seems working  well, as described in Sect. 3 and 4.\\
   We point out that helical motion  of radio superluminal knots might also
    exist. For example, model-simulation of  the kinematics of the 
   superluminal components C10 and C11 of OJ287 in terms of helical models
   has been tried in Qian (\cite{QiXiv18}), which are  shown in Figure 3. 
   These helical trajectories have very small pitch angles and  could not be
   easily discovered. It should be noted that in the precessing jet-nozzle
    models (Qian et al. \cite{Qi19}) the flux density  curves of the optical 
   and radio knots can be explained in terms of  lighthouse effect caused by
   their helical motions, but different knots may move along  different
   helical trajectories. This is different from the Doppler boosting and 
   beaming effects caused by the precession of the whole jet 
   (with a $\sim$12\,yr period) which do not contribute to the
   radio and optical flaring activities with timescales of ten days or so.
   \begin{table}
    \caption{Parameters for model simulation of the R-band light curves of the
    periodic outburst in December/2015 (peaking at JD2457360) and the 
    March/2016 outburst (peaking at JD2457450). $\Gamma$ -- Lorentz factor
    of the superluminal optical knot, ${\delta}_{max}$ -- maximum Doppler
    factor, ratio ${\delta}_{max}$/${\delta}_{min}$, 
   $\rm{S_{int}}$($\rm{mJy}$) -- intrinsic flux density of the optical knot,
    base-level (underlying jet) flux density $\rm{S_b}$=3.5$\rm{mJy}$ 
   at R-band, FWHM (full width at half maximum) of the modeled light curve
     (day). $t$\,=\,flare time\,=\,day-2457000.}
    \begin{flushleft}
    \centering
    \begin{tabular}{llllll}
    \hline
    $t$ & $\Gamma$ & ${\delta}_{max}$ & ratio & $\rm{S_{int}}$ & FWHM \\
    \hline
    360 & 9.5 & 18.88 & 4.11 & 1.16$\times{10^{-4}}$ & 5.9 \\
    450 & 9.5 & 18.88 & 4.11 & 1.16$\times{10^{-4}}$ & 5.9 \\
     \hline
    \end{tabular}
    \end{flushleft}
    \end{table}
    \begin{figure*}
   \centering
   \includegraphics[width=6cm,angle=-90]{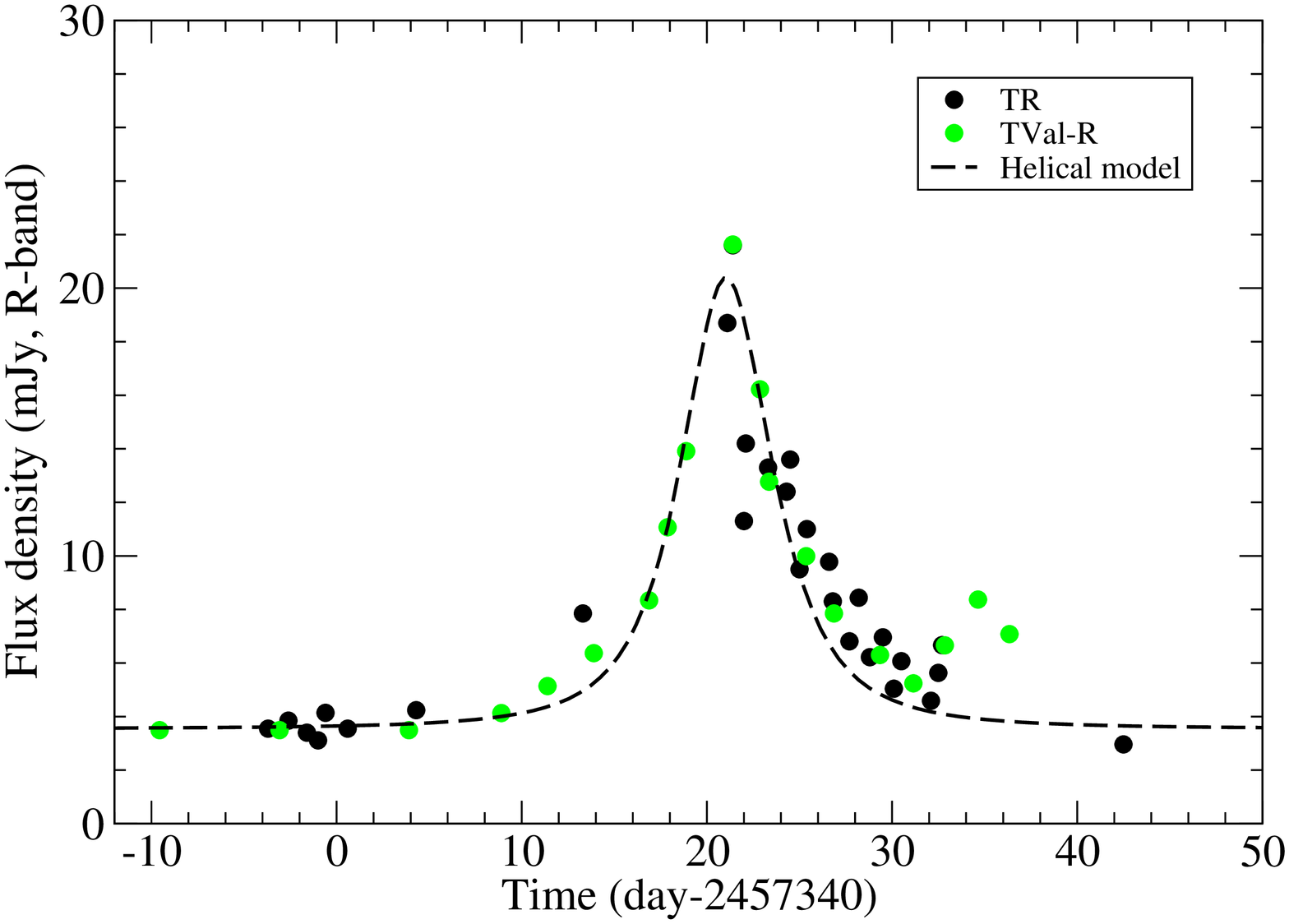}
   \includegraphics[width=6cm,angle=-90]{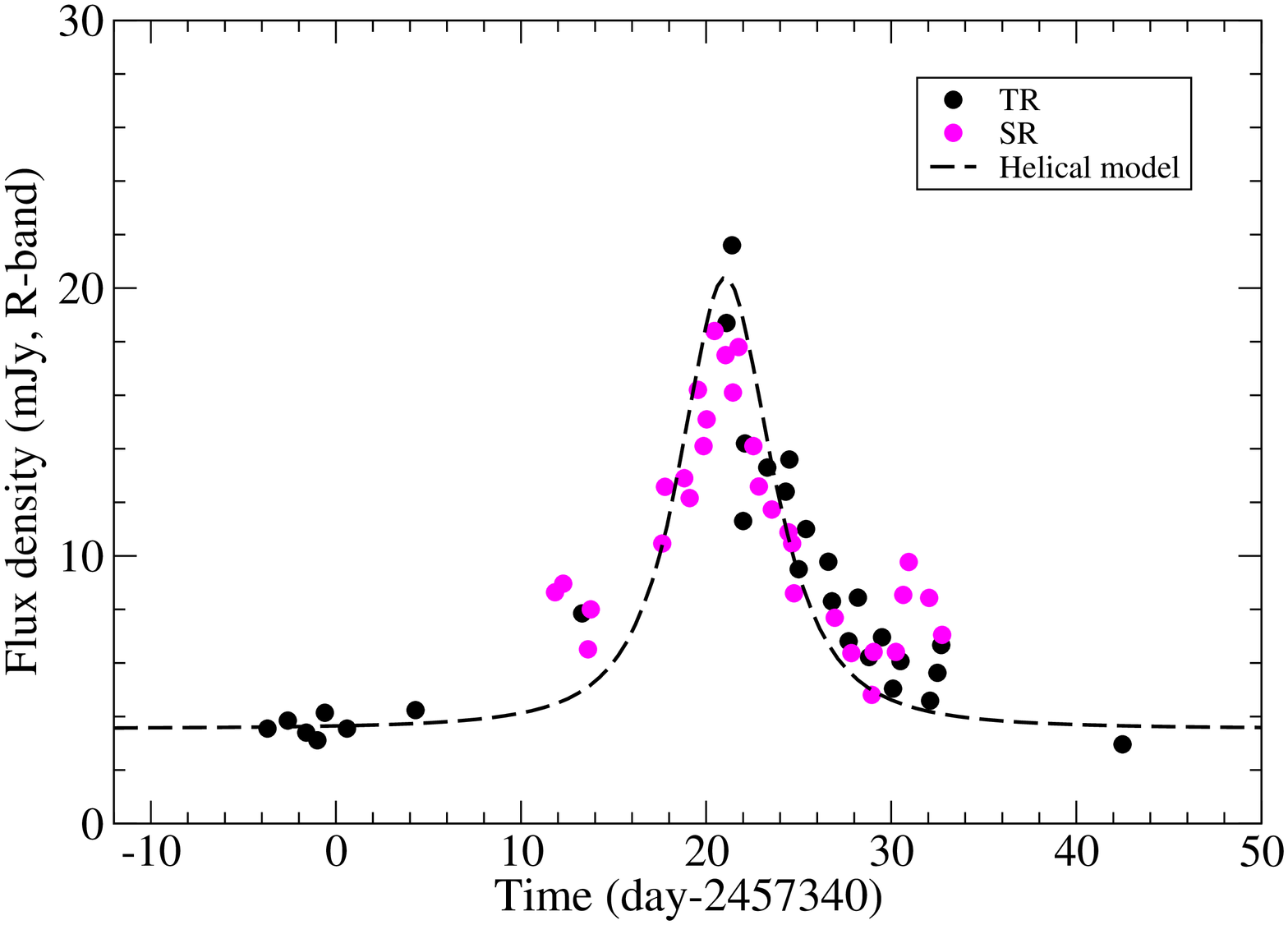}
   \caption{Left panel: model simulation of the R-band light-curves for the
   outburst in December/2015 observed by Kushwaha et al. (\cite{Ku18}; 
   labeled by "TR") and observed by Valtonen et al. (\cite{Va16}; labeled 
   by "TVal-R"). Right panel: model simulation of the R-band light curves for
   the combination of December/2015 outburst and March/2016 outburst
    (Kushwaha et al. \cite{Ku18}; labeled
   by "SR"). The light curve of the March/2016 outburst has been shifted 
   in time backward  by 89.4 days. The combination of the light-curves provides
   a sufficient time coverage  to clearly exhibit the symmetric profiles with
   similar rising and declining time scales.}
     \end{figure*}
    \begin{figure*}
    \centering
    \includegraphics[width=5.5cm,angle=-90]{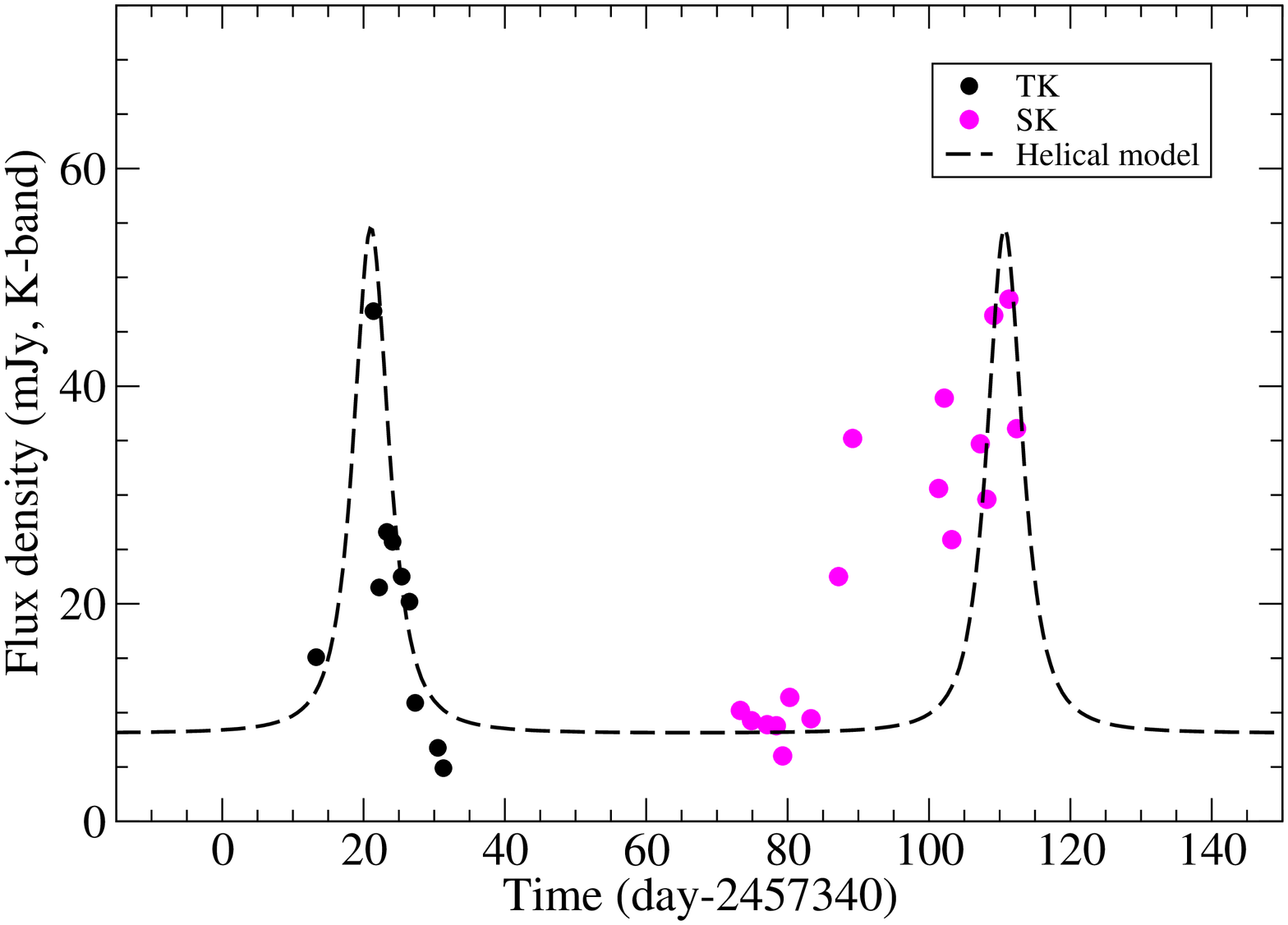}
    \includegraphics[width=5.5cm,angle=-90]{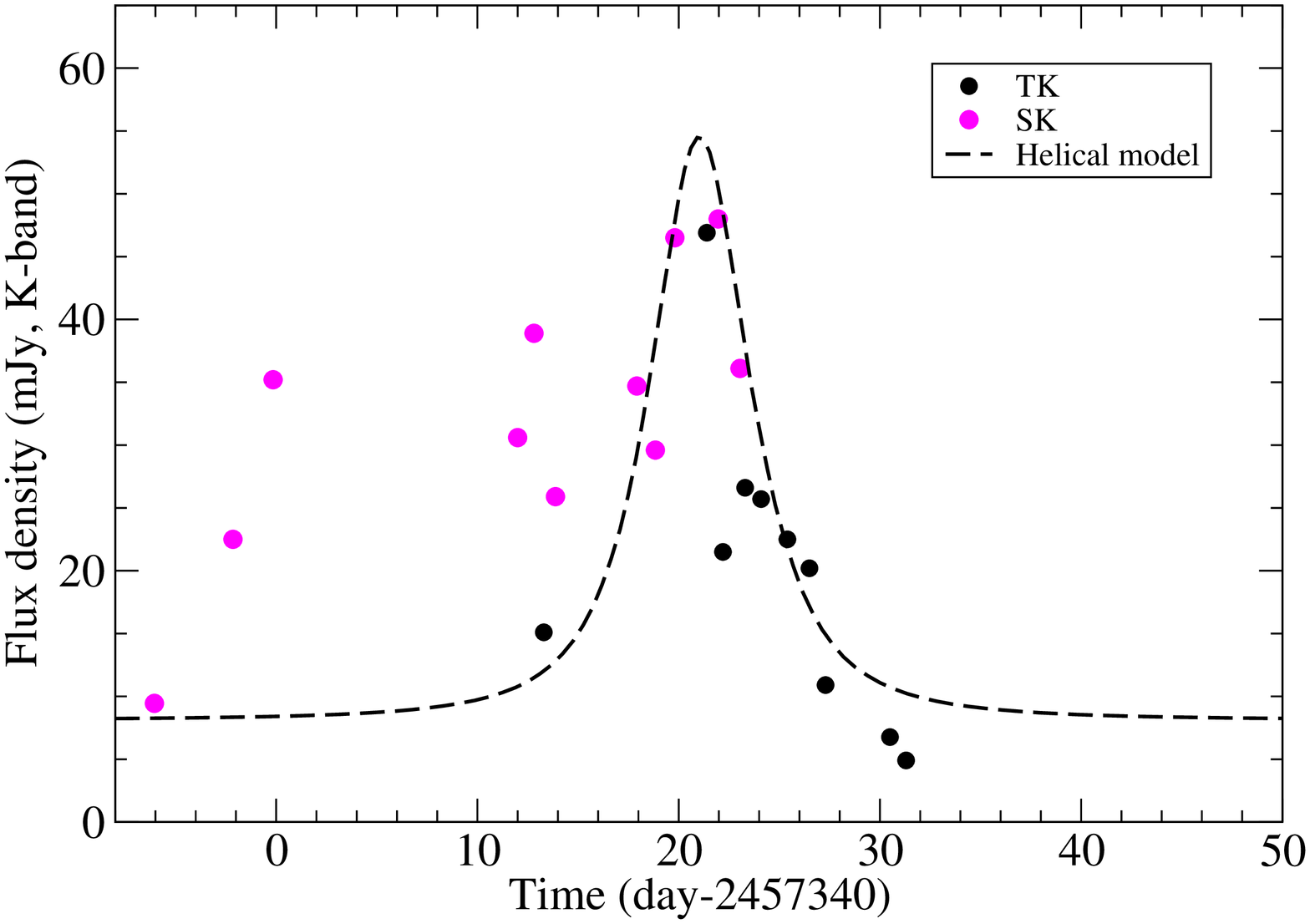}
    \caption{Model simulation of the K-band light curves 
    for the outbursts in December/2015 (labeled by ``TK'') and in March/2016 
    (labeled by ``SK''). In the right panel the observing time
     of the March/2016 outburst has been shifted backward by 89.4 days.
      The light curves are very well simulated for both outbursts by the 
   lighthouse model.}
    \end{figure*}
    \begin{figure*}
    \centering
    \includegraphics[width=6cm,angle=-90]{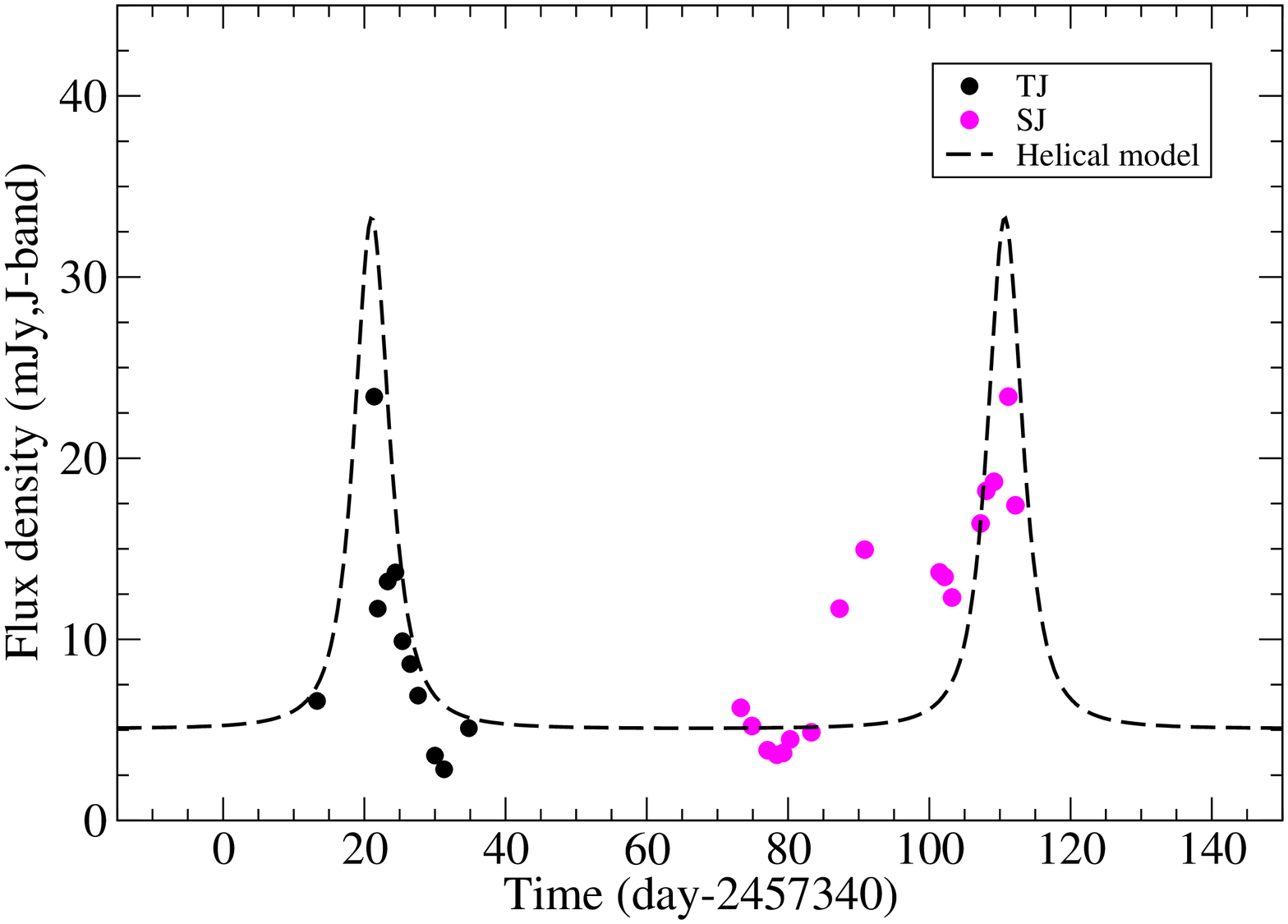}
    \includegraphics[width=6cm,angle=-90]{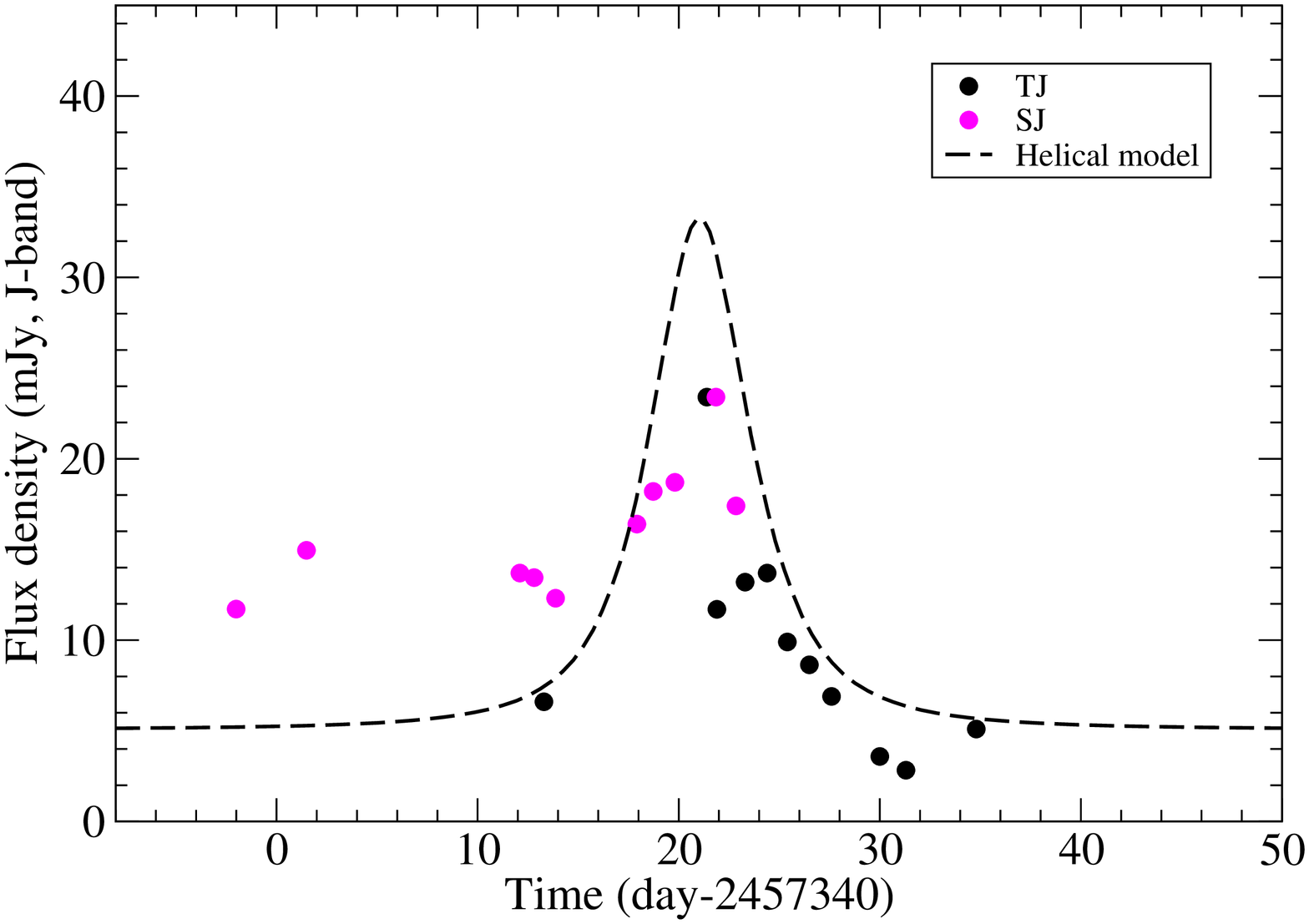}
    \includegraphics[width=6cm,angle=-90]{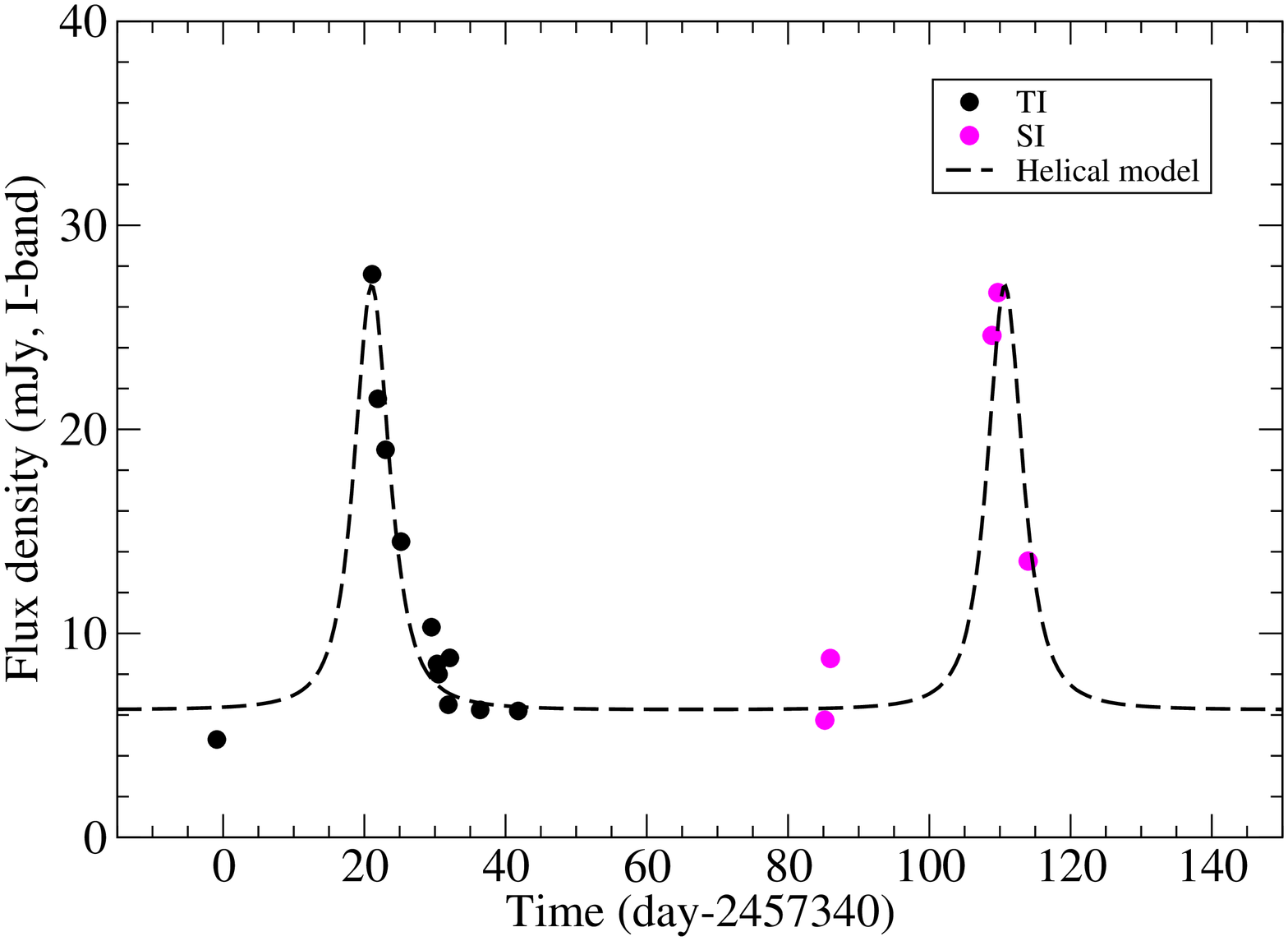}
    \includegraphics[width=6cm,angle=-90]{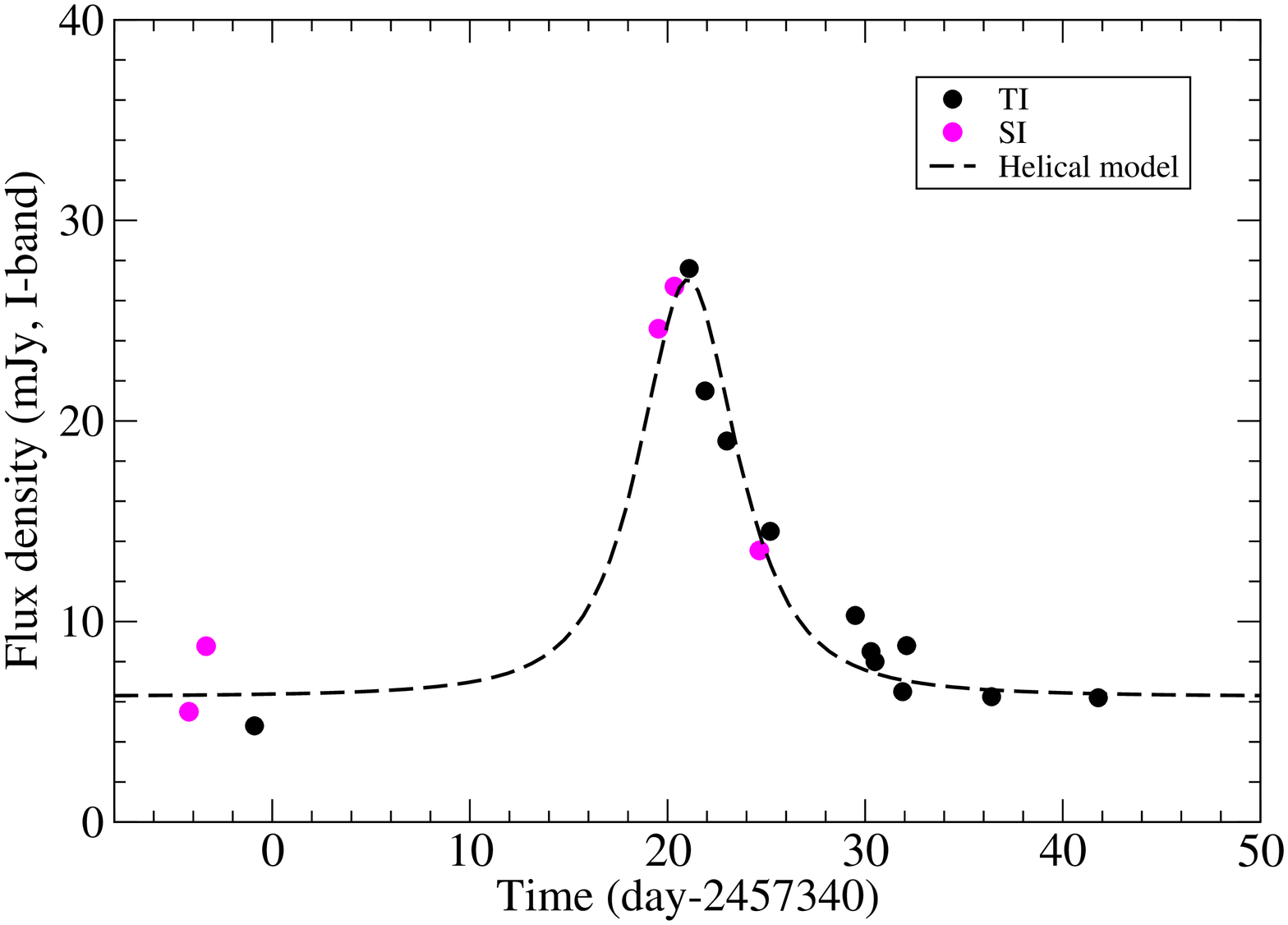}
    \caption{Model simulation of the J-band (top panels) and I-band 
    (bottom panels) light curves for the outbursts in December/2015 (labeled by
    ``TJ'' and ``TI'') and in March/2016 (labeled by ``SJ'' and ``SI'').
    In the right panels the 
    observing time of the March/2016 outburst has been shifted backward 
    by 89.4 days. The observed I-band light curves are very well fitted by the 
    helical motion model. The peak of the model light curve for the J-band
    is much higher than the observed one and this should have been expected,
    because the assumed model spectral index at J-band $\alpha$=0.8
    is larger than the observed one.}
    \end{figure*}
    \begin{figure*}
    \centering
    \includegraphics[width=6cm,angle=-90]{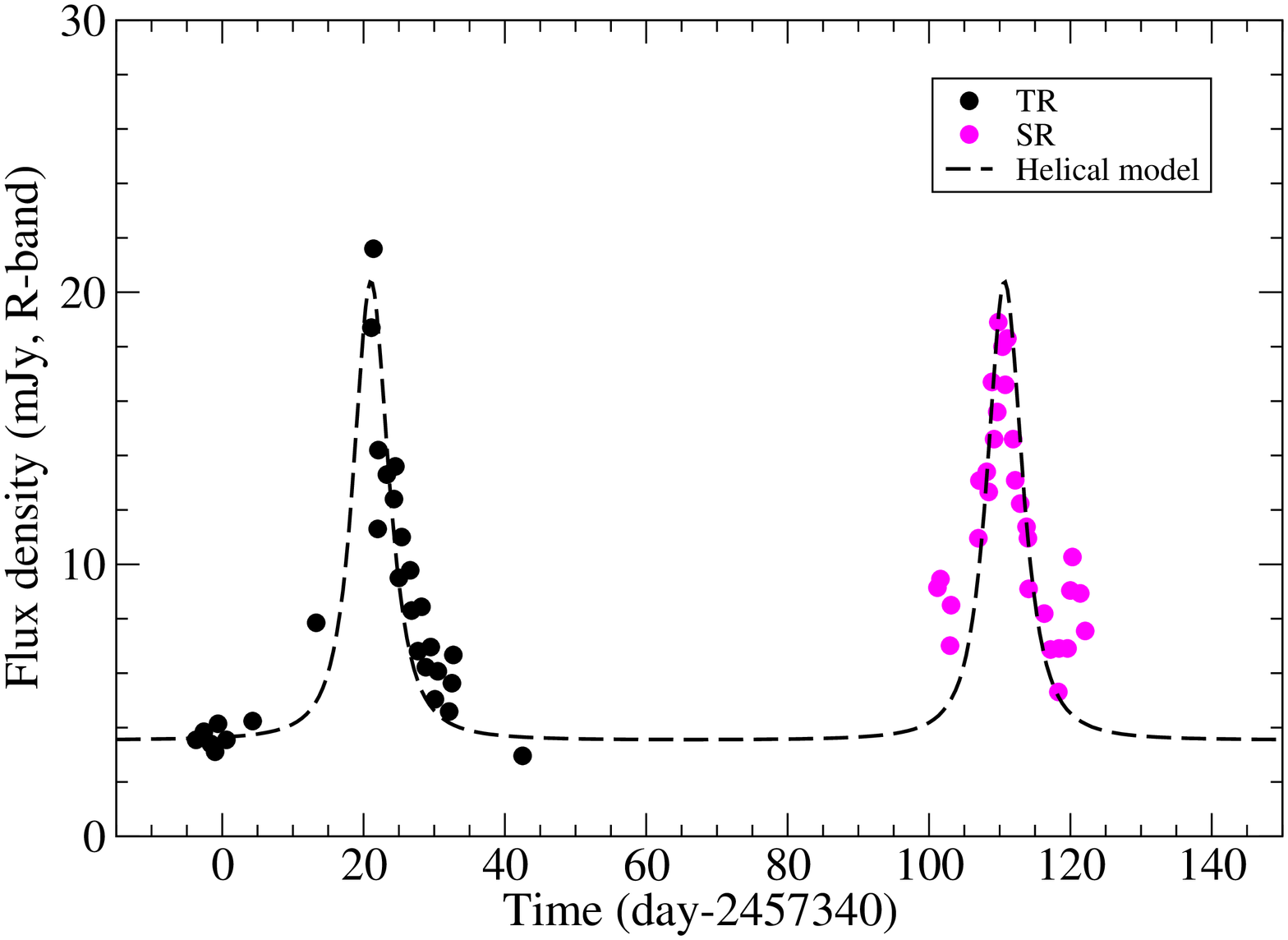}
    \includegraphics[width=6cm,angle=-90]{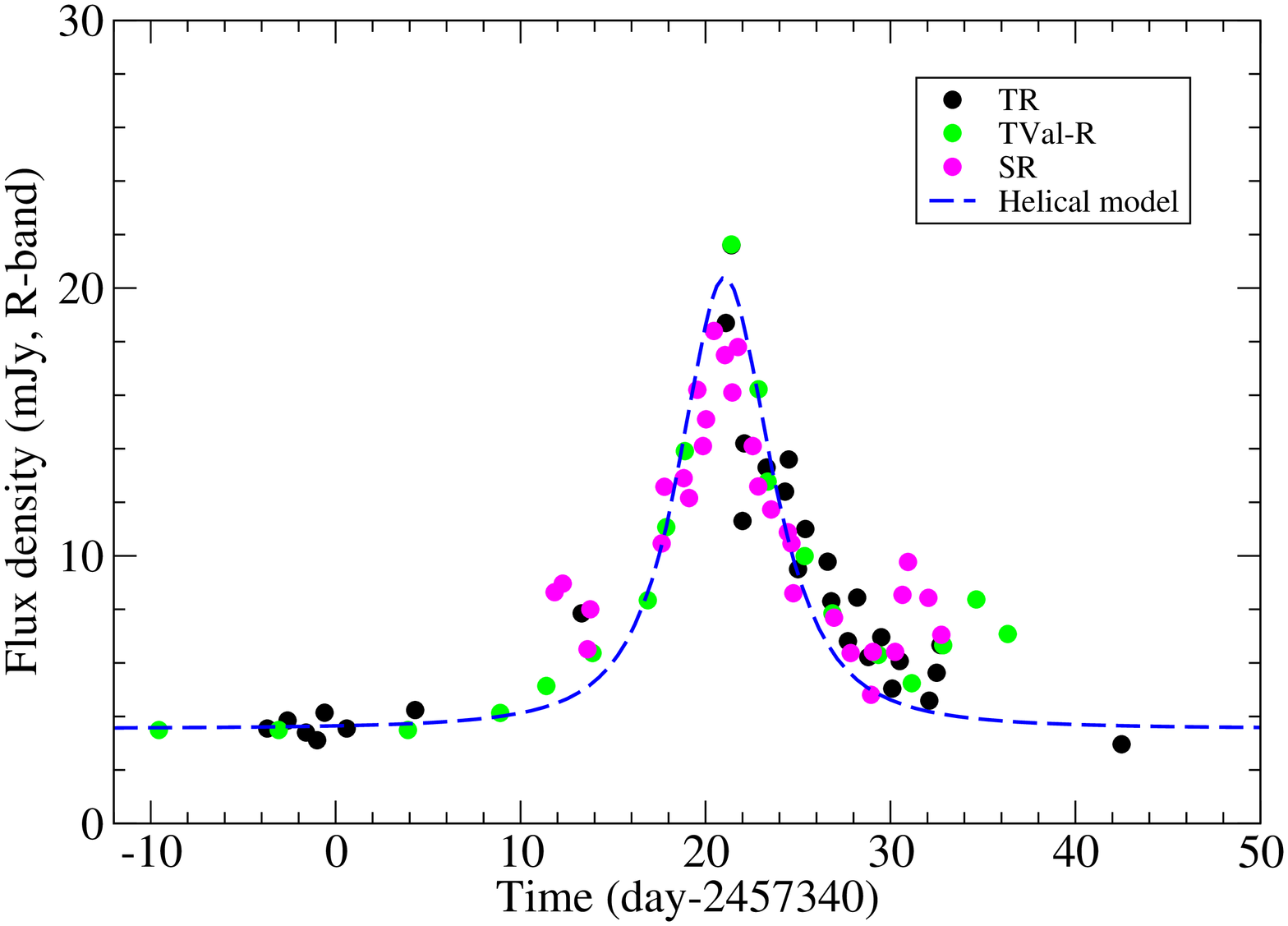}
   \includegraphics[width=6cm,angle=-90]{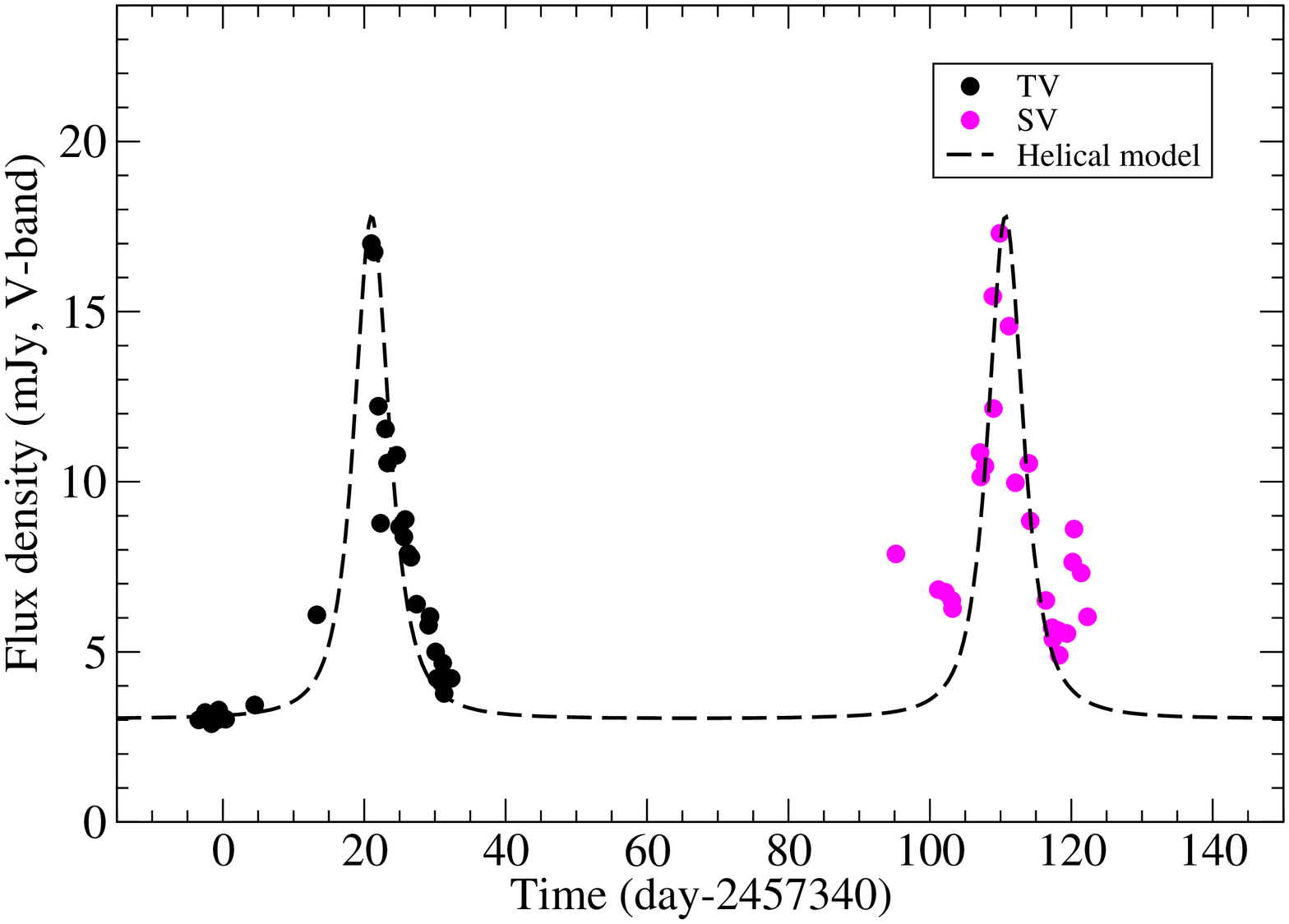}
    \includegraphics[width=6cm,angle=-90]{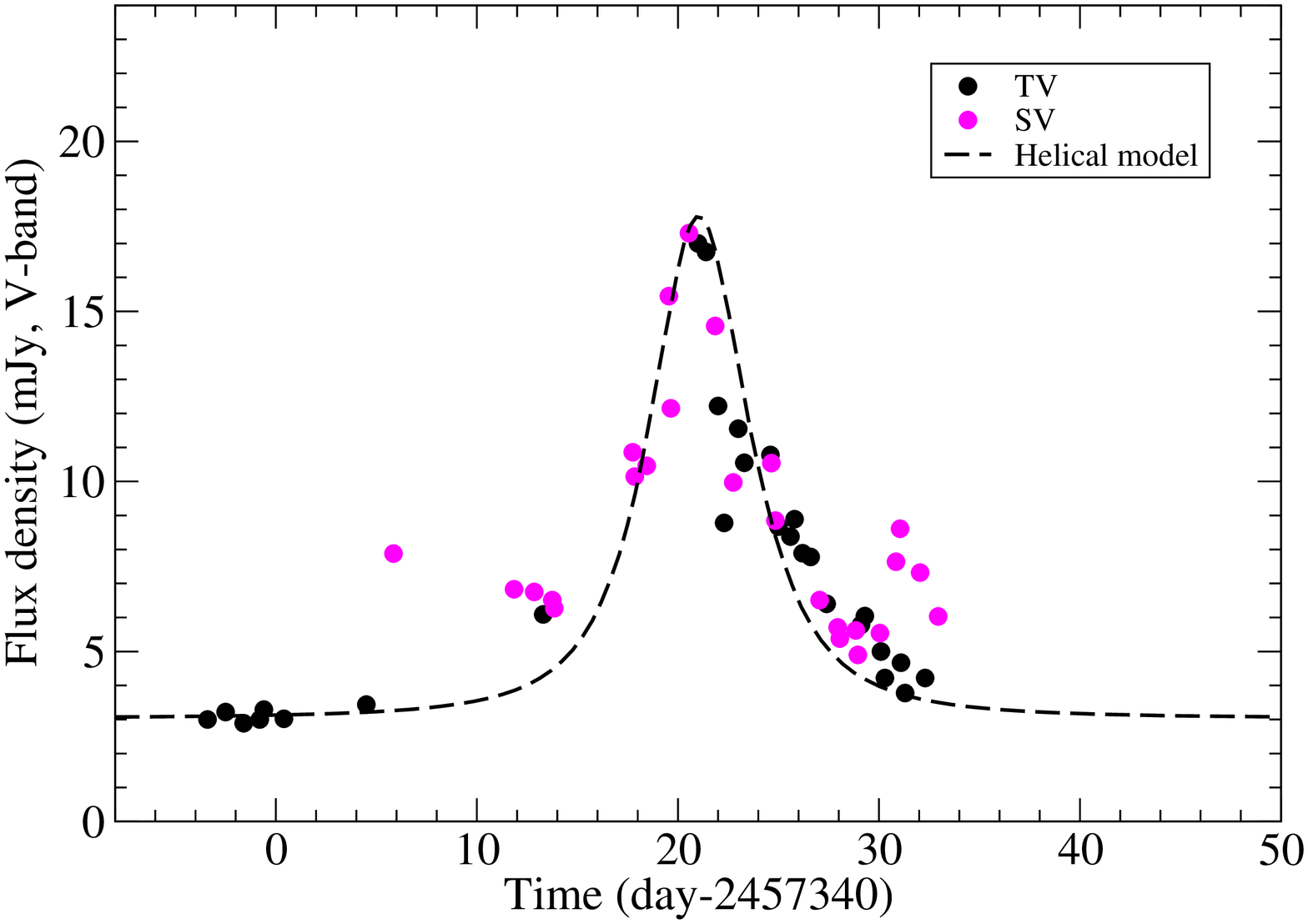}
    \caption{Model simulation of the R-band (top panels) and V-band 
    (bottom panels) light curves for the outbursts in
    December/2015 and in March/2016. In the right panels the observing time 
    of the March/2016 outburst has been shifted backward by 89.4 days. 
    The light curves at both bands are very well simulated by the helical 
    motion model. The good fits to the combined light curves (right panels)
    with a  common helical motion model indicate the strong similarity 
    in optical variations between the December/2015 and the synchrotron 
    outburst in March/2016 and their common radiation mechanism.}
    \end{figure*}
    \begin{figure*}
    \centering
    \includegraphics[width=6cm,angle=-90]{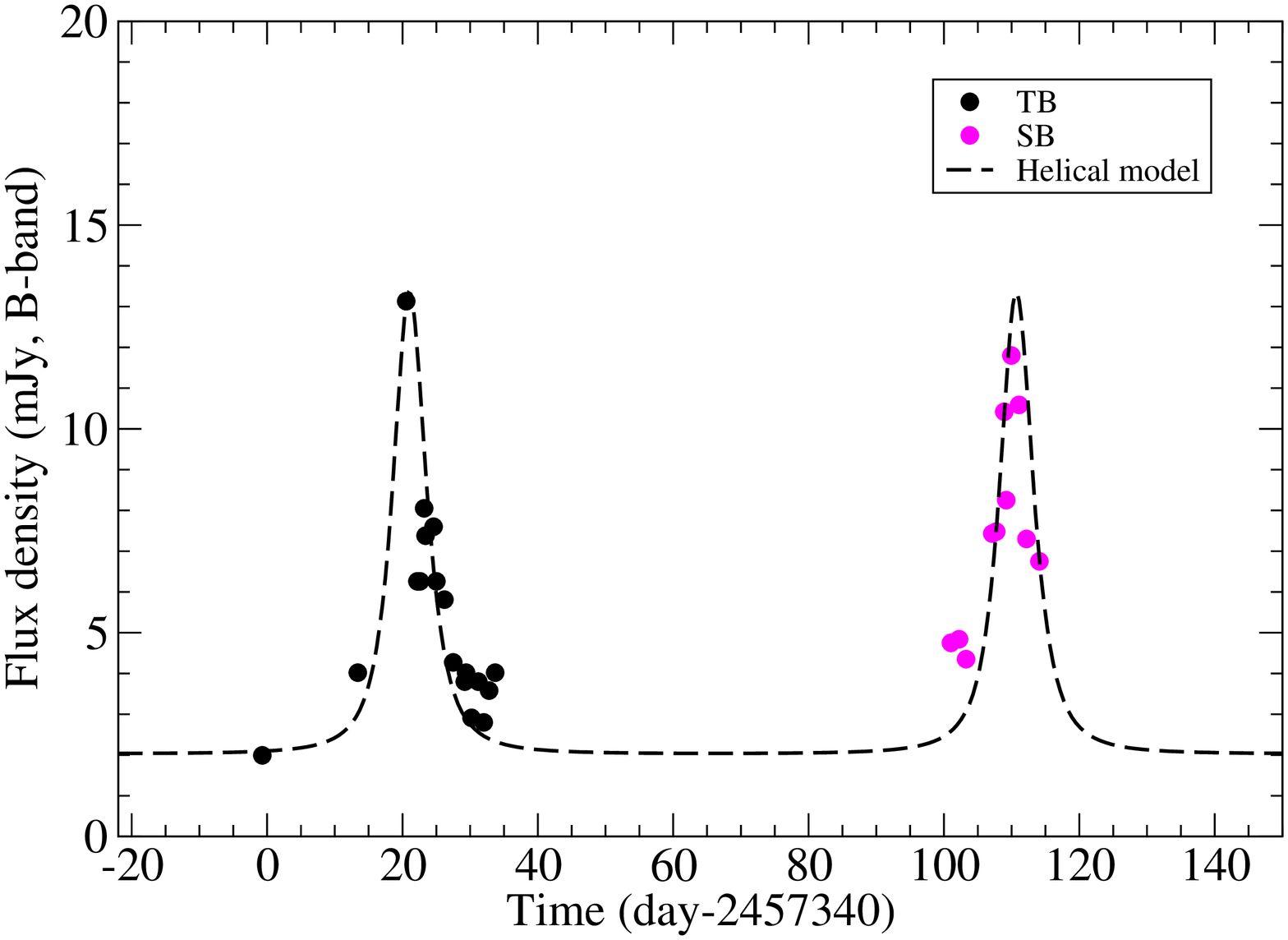}
    \includegraphics[width=6cm,angle=-90]{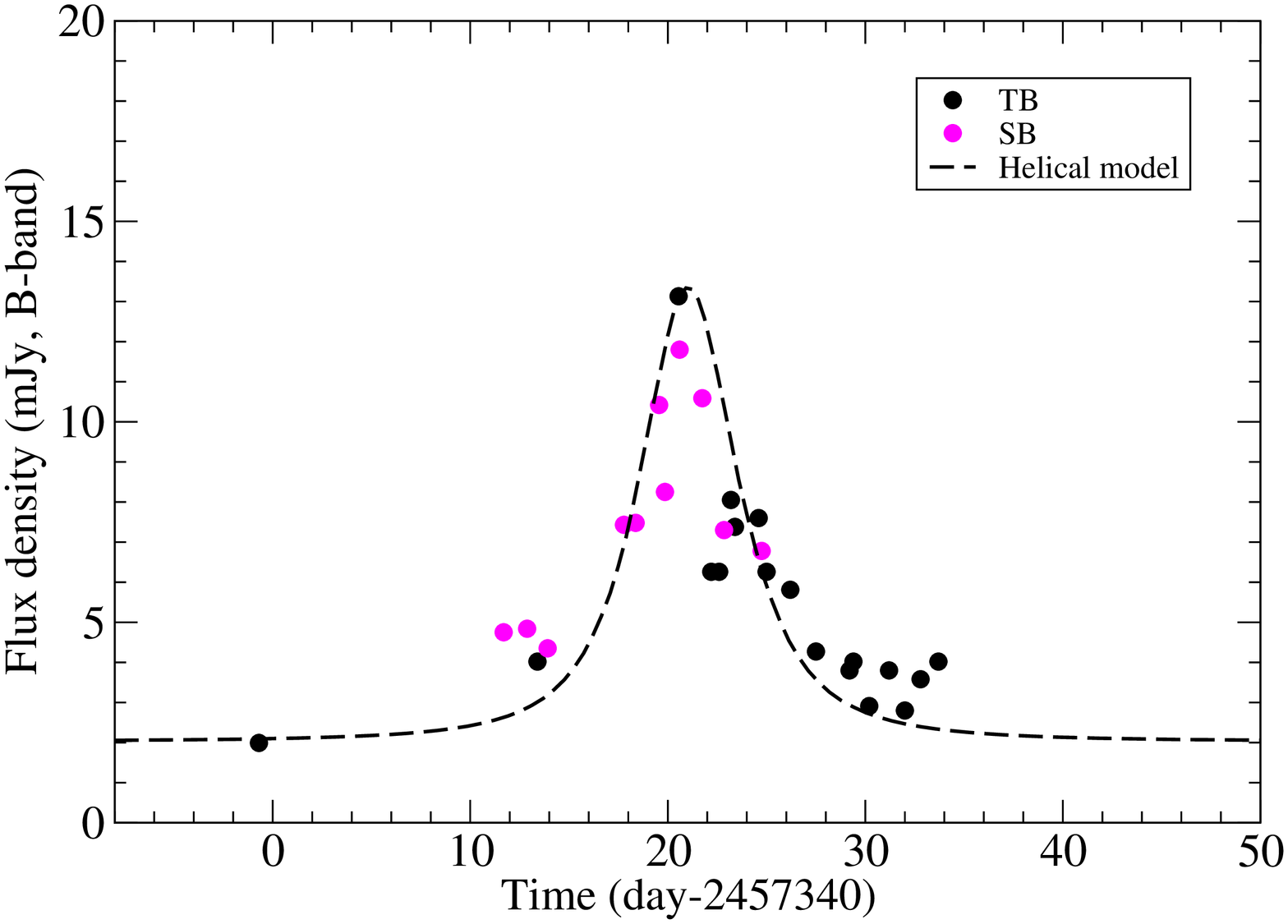}
     \includegraphics[width=6cm,angle=-90]{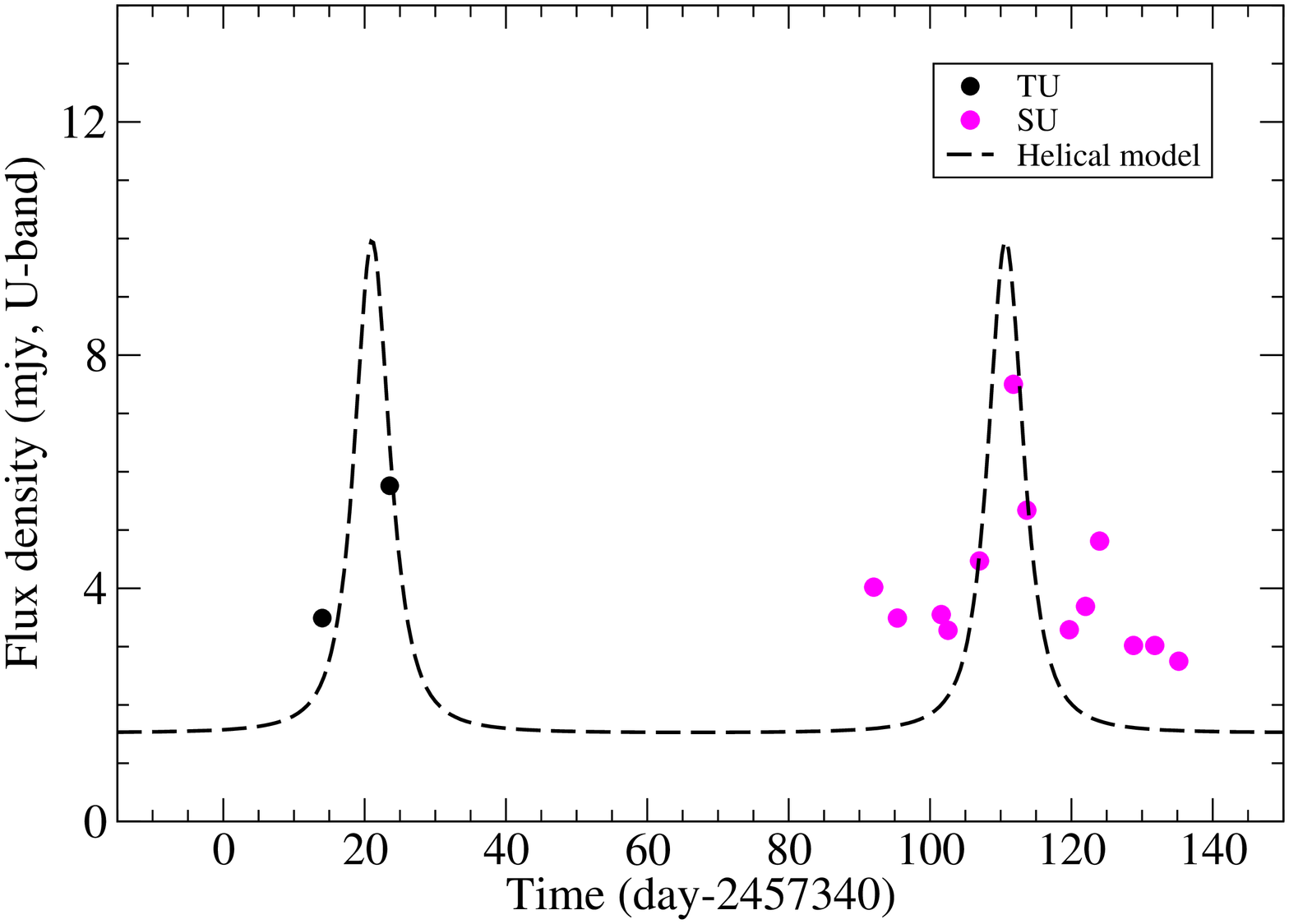}
    \includegraphics[width=6cm,angle=-90]{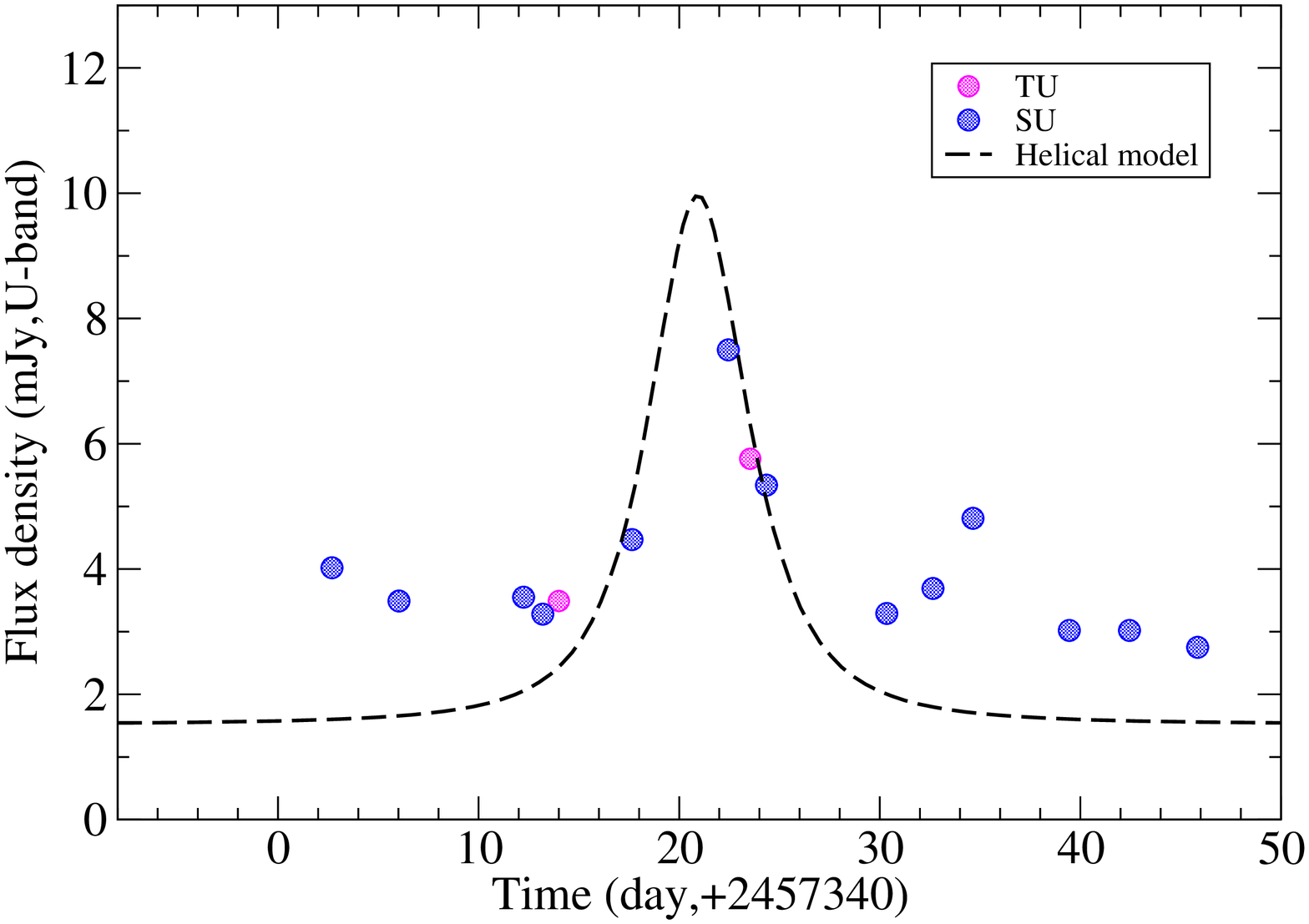}
    \caption{Model simulation of the B-band (top panels) and  U-band 
    (bottom panels) light curves for the outbursts in December/2015 and 
   in March/2016. In the right panels the observing time of the March/2016 
    outburst has been shifted backward by 89.4 days. The good fits of the 
    combined light curves in both bands demonstrate  the applicability of 
    the helical-motion model in the high-frequency region, revealing
    the December/2015 outburst has a variability behavior very similar to
    that of the synchrotron outburst in March/2016.}
    \end{figure*}
        \subsection{Lighthouse effect and Doppler boosting} 
   We will explain the multi-wavelength light curves of the optical
   outbursts in terms of lighthouse effect. For simplicity,  it is  assumed 
   that the superluminal optical knots move along helical trajectories around
   the rectilinear jet axis which precesses around the precession axis, 
   as sketchily shown in Figure 1 (left panel). In this case the 
   lighthouse effect results in a symmetric light curve via Doppler 
   boosting per revolution.  In addition, we assume that the observed 
   flux density $\rm{S_{obs}}$ of the 
   optical outbursts at any frequency
   consists of two constituents: a steady base-level ($\rm{S_b}$) and the 
   flaring part $\rm{S}$(t):
    \begin{equation}
   {\rm{S_{obs}}}(t)={\rm{S(t)}}+{\rm{S_b}}
    \end{equation}
   Using relativistic jet models, the evolution of the flux 
   density of superluminal optical knots can be written as:\\ 
   $\rm{{S}(t)}$=${\rm{S_{int}}}$$\times$${\rm{{\delta}(t)}^{p+\alpha}}$. 
   $\rm{S_{int}}$ is the intrinsic flux density (in the comoving frame of 
   the optical knots). For moving optical knots p=3 (Blandford \& K\"onigl
    \cite{Bl79}) and $\alpha$ is the spectral 
   index. In our model simulation $\rm{S_{int}}$ is assumed to be constant
   \footnote{${\rm{S_{int}}}$=constant is a simplified assumption. Taking the 
    rising and declining parts of the outbursts into account would result in 
   truncations of the model light curves at both start- and end-points,
   more clearly separating the contributions from consecutive subbursts.} and 
   the spectral index in V-band equals to 1.0. The broken power-law spectrum in
   the NIR-optical-UV bands is
   assumed as: $\alpha$=0.8 in the NIR\,--\,optical (V) bands and 
   $\alpha$=1.3 in the optical (V)\,--\,UV bands. The steady base-level
    spectrum $\rm{{S_b}}$($\nu$) is listed in Table 3 and shown in Figure 2
    (middle panel). The modeled intrinsic flux densities $\rm{S_{int}}{(\nu)}$ 
    for the optical knot are shown in Figure 2 (right panel) 
   for I-, V- and U-bands as a demonstrating example. The multi-wavelength
   light curves of the optical knot are only determined by the Doppler
    boosting.\\
    In the simulation of the outburst light curves the modeled flux 
    density of the outbursts (at V-band) will vary in the range from 
   $\rm{{S_{int}}{{\delta}^4}_{min}}$ to $\rm{{S_{int}}{{\delta}^4}_{max}}$,
    while the total flux density in the range from 
    $\rm{S_b}$+$\rm{{S_{int}}{{\delta}^4}_{min}}$
    to $\rm{S_b}$+$\rm{{S_{int}}{{\delta}^4}_{max}}$. Because the 
    base-level flux $\rm{S_b}$ does not
   vary during the outbursts, the variability amplitude of the total
    flux will be smaller than that of the flaring components. 
    This shows the characteristic feature of our 
   precessing nozzle model distinct from the precessing jet models 
   usually used in literature, where the precession of the entire jets will
   also cause the variations in the base-level flux density. 
     \subsection{Selection of parameters for simulation} 
    In this work we shall use simple approaches to make model simulation 
   of the observed light curves for all the outbursts concerned, that is, 
   all the model parameters listed in Table 1 are assumed to be constant.
    The parameters describing the spectral features of the outbursts 
    (the broken power-law spectra) are also taken to be constant.  
    Notwithstanding this simplification, the variability behaviors of all
   the outbursts  (both periodic and non-periodic) can be well 
   interpreted in terms of our lighthouse scenario. At the same time,
    there remains a wide scope for  choosing the model parameters to improve
   the simulation for any individual outburst. For example, an adoption of a
   different value for $\omega$ can consider the helical motion (and lighthouse
    effect) at a different precession phase. Changes in parameters $A_0$ and 
   d$\phi$(\rm{z})/dz can be used to investigate various patterns of  
   helical motion of the superluminal optical knots. Specifically, a local
  slight change of the spectral index at J-band could result in a better fit to 
   the J-band flux densities observed in the December/2015 and March/2016 
   outbursts (see Fig.6, upper panels). In addition, intraday
   variations (IDV) could also be included for explaining the rapid 
   variations, which might be  due to interstellar
    scintillation or turbulent fluctuations in the emitting sources
    (Qian et al. \cite{Qi91b},
   Melrose \cite{Me94}, Marscher et al. \cite{Ma08}, Marscher \cite{Ma14}).
   It is found from the light curve simulations that intraday spectral 
   variations might be an important ingredient, which could  result
    in the data-points deviating from the model light curves.\\
    Therefore through adjusting the model parameters, the model-fits to
   the light curves of all the outbursts discussed in this paper could be
   further improved.
    \subsection{Relevant MHD theories}
    Most astrophysicists believe that relativistic jets are formed by rotating
    magnetic fields in the magnetospheres of the black-hole/accretion-disk 
    systems, as the magnetohydrodynamic (MHD) theories of jet formation 
    indicate (Blandford \& Znajek \cite{Bl77}, Blandford \& Payne \cite{Bl82},
    Camenzind \cite{Cam90}, Beskin \cite{Be10}, Meier \cite{Mei01}, Vlahakis \&
    K\"onigl \cite{Vl04}). However, few observations have provided direct and 
    compelling evidence for helical magnetic fields and helical motions on 
    parsec scales (e.g., Gabuzda  et al. \cite{Ga04}, \cite{Ga15}).
    Now, as demonstrated in this work,  the 
    prevailing symmetric properties of the light curves of the optical 
    outbursts observed in OJ287 may be recognized as a favorable evidence of 
    helical motion of its superluminal optical knots along helical magnetic
    fields in the relativistic jet. This would also help to investigate the
    helical motion of superluminal knots in other blazars. However, for
    OJ287, the interpretation of the  
    quasi-periodicity in its variability behavior and the timing of the 
    double-peaked  outbursts remains to be a challenge.
    \subsection{A brief summary}
     For understanding the phenomena observed in blazar OJ287, comparison of 
    the emission properties of the periodic outbursts (so claimed ``disk-impact
    outbursts'') and the non-periodic optical outbursts (normal synchrotron 
    outbursts) may be an appropriate approach.
     Based on the above assumptions we will be able to
    model simulate the multi-wavelength light curves of both the December/2015
    and March/2016 using a very simple model, which only involves a 
    superluminal optical knot (with a steady broken power-law synchrotron
    spectrum) moving along a steady helical trajectory. As shown by the model
    simulation results given in the  next section, this simplified model is 
    already sufficient to explain the most of the basic properties of the 
    temporal and spectral variations of the two optical outbursts, showing 
    their very similar variability behaviors and the nature of their 
    optical/UV emission.  Both the  optical outbursts in December/2015 and
    March/2016 can be well interpreted as produced by the lighthouse effect 
    due to one superluminal optical knot moving along a helical trajectory
    through two helical revolutions: the 
    lighthouse effect firstly produces the December/2015 outburst and then 
    the March/2016 outburst 90-days later. This would imply, as suggested 
    in Qian (\cite{Qi15}), that there may exist a stable and perfect
    collimation/acceleration zone (with strong magnetic fields) in OJ287,
     providing the necessary 
    physical conditions (injection of relativistic electrons and helical motion)
    to cause such a behavior of optical outbursts for long times, although
    in most cases only one outburst is caused by one helical revolution due to
    the opening of the jet or intrinsic dimming of the optical knots in the
    outer parts of the jet.\\
    We will also model simulate the light-curves of other five periodic
    optical outbursts in 1983.00, 1984.10, 1994.75, 2005.76, and 2007.70, and
    for a few isolated moderate outbursts (in 1993.93, 1994.17 and 
    at JD2457380), showing the common properties of these outbursts 
    and their common mechanism for optical radiation production.\\
    All the data on the light curves of the optical outbursts used in this work
    were collected from: Valtonen et al.(\cite{Va16}, \cite{Va17}, \cite{Va08}),
   Kushwaha et al. (\cite{Ku18}),
   Valtaoja et al (\cite{Val20}) and Sillanp\"a\"a et al. (\cite{Si96a}, 
   \cite{Si96b} and  private communication).\\
   In this work, we adopt a $\Lambda$CDM cosmological model with the
   parameters as: ${\Omega}_m$\,=\,0.27, ${\Omega}_{\Lambda}$\,=\,0.73 and
   $\rm{H_0}$=71\,km\,${\rm{s}}^{-1}$\,${\rm{Mpc}}^{-1}$ (Spergel et al.
    \cite{Sp03}, Komatsu et
    al. \cite{Ko09}). 1\,mas\,=\,4.5\,pc (Hogg, \cite{Ho99}).
    \begin{figure*}
     \centering
     \includegraphics[width=6cm,angle=-90]{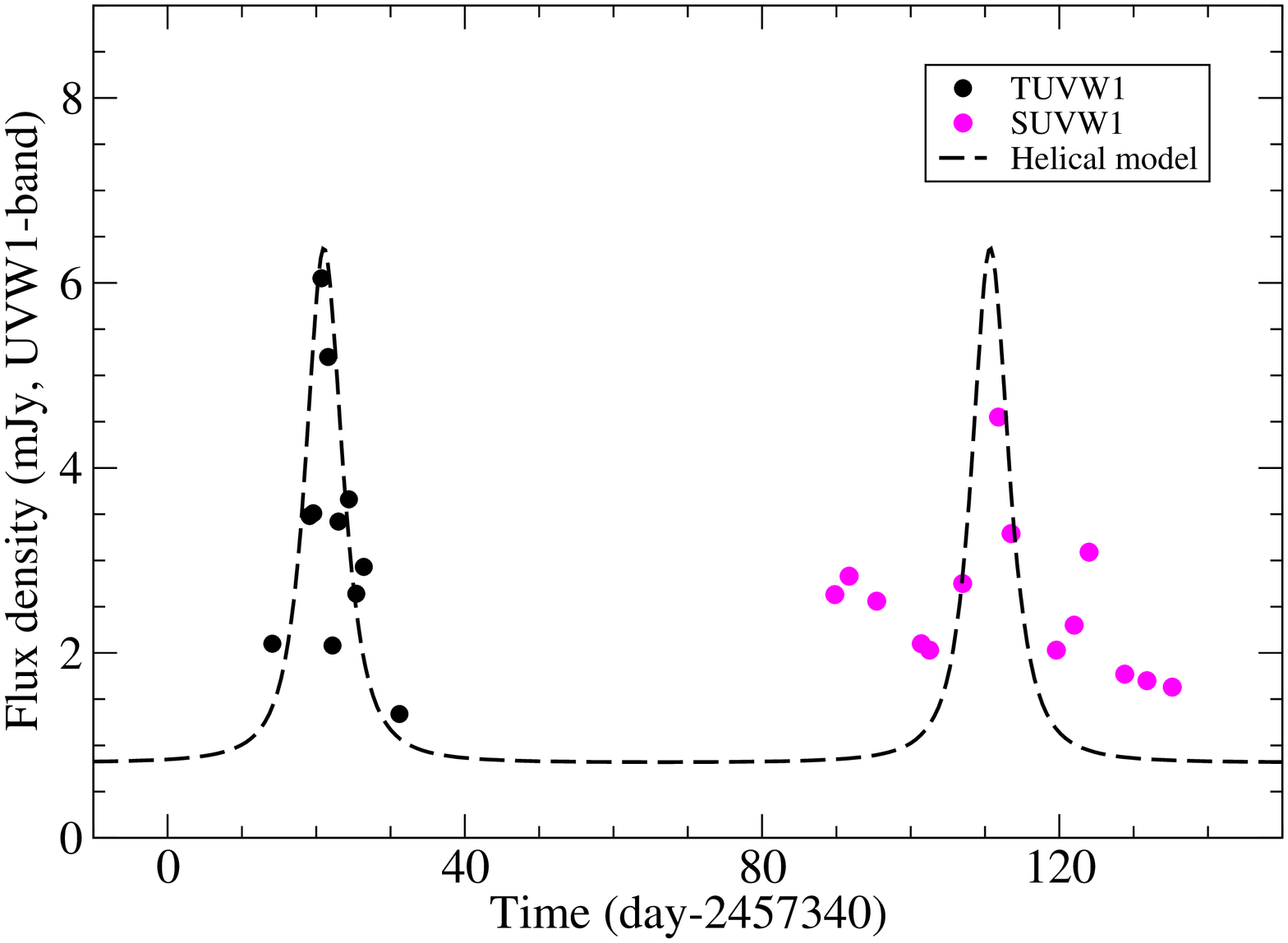}
     \includegraphics[width=6cm,angle=-90]{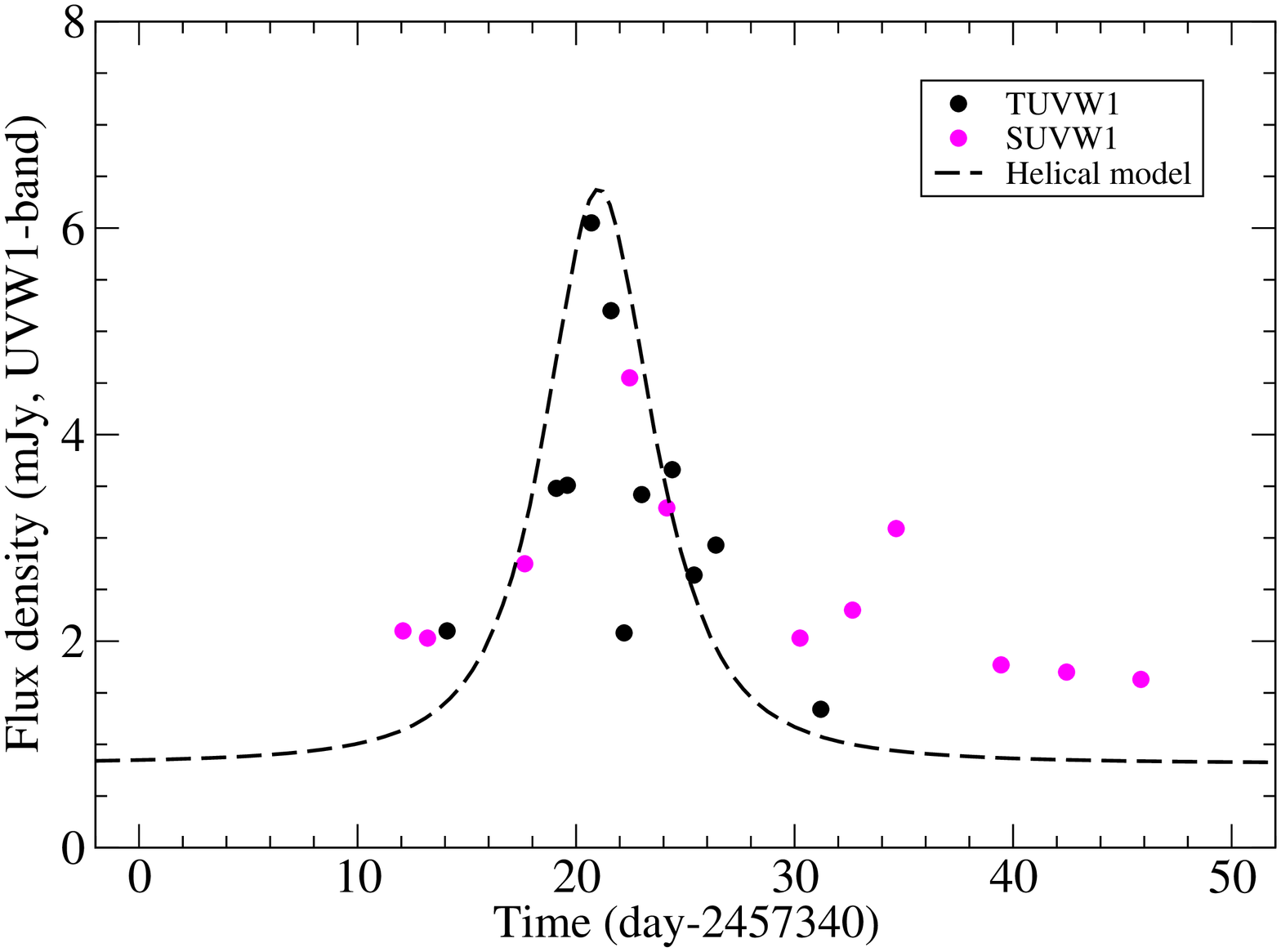}
     \includegraphics[width=6cm,angle=-90]{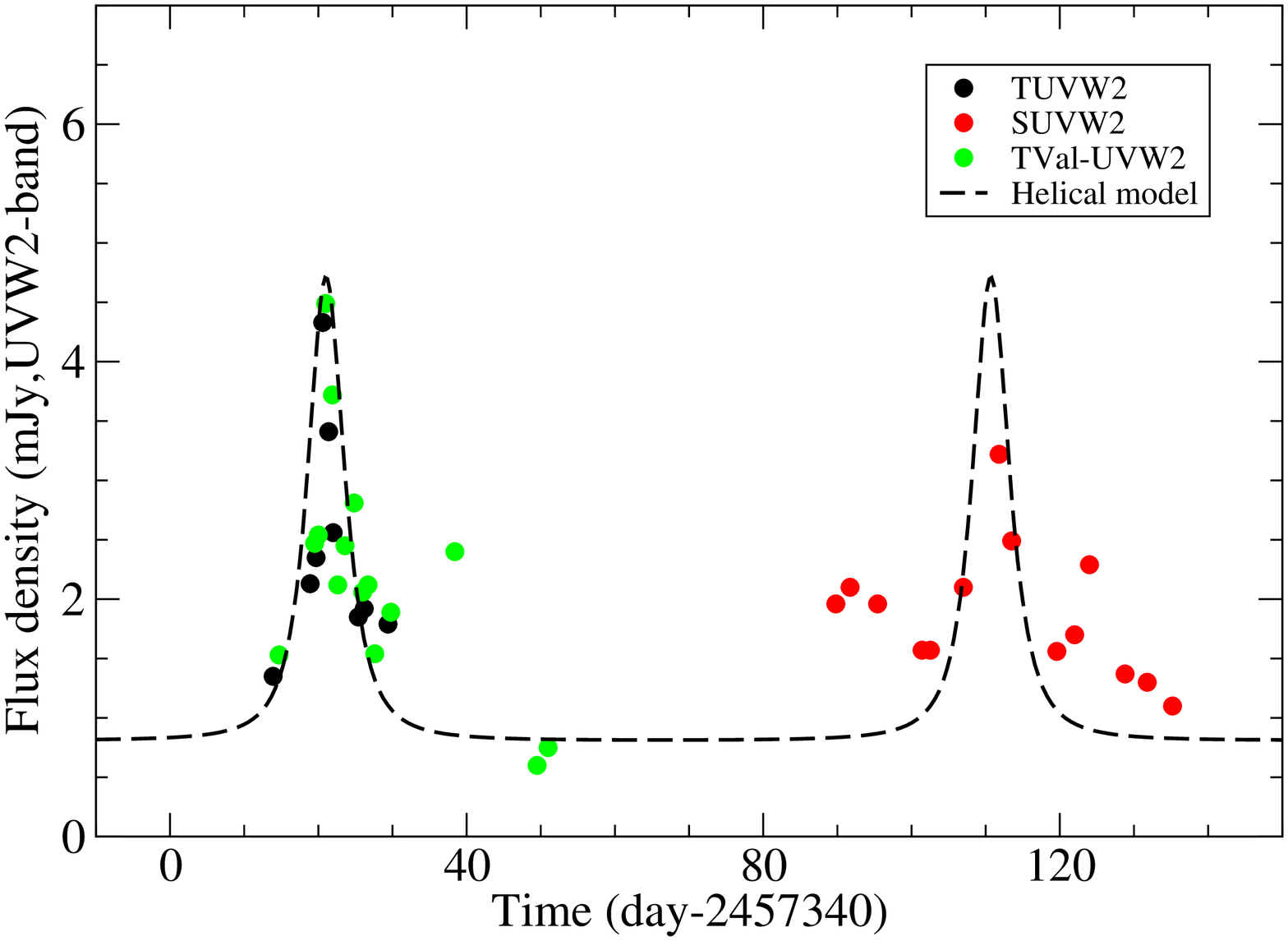}
     \includegraphics[width=6cm,angle=-90]{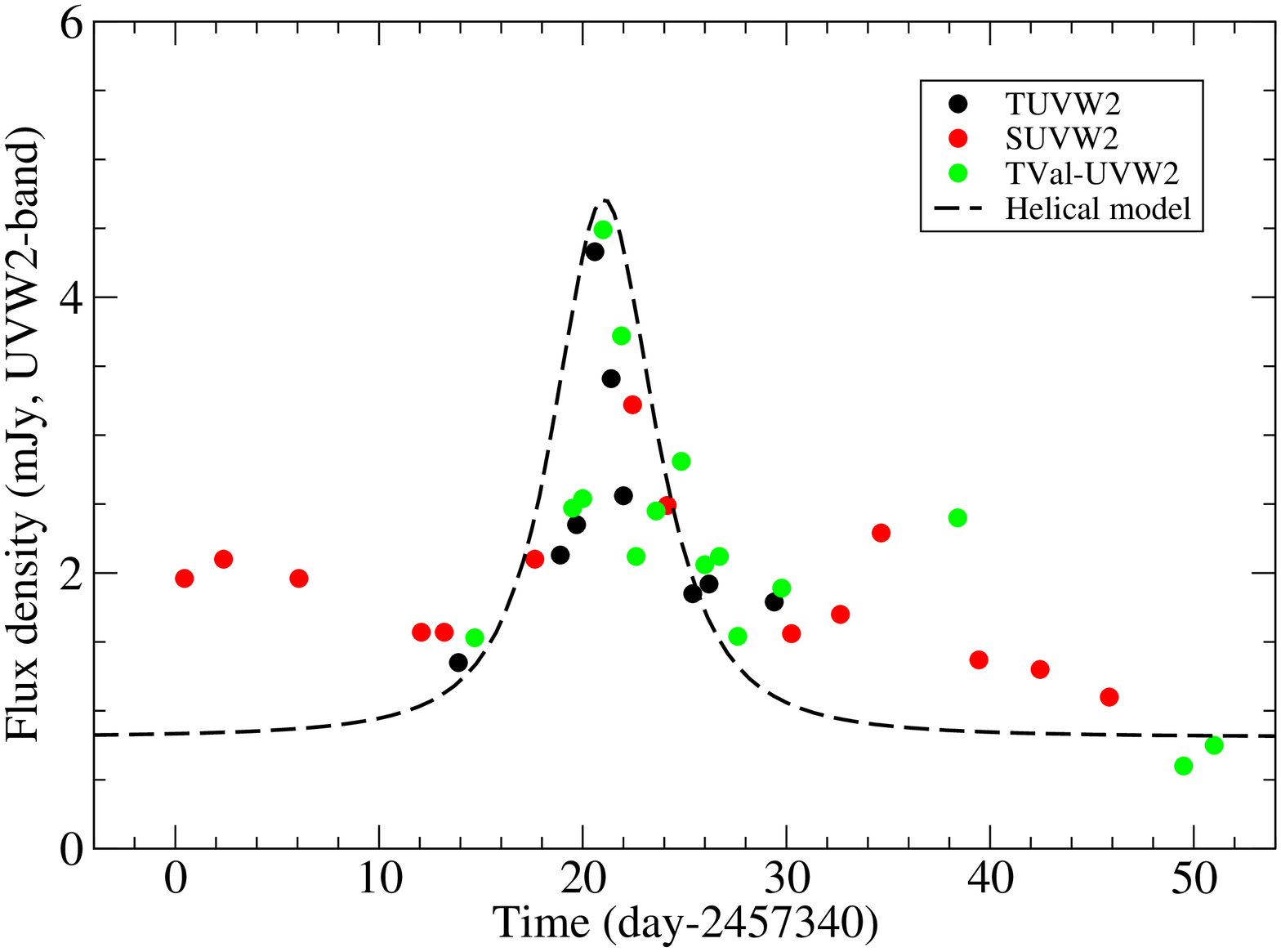}
     \includegraphics[width=6cm,angle=-90]{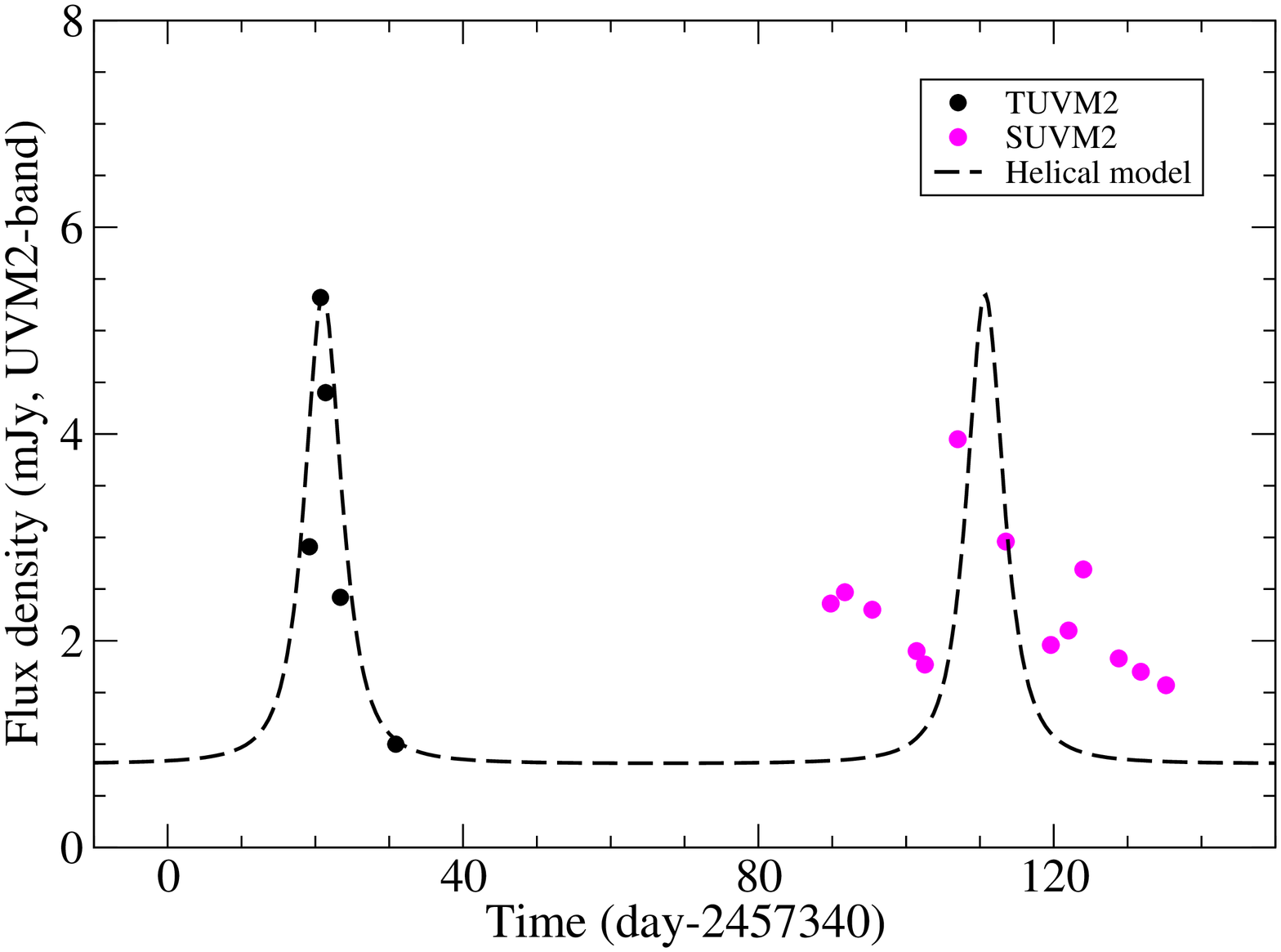}
    \includegraphics[width=6cm,angle=-90]{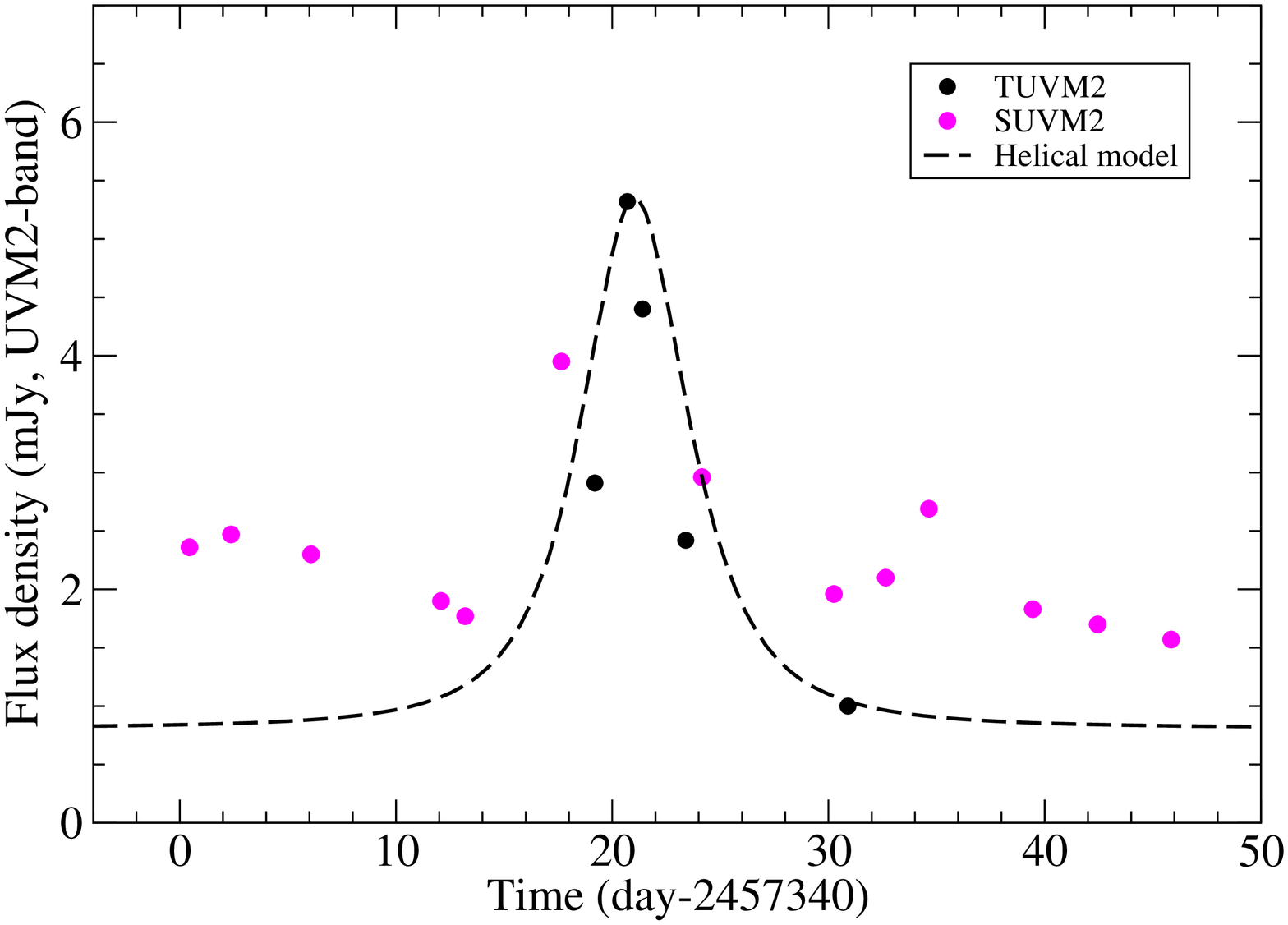}
    \caption{Model simulation of the UVW1-band (top panels), UVW2-band 
   (middle panels) and UVM2-band (bottom panels) light curves for the 
    outburst in December/2015 and in March/2016. For the UVW2-band the
    data-points observed by Valtonen et al. (\cite{Va16}) are also
    incorporated (labeled by TVal-UVW2 in green). In the right panels
    the observing time of the March/2016 outburst has been shifted backward 
    by 89.4 days. The good fits of the combined UV-band light curves 
    (right panels) indicate the applicability of the lighthouse model
    in the UV-bands
     and the December/2015 outburst still has its variability behavior 
    similar to that of the synchrotron outburst in March/2016.}
    \end{figure*}
     \begin{figure*}
     \centering
    \includegraphics[width=8cm,angle=-90]{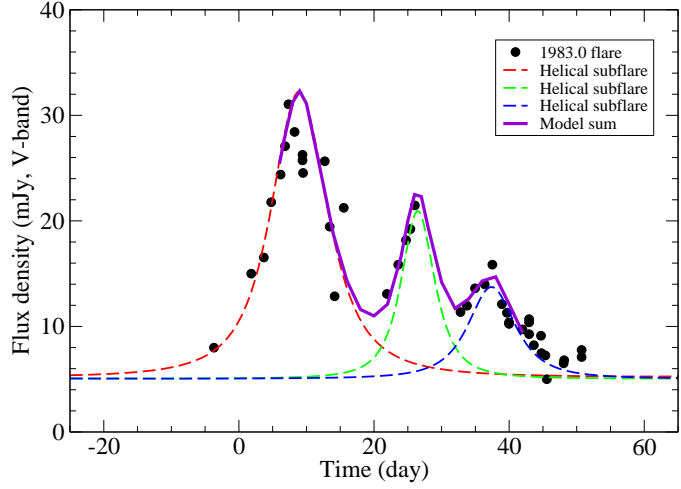}
    \caption{Model simulation of the light curve for the 1983.00
    periodic optical outburst, which
   has been decomposed into three subbursts with symmetric profiles. The violet 
    line denotes the model-fit of the total flux density curve. 
   The origin of the time-axis is 1983.00. Data are taken from 
   Valtonen et al. (\cite{Va08}).}
    \end{figure*}
   \begin{figure*}
   \centering
   \includegraphics[width=8cm,angle=-90]{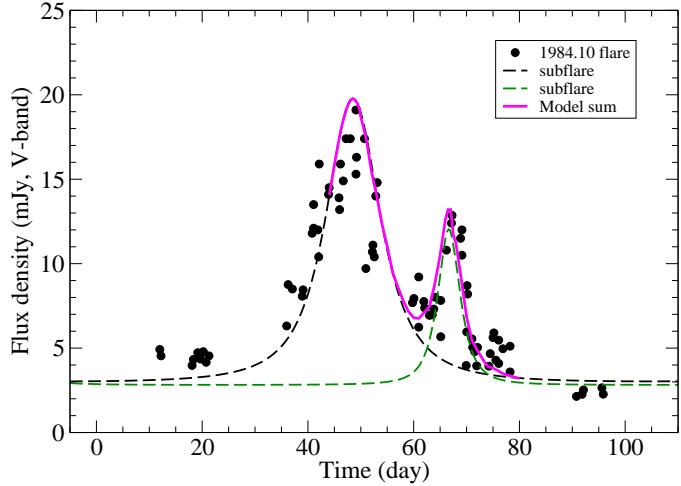}
   \caption{Model simulation of the light curve for the 1984.10
    periodic optical outburst, which
   has been decomposed into two subbursts with symmetric profiles. The magenta
   line represents the model-fit of the total flux density curve. The dashed 
   lines denote the model-fits of the two subbursts, respectively. The origin
   of the time-axis is 1984.05. Data are taken from Valtaoja et al. 
   (\cite{Val20}).}
   \end{figure*}
    \begin{figure*}
    \centering
    \includegraphics[width=8cm,angle=-90]{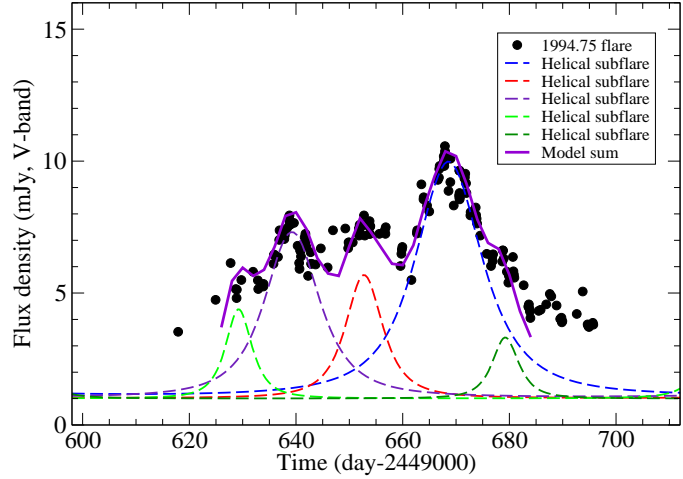}
  \caption{Model simulation of the light curve for the 1994.75 periodic
    optical outburst which has
  been decomposed into five subbursts with symmetric profiles. The violet 
   line denotes the model-fit of the total flux density curve.
     Data are taken from Sillanp\"a\"a et al. (\cite{Si96a};
     OJ-94 project).}
    \end{figure*}
    \begin{figure*}
    \centering
    \includegraphics[width=6cm,angle=-90]{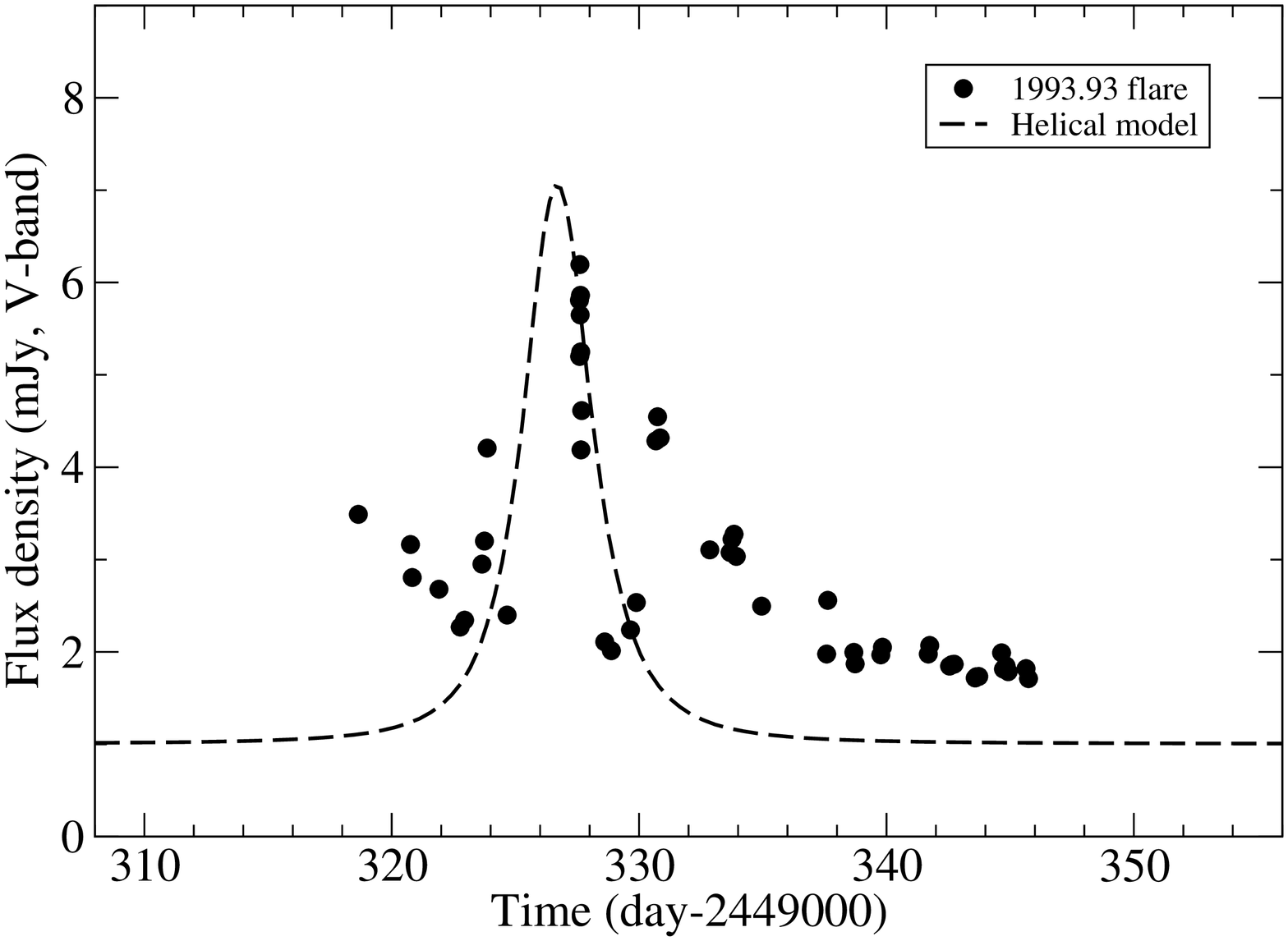}
   \includegraphics[width=6cm,angle=-90]{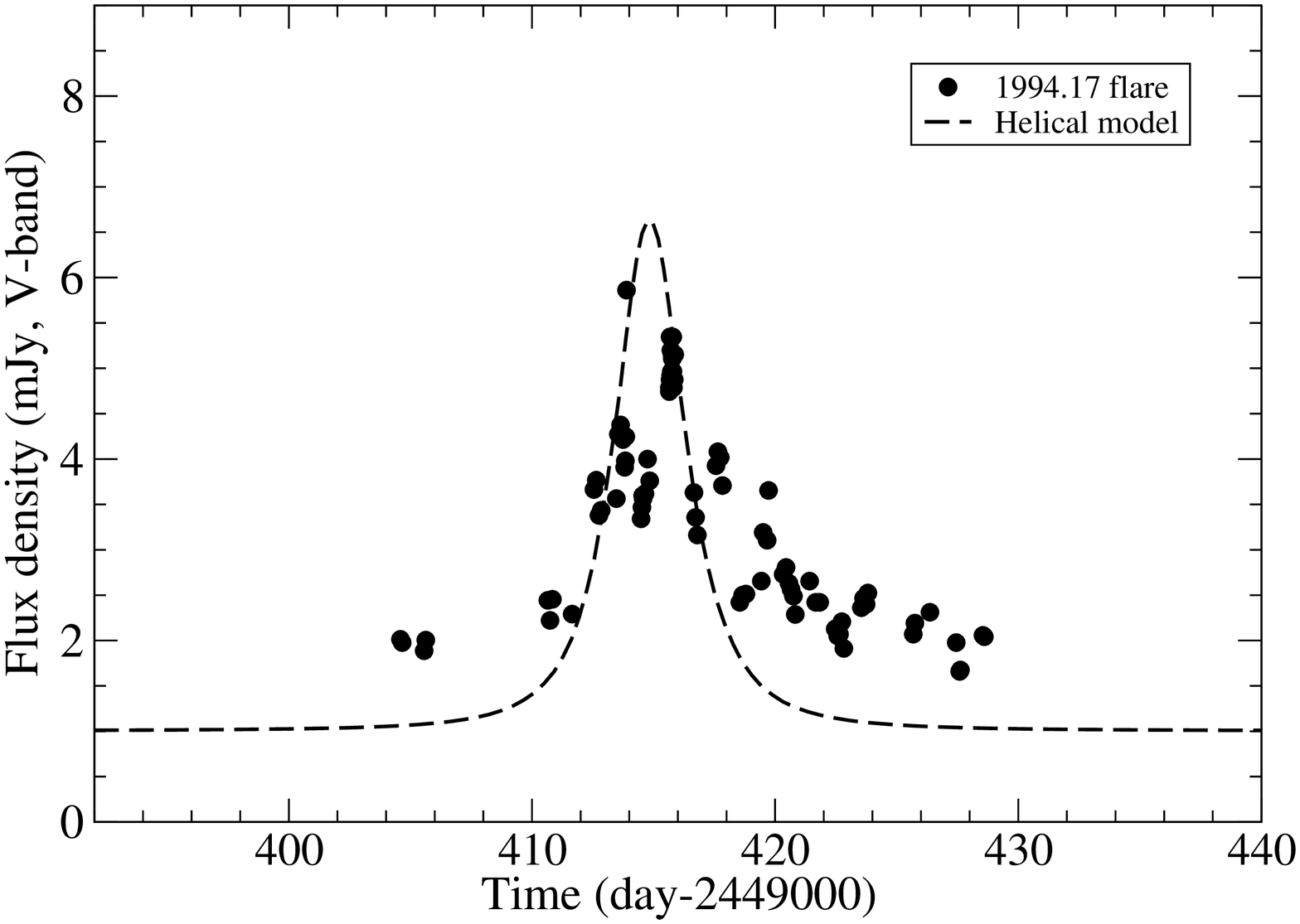}
    \caption{Model simulation of the light curves for two isolated moderate
    outbursts (in 1993.93 and 1994.17) with each having a symmetric profile.
     Data are taken from Sillanp\"a\"a et al. (\cite{Si96a}, OJ-94 project).}
    \end{figure*}
     \begin{figure*}
    \centering
    \includegraphics[width=8cm,angle=-90]{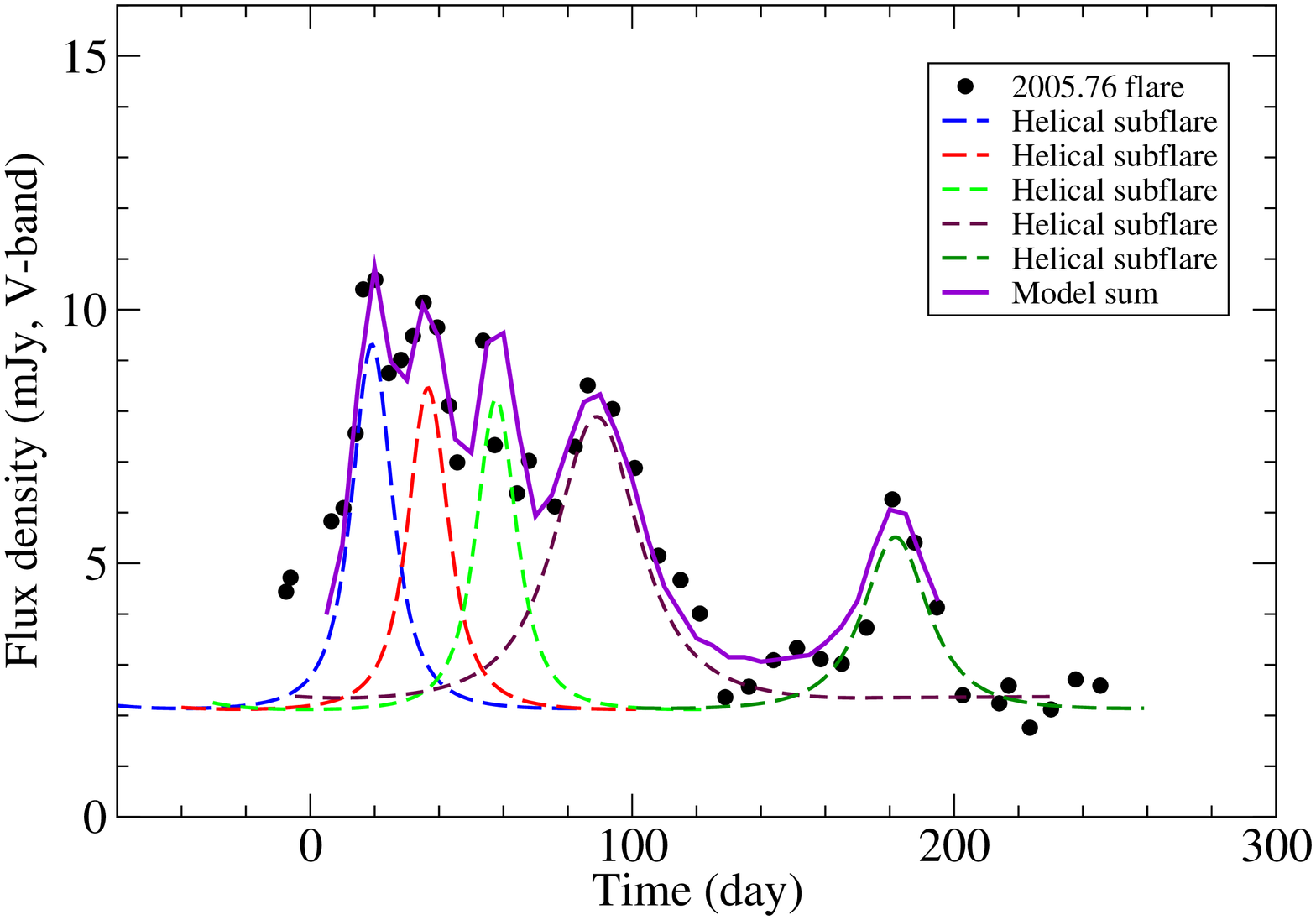}
    \caption{Model simulation of light curve for the 2005.76 periodic 
   optical outburst which is decomposed into five subbursts with symmetric 
   profiles. The violet line denotes the 
   model-fit of the total flux density curve. Time-axis denotes [day-2005.76].
   Data are taken from Valtonen et al. (\cite{Va08}).}
    \end{figure*}
    \begin{figure*}
    \centering
    \includegraphics[width=8cm,angle=-90]{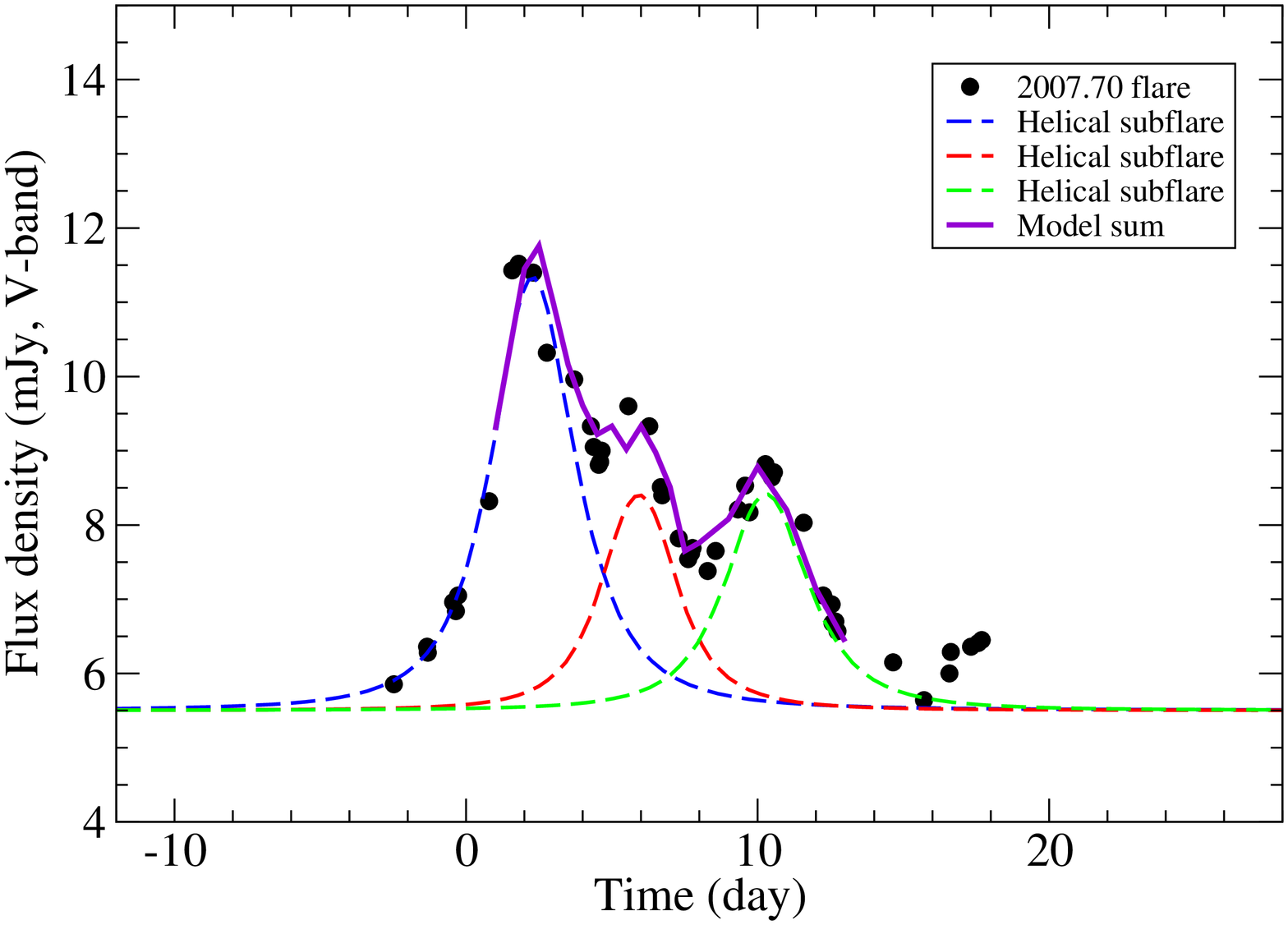}
    \caption{Model simulation of the light curve for the 2007.70 periodic
    optical outburst which is decomposed into three subbursts with 
   symmetric profiles. The violet line denotes
   the model-fit of the total flux density. Time-axis denotes [day-2007.70].
    Data are taken from Valtonen et al. (\cite{Va08}).}
    \end{figure*}
   \begin{figure*}
   \centering
   \includegraphics[width=6cm,angle=-90]{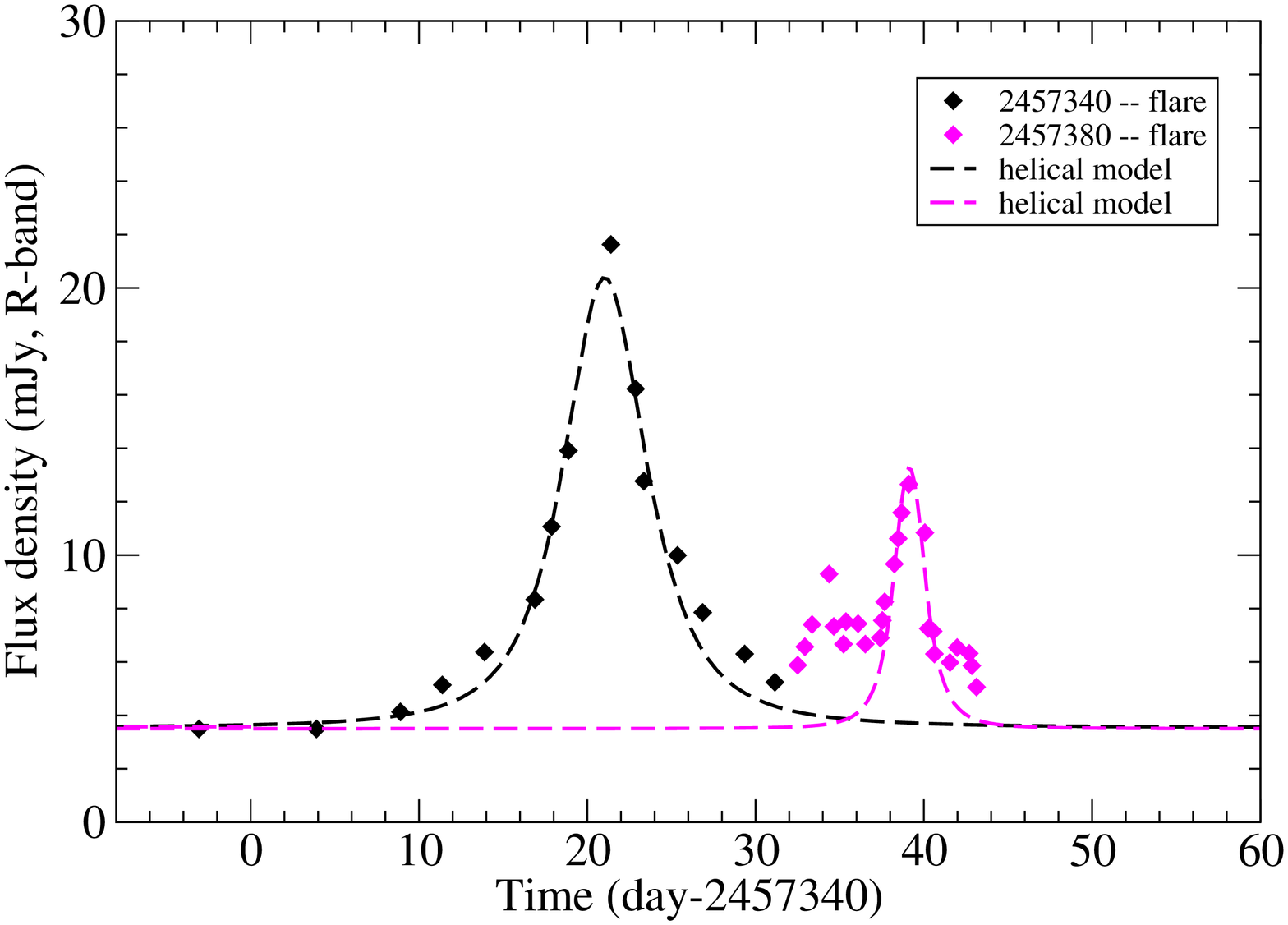}
   \includegraphics[width=6cm,angle=-90]{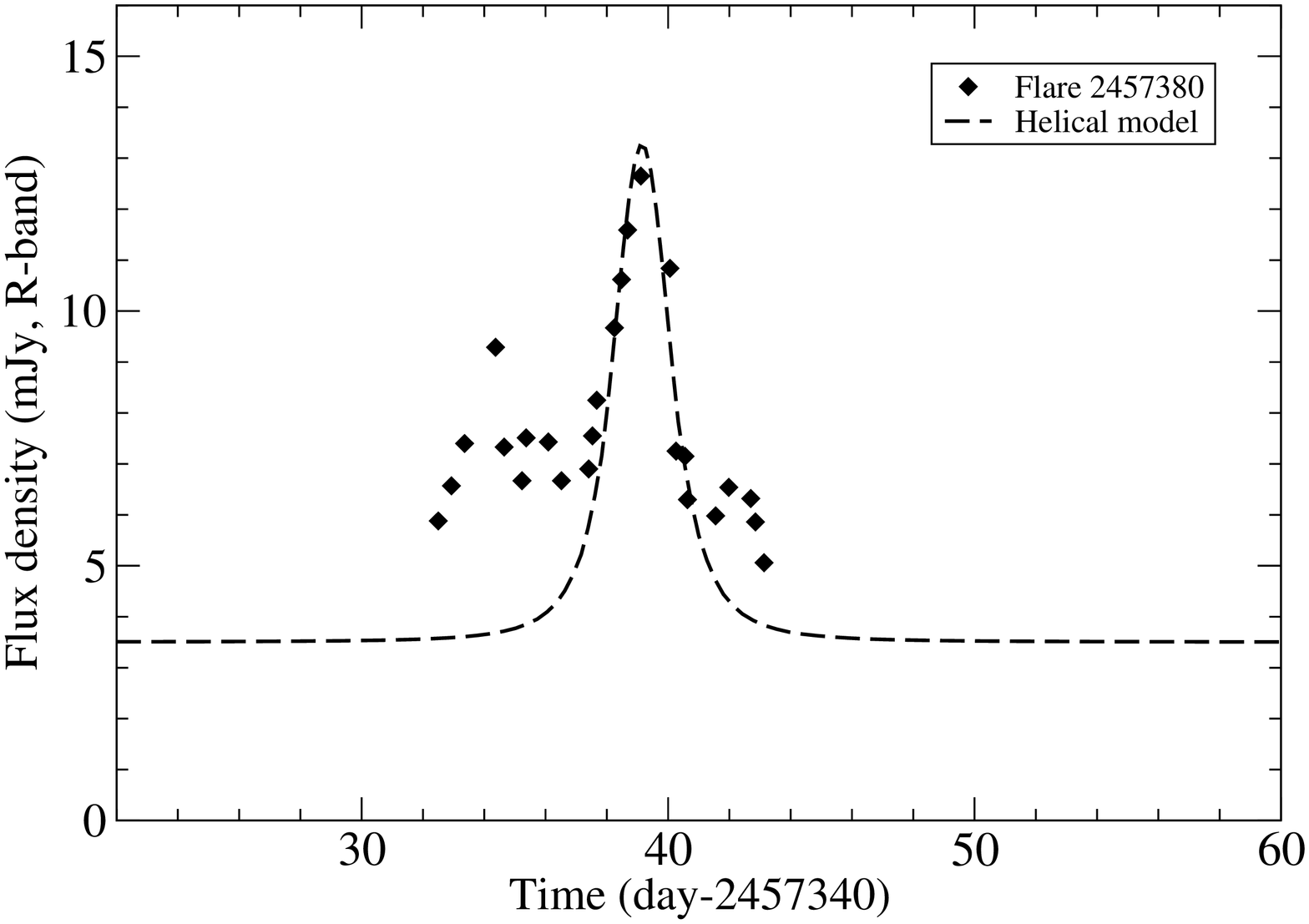}
   \caption{Model simulation of the light curve for the  isolated non-periodic
    optical burst peaking at JD2457380 (right panel, R-band, highly polarized
    with a polarization degree of $\sim$40\%) and its comparison with the
   model simulation of the light curve of the  major periodic optical 
   outburst peaking at JD2457360 (left panel). Both the light curves show
    very similar symmetric profiles and are well fitted by the helical 
   motion model. The data points for the major  outburst well emulate its 
   observed light curve after the small-amplitude fluctuations on it were
    removed. Two small spikes (at JD2457364  and JD2457368) superposed 
   on the major outburst are not displayed.} 
   \end{figure*}
  \section{Simulation results of multi-wavelength light curves}
    The model simulation results of  the multi-wavelength light curves
   (in NIR-optical-UV bands) 
   for the periodic optical outburst in December/2015 (peaking at JD2457360) 
   and the non-periodic synchrotron outburst in March/2016 
    (peaking at JD2457450) are shown 
   in Figures 4--9. The relevant model parameters are listed in Table 4. \\
     Firstly, in Figure 4 (left panel), we display the model simulation 
   of the R-band light curves of the December/2015 outburst observed 
   by Valtonen et al. (\cite{Va16}; labeled by ``TVal-R'') and by Kushwaha 
   et al. (\cite{Ku18}; labeled by ``TR''). It can be seen that the combination
   of the data-points from Valtonen et al. (during the rising phase) and
    Kushwaha et al. (during the decaying phase) constructs a well-defined
    profile simulated by the modeled symmetric light curve. In the right panel
    of Figure 4, the R-band light curve of the March/2016 outburst (labeled 
    by ``SR'') observed by Kushwaha et al. has been incorporated in the 
    simulation with its time-axis being shifted backward by 89.4 days.
    It can be seen that the R-light curves of both the December/2015 and
    March/2016 outbursts can be well simulated in terms of a common 
    symmetric profile. Since the March/2016 outburst is definitely a
    non-thermal (synchrotron) flare with high optical polarization degrees,
    the strong similarity between the variability behaviors of the 
    December/2015 and March/2016 outbursts leads us to recognize that 
   the December/2015 outburst\footnote{The December/2015 outburst peaking
     at JD2457360 was identified as the periodic ``impact thermal flare'' 
   in the disk-impact model.} is also a synchrotron flare generated in 
   the relativistic jet. Thus both outbursts can be interpreted in terms of
    lighthouse effect due to the helical motion of one superluminal optical
   component during two helical revolutions. \\
    Now we present the simulation results of the multi-wavelength
    light curves for each waveband individually.
   \begin{itemize}
   \item The model simulation results of the K-band light curves for both
    outbursts in December/2015 and March/2016 are shown in Figure 5. Here
    two panels are presented: the left panel displays the simulation of the 
    two light curves in time sequence and the right panel shows the simulation
    of the combined light curve. It can be seen that the K-band light curves of
    both outbursts are well fitted by the helical motion model, implying that
    the outburst in December/2015 has its variability behavior very similar
    to that of the synchrotron outburst in March/2016.
   \item  The model simulation results of the J-band and I-band light curves 
    for the outbursts in December/2015 and March/2016 are shown
   in Figure 6. The upper panels show  that the observed J-band peak
    is much lower than the model light curve. This result should have
   been expected, because the assumed model spectral index at J-band 
   ${\alpha}$=0.8 is larger than the 
   observed one (Kushwaha et al.
    \cite{Ku18}). The observed I-band light curves for both outbursts 
   are well fitted by the   helical motion model. 
   \item The model simulation of the R-band light curves for both 
   outbursts in December/2015 and in March/2016 are shown in Figure 7
    (top panels). The
   left panel displays the simulation of the light curves for both outbursts 
   in time sequence and the right panel shows the simulation of the
     three light curves  (labeled by ``TR'', ``SR'' and ``TVal-R'') in a
    combined form. It can be seen that the R-band light curve of the outburst
   in December/2015 has its variability behavior very similar to that
   of the synchrotron outburst in March/2016: both are well fitted by the
   helical motion model with  symmetric profiles
   having similar rising and decaying time-scales.\\
     Similarly, the model simulation results of the V-band
    light curves for the outbursts in December/2015 and in March/2016 
    are displayed in Figure 7 (bottom panels). 
    It can be seen that both the observed V-band light curves (whether 
    presented in time sequence or in combined form) are very well fitted
    by the helical motion model. Thus at both R- and V-bands the outburst in
    December 2015 has its variability behavior very similar to that of 
    the synchrotron outburst in March/2016, implying that their emission
    may originate from a similar mechanism. In this work we suggest that
    both outbursts are produced by the helical motion of one superluminal 
    optical knot via lighthouse effect through two helical revolutions, 
    although we cannot exclude the possibility that they might be independent
    flares with very similar temporary and spectral variations.
   \item The model simulation results of the B-band and U-band light curves
    for the outbursts in December/2015 and in March/2016 are shown in Figure 8.
    It can be seen that the observed light curves for both outbursts are
    well fitted by the helical motion model, showing that even in high
    frequency regions (B- and U-bands) the outburst in December/2015 has 
    a variability behavior similar to that of the synchrotron outburst in
    March/2016. The symmetry of the outburst profiles characteristic 
    in the low frequency region (K-band to V-band, modeled spectral index 
   $\alpha$\,=\,0.8) persists in the high  frequency region 
   (modeled spectral index $\alpha$\,=\,1.3).
   \item The model simulation results of the UVW1-, 
   UVW2- and UVM2-band light curves for the outbursts in December/2015 and 
    in March/2016 are shown in Figure 9.  It can be seen that the light
     curves observed at the three bands for both outbursts were well fitted
    by the helical motion model. Thus the
   outburst in December/2015 has its variability behavior very similar to that
   of the synchrotron outburst in March/2016 in the UV-bands. But it should 
   be noticed that rapid (intraday) variability due to interstellar
    scintillation (or extinction) or turbulent fluctuations in optical knots 
   (Qian et al. \cite{Qi91b}, Marscher et al. \cite{Ma08}, 
   Marscher \cite{Ma14}) might cause scattering of the observational 
   data-points. 
   \end{itemize}
    Based on the model simulation of the multi-wavelength light curves for 
  the periodic outburst in  December/2015 and the non-periodic synchrotron
   outburst  in March/2016 as shown in Figure 4\,--\,9,  we come to the
   conclusion that the December/2015 outburst, which was claimed to be a
   bremsstrahlung flare, has its variability behavior (both temporary and
    spectral) very similar to the 
   synchrotron flare in March/2016: both have similar peaking flux densities
    and  spectral features, and their flux density curves having symmetric 
   profiles with similar rising and decaying time-scales. This would imply
   that the December/2015 outburst may also be  a synchrotron flare.
    \section{Light curve structure of periodic optical outbursts}
   As argued in the previous section, the symmetry in the optical light curves
   and the similarity between the periodic and non-periodic optical outbursts 
   may be important for understanding the optical variations in OJ287. In order
   to further investigate this behavior and clarify the nature of the phenomena
   in OJ287, we will make model simulation of the light curves for the five
   periodic optical outbursts observed in 1983.00, 1984.10, 1994.75, 2005.76
   and 2007.70, and show that their light curves can be decomposed into a 
   number of subflares (or elementary flares) with symmetric profiles.
    In addition, we will make model simulation for some well-resolved
   (or isolated) non-periodic optical bursts to reveal the similarity 
   in optical variations between the periodic outbursts 
   (claimed as thermal flares)
    and non-periodic outbursts (recognized as synchrotron flares). 
   In combination with the results presented in Sect.3 for the 
   December/2015 and 
   March/2016 outbursts, it can be seen that the consistency in the  
   variability behavior of all these outbursts provides 
   important clues for understanding the 
   nature of their emission, affording persuasive evidence that the entire 
   phenomena observed in blazar OJ287 may be caused by lighthouse effect 
   due to the helical motion of superluminal 
   optical knots (plasmons or shocks) within the  relativistic jet.\\
    \begin{table}
    \caption{Base-level flux densities (V-band) for the six periodic optical
     outbursts.}
    \begin{flushleft}
    \centering 
    \begin{tabular}{ll}
    \hline
    Flare & $\rm{S_b}$(mJy)\\ 
    \hline
    1983.00 & 5.0 \\
    1984.10 & 2.8 \\
    1994.75 & 1.0 \\
    2005.76 & 2.0 \\
    2007.70 & 5.5 \\
    2015.87 & 3.5 \\
    \hline
    \end{tabular}
    \end{flushleft}
    \end{table}
    For simplicity and easy comparison, we will apply the same approaches as
    used in Sect.3 to make model simulation of 
   the light curves for the five periodic outbursts. The outbursts are assumed
   to be composed of two components: the underlying jet emission (or 
    base-level component) and the flaring component. The base-level component
     can be taken to be constant during 
   individual outbursts, but may change on longer time-scales due to the jet 
   precession and intrinsic variations in the underlying jet (jet-parameters 
   and bulk Lorentz factor).
   For the six periodic outbursts simulated, the base-level flux densities
   (at V-band)  are listed in Table 5.
   \subsection{Periodic optical outburst in 1983.00}
    The outburst in 1983.00 is the first optical flare of the 1983\,--\,1984 
    double-peaked outbursts. 
    The simulation results for the 1983.00  periodic outburst are shown in
    Figure 10. Its total flux density curve has 
    been decomposed into three subflares. The first flare was identified
    as the ``impact (thermal) outburst''. Obviously, its light curve does not
    have the ``standard shape'' expected for impact outbursts. It looks like 
    a single elementary flare and its light-curve can be very well fitted
     with a symmetric profile in terms of the helical motion
    (or lighthouse) model. Similarly, the light-curves of the other two 
   subflares can also be well fitted. The  model parameters 
   for the three subflares are given in Table 6.  \\
    \begin{table}
    \caption{Model simulation results for the 1983.00 periodic optical
     outburst (V-band). $\rm{S_b}$\,=\,5\,mJy. 
   $\Gamma$\,--\,Lorentz factor of the
    superluminal optical knot, $\rm{{\delta}_{max}}$\,=\,
    maximum Doppler factor, 
     ratio\,=\,$\rm{{\delta}_{max}}$/$\rm{{\delta}_{min}}$,
    $\rm{S_{int}}$(mJy)\,--\,intrinsic 
    (comoving) flux density. $t$(day)\,=\,flare time\,=\,day-1983.00.
    FWHM (day)\,=\,full width at half maximum of the model light curve.}
    \begin{flushleft}
    \centering
    \begin{tabular}{llllll}
    \hline
   t & $\Gamma$ & $\rm{{\delta}_{max}}$ & ratio & $\rm{S_{int}}$ & FWHM \\
    \hline
    9 & 8.0 & 15.90 & 3.20 & 4.27$\times{10^{-4}}$ & 10.0 \\
   26 & 9.5 & 18.84 & 4.10 & 1.26$\times{10^{-4}}$ & 5.1 \\
   37 & 8.5 & 16.89 & 3.49 & 1.07$\times{10^{-4}}$ & 7.1 \\
   \hline
   \end{tabular}
   \end{flushleft}
   \end{table}
   \begin{table}
   \caption{Model simulation  results for the 1984.10 periodic
    optical outburst. Explanation of the parameters and units as in Table 6. 
   $\rm{S_b}$=2.8\,mJy.}
   \begin{flushleft}
   \centering
   \begin{tabular}{llllll}
   \hline
    t & $\Gamma$ & $\rm{{\delta}_{max}}$ & ratio & $\rm{S_{int}}$ &FWHM \\
   \hline
   48 & 7.5 & 14.90 & 2.93 & 3.44$\times{10^{-4}}$ & 12 \\
   67 & 10.0 & 19.9 & 4.45 & 5.93$\times{10^{-5}}$ & 5 \\
   \hline
   \end{tabular}
   \end{flushleft}
   \end{table}
    \subsection{Periodic optical outburst in 1984.10}
    Interestingly, the second optical flare of the 1983\,--\,1984
    double-peaked outbursts (during 1984.1--1984.3; see Fig.6
    in Valtaoja et al. \cite{Val20}) 
    clearly exhibits its structure  consisting of two rather separated 
    outbursts peaking at 1984.18 and 1984.23 , respectively. 
    As like the 1983.00 outburst the major flare (peaking at 1984.18 with
    peak flux=$\sim$20\,mJy at V-band) is a single elementary one and 
    its flux density curve  can be very well simulated by a 
    symmetric profile and is displayed  in Figure 11. The relevant model
    parameters for the two subflares are listed in Table 7.
    Thus for the 1983\,--\,1984 double-peaked outbursts, both the major
    flares exhibit the symmetry in their light curves.   As discussed below,
    the major flares of the outbursts in 2007.70 and 2015.87 (Figs.15 and 16)
    also reveal this characteristic feature. Symmetry in the optical outburst
    light curves can not be explained in terms of the disk-impact scenario,
    where the flux density curves of ``impact outbursts'' emitted from evolving
    gas-bubbles would have a non-symmetric pattern with a rapid rising and
    slower decaying  phases.
     \subsection{Periodic optical outburst in 1994.75}
   The 1994--1996 periodic double-peaked outbursts were intensively monitored 
   via international cooperation of the  ``OJ-94 project'' (Takalo 
   \cite{Tak96a}, Takalo  et al. \cite{Tak96b}), starting at 1994.75
    and 1995.84, respectively. During this period the optical flares
    may be related to the ejection of superluminal radio (15\,GHz)
    components C1, C2, C3 and C4 (Britzen et al. \cite{Br18},
   Qian \cite{QiXiv18}, Tateyama et al. \cite{Ta99}).\\
   The simulation results for the first outburst in 1994.75
   \footnote{In Lehto \& Valtonen (\cite{Le96}) and Valtonen \& Lehto
    (\cite{Va97}) this outburst was identified as the "disk-impact
   thermal outburst". Recently Dey et al. (\cite{De18}) suggested that 
   its starting time should be changed to 1994.60 and the corresponding "impact 
   outburst" was missed due to lack of monitoring observations.} 
   (peaking at 1994.86)  are shown in Figure 12. The observational 
   data are taken from Sillanp\"a\"a et al. (\cite{Si96a},
   OJ-94 project; private communication). Its total flux density curve has been
    decomposed into five subflares simulated with symmetric profiles.\\
    Two  rather isolated  moderate outbursts peaking at JD2449326
    and 2449415 were also model simulated and the results are shown in
    Figure 13. The relevant model parameters are described in Table 8.
    The successful simulation of the light curve for the 1994.75 outburst
    might have demonstrated that any complex outburst observed in OJ287
     can be decomposed into a few elementary flares and explained 
   in terms of the proposed helical motion (or lighthouse) model.   
   \subsection{Periodic optical outburst in 1995.84}
     The outburst starting at 1995.84 was the second one of the 
   1994\,--\,1996 double-peaked outbursts. Its light curve exhibits a 
    complex structure, showing some distinct features which may be 
   meaningful for understanding the phenomena in OJ287 (Valtaoja et al.
    \cite{Val20}).\\
    Firstly, the ``impact flare'' (during 1995.84--1995.90) identified by the 
   disk-impact model (Dey et al. \cite{De18}) was a small one with its 
   peak flux density $\simeq$5\,mJy, much weaker than the follow-up flares 
    with peak flux of $\sim$12\,mJy (Valtaoja et al. \cite{Val20}). This
   would pose a problem: why a strong disk impact 
   \footnote{According to the disk-impact model (Dey et al. \cite{De18}),
    the 1995.84 outburst occurred at a distance of $\sim$3800\,AU from 
   the primary hole and should be produced by a strong impact of the secondary
    hole onto the primary disk.} could only produce a small ``thermal''
   optical flare, but resulted in the production of strong follow-up 
   synchrotron flares. Obviously, it seems that some other physical processes
   would  have played their roles. \\
     Secondly,  it is  noticed that the 1995.84 flaring event (during 
    1995.84\,--\,1996.12) had its light curve very 
    similar to that of the 1994.75 outburst and could also be
    decomposed into a number of subbursts simulated with symmetric profiles. 
    The symmetry of this double-peaked outbursts  was
    firstly discovered by Sillanp\"a\"a et al. (\cite{Si96b}), which 
    essentially reflects the symmetry existing in the light curves of 
   their subbursts with similar rising and declining time-scales. 
   Thus both the double-peaked outbursts during 1994\,--\,1996 period 
   can be interpreted in terms of our helical motion (lighthouse) model.
    \footnote{Due to lack
    of relevant data the 1995.84 outburst was not simulated in this work.} \\
    Thirdly, the most interesting feature of the 1995.84 outburst may be 
    the close connection between the radio and optical variations.
    According to Valtaoja et al. (\cite{Val20}, Fig.8 therein), its radio
    variations at 22/37\,GHz were very similar to the V-band optical
    variations: both variations are  not  only simultaneous but also 
     have similar envelopes. Even a few radio emission peaks can be
     recognized to be concurrent with the optical peaks. This strict
     simultaneity of the radio  and optical variability seems  important 
    and may have provided some significant clues to  the physical processes
    producing the radio/optical outbursts in OJ287. Unfortunately, this 
   radio-optical connection has not been explained since its discovery.
    In combination with the model simulation of the light curves for the 
   December/2015 and March/2016 outbursts in Sect.3, we would come to the 
   conclusion that this close connection between the radio and optical 
   variability can not be explained in terms of disk-impact scenario and 
   shock-in-jet models, and can only be explained in terms of
   lighthouse effect due to the helical motion of superluminal knots, but
   requiring  some special structure of the emitting source.\\
    \begin{table}
   \caption{Model parameters for the simulation of the 1994.75 
   periodic optical outburst and two 
   isolated moderate bursts (at V-band). $\rm{S_b}$=1.0\,mJy. 
   $t$\,=\,flare time\,=\,day--1994.75. Bursts at JD2449325 and JD2449415 are
   isolated moderate flares. The major periodic outburst has been
    decomposed into five subbursts. Explanation of the parameters and units
    as in Table 6.}
   \begin{flushleft}
   \centering
   \begin{tabular}{llllll}
   \hline
   $t$ & $\Gamma$ & ${\delta}_{max}$ & ratio & $\rm{S_{int}}$ & FWHM \\
   \hline
   325 & 11.5 & 22.83 & 5.57 & 2.23$\times{10^{-5}}$ & 3.3 \\
   415  & 11.3 & 22.44 & 5.41 & 2.22$\times{10^{-5}}$ & 3.3 \\
   \hline
   630 & 9.5 & 18.88 & 4.11 & 2.67$\times{10^{-5}}$ & 5.7\\
   640 & 7.5 & 14.90 & 2.93 & 1.28$\times{10^{-4}}$ & 11.6 \\
   650 & 8.5 & 16.89 & 3.49 & 5.75$\times{10^{-5}}$ & 8.2 \\
   670 & 7.0 & 13.90 & 2.68 & 2.41$\times{10^{-4}}$ & 14.8\\
   680 & 9.5 & 18.88  & 4.11 & 1.83$\times{10^{-5}}$ & 5.8\\
   \hline
   \end{tabular}
   \end{flushleft}
   \end{table}
   \subsection{Periodic optical outburst in 2005.76}
    The double-peaked outbursts during the period of 2005--2007 were 
    extensively observed by Valtonen et al. (\cite{Va08}) and Villforth et al.
    (\cite{Vil10}). The ejection of superluminal radio components C11, C12,
    C13L, C13U and C14 (Qian \cite{QiXiv18}) may be connected with these 
   optical flaring events.\\
     The simulation results for the light curve of the first 
   outburst starting in 2005.76\footnote{It was identified as the ``impact 
   (bremsstrahlung) flare'' by Valtonen et al. (\cite{Va08}).} are shown 
   in Figure 14. Its total flux density curve has been decomposed into  
   five subflares, which are well simulated with symmetric profiles in terms 
   of the lighthouse model. The model parameters are given in Table 9.\\
    According to the
     disk-impact model, the 2005.76 and 1994.75 outbursts occur at 
    quite different distances (Dey et al. \cite{De18}):
   $\sim$12,000AU (with the secondary-hole velocity of 0.17c) and 
    $\sim$7,000AU (with the secondary-hole velocity of 0.10c), 
    respectively.  As shown in Fig.14 
    and in Fig.12, while their light curves have similar multi-component 
    structures, the peak flux density of the 2005.76 outburst ($\sim$9.0\,mJy)
    is much higher than that of the 1994.75 outburst ($\sim$5.0\,mJy). 
    This is inconsistent with the prediction of the disk-impact model:
    the strength of periodic optical outbursts is mainly dependent on
     the impact-distance
     and the secondary-hole velocity. It seems that some other 
    ingredients could exist to determine the strength of 
     the flaring activity, e.g., variations in the circumbinary disk 
     and accretion rates onto the binary holes.
   \begin{table}
   \caption{Model simulation results for the 2005.76 outburst. Its total flux 
   density curve was decomposed into five sub-flares with symmetric profiles.
   $\rm{S_b}$=2.0\,mJy. $t$\,=\,flare time\,=\,day--2005.76. Explanation of 
   the parameters and units  as in Table 6.}
   \begin{flushleft}
   \centering
   \begin{tabular}{llllll}
   \hline
   $t$  & $\Gamma$ & ${\delta}_{max}$ & ratio & $\rm{S_{int}}$ & FWHM \\
   \hline
   20 & 7.0 & 13.90 & 2.68 & 1.81$\times{10^{-4}}$ & 14.6\\
   37 & 7.0 & 13.90 & 2.68 & 1.73$\times{10^{-4}}$ & 14.6\\
   58 & 7.0 & 13.90 & 2.68 & 1.63$\times{10^{-4}}$ & 14.6\\
   90 & 5.5 & 10.89 & 2.03 & 4.18$\times{10^{-4}}$ & 31.7\\
   182 & 6.0 & 11.90 & 2.23 & 1.76$\times{10^{-4}}$ & 25.8\\
   \hline
   \end{tabular}
   \end{flushleft}
   \end{table}
   \subsection{Periodic optical outburst in 2007.70}
   The outburst starting at 2007.70 is the second flare of the double-peaked
   outbursts during 2005\,--\,2007 period.
    The model simulation results for this periodic optical outburst 
    are shown in Figure 15.\footnote{The whole outburst was observed at 
    R-band by Villforth et al. (\cite{Vil10}) during 
    September/2007\,--\,February/2008.
    Its entire light curve comprises of at least seven individual subflares,
    overlapping on each other and  forming a very complex structure with a time
    scale of about five months. The optical flare
    discussed here is the first one which was identified as the ``impact 
    (bremsstrahlung) flare'' by Valtonen et al. (\cite{Va08})}. 
   Its total flux density curve has been decomposed into three subbursts
    which are all well simulated with symmetric profiles in terms of the 
   lighthouse model. The model parameters are
    given in Table 9. Although the declining part of the major outburst
    is mixed with the rising part of the secondary burst, its rising-peaking
     part still clearly demonstrates the trend of its symmetric profile.\\
    In combination with the results given in Sect. 4.5, we found that
    the total
    flux density curves of both periodic outbursts (in 2005.76 and 2007.70)
    can be well interpreted in terms of the lighthouse model with
    symmetric profiles.\\
    It can be seen from Figures 14 and 15 (also see Tables 9 and 10) 
    that the subflares of 2005.76 outburst
    all  have timescales much longer than those for the subflares of 2007.70
    outburst, this might be related to their different impact distances 
    as expected by the disk-impact model:
    the 2005.76 outburst occurred at $\sim$12,000\,AU and the 2007.70 outburst 
    at only $\sim$3,000\,AU. However, the peak flux density of the 2005.76 
    outburst ($\sim$9.0\,mJy) is much higher than that of the 2007.70
    outburst ($\sim$6.5\,mJy), which seems inconsistent with the expectation 
    of this model.     
   \begin{table}
   \caption{Model simulation results for the 2007.70 outburst. Its total flux
   density curve is decomposed into three subbursts with symmetric profiles. 
   $\rm{S_b}$=5.5\,mJy. $t$\,=\,flare time\,=\,day--2007.70. Explanation of 
   the parameters and units as in Table 6.}
   \begin{flushleft}
   \centering
   \begin{tabular}{llllll}
   \hline
   $t$ & $\Gamma$ & ${\delta}_{max}$ & Ratio & $\rm{S_{int}}$ & FWHM \\
   \hline
   2 & 11.4 & 22.64 & 5.49 & 2.22$\times{10^{-4}}$ & 3.4\\
   6 & 12.0 & 23.82 & 5.97 & 9.00$\times{10^{-6}}$ & 3.0\\
   10 & 11.4 & 22.64 & 5.49 & 1.12$\times{10^{-5}}$ & 3.4\\
   \hline
   \end{tabular}
   \end{flushleft}
   \end{table}
   \subsection{Periodic optical outburst in 2015.87}
    In Sect.3 we have presented the model simulation results of the 
   multi-wavelength light curves for the major periodic optical outburst in 
   December/2015 (peaking at JD2457360) and shown that its multi-wavelength
    light curves all have symmetric profiles and can
   be interpreted in terms of light-house effect
   due to the helical motion of a superluminal optical knot. Here we present
    the model simulation of the light curve (at R-band) for an isolated optical 
    flare peaking at JD2457380 (Valtonen et al. \cite{Va16}), which is  
     shown in Figure 16 (left panel). For comparison, the modeled light curve
    for the major outburst is also displayed (right panel). The model
     parameters are given in Table 11.\\
    It can be seen from Figure 16 and Table 11 that the isolated non-periodic
      outburst
    peaking at JD2457380 has a symmetric light curve similar to that of 
   the December/2015 outburst and both can be well fitted by the helical 
   motion model but with different bulk
   Lorentz factors: $\Gamma$=13.5 for the non-periodic flare and $\Gamma$=9.5
   for the major outburst. It should be noted that the non-periodic flare
   (peaking at JD2457380) is a non-thermal (synchrotron) flare with
   polarization degree of $\sim$40\% (Valtonen et al. \cite{Va17}). 
   Therefore, the similarity in the light curve patterns between the 
   December/2015 outburst and this non-thermal flare \footnote{This non-thermal
   flare appeared at JD2457380, only 20\,days after the appearance of the 
   December/2015 outburst.}further prove the suggestion that
   the December/2015 outburst may originate from synchrotron process.\\
    In addition, we notice that the December/2015 outburst has its light curve
   structure similar to that of the 1984.10 outburst (Figures 11 and 16) and 
   they have similar strengths: peak flux density of $\sim$14.5\,mJy 
   for the December/2015 outburst and $\sim$\,17.2mJy for the 1984.10 outburst.
    This seems in contradiction with the expectation of the disk-impact model.
    According to the 
   disk-impact model, the December/2015 outburst should be
   much weaker than the 1984.10 outburst, because it appeared at 
   impact-distance of $\sim$18,000\,AU much farther than the 1984.10
   outburst (at impact distance of  $\sim$5,000\,AU).  
    This seems to demonstrate that there may exist some additional ingredients
   (or processes) which determine the strength of the outbursts, e.g., 
   variations in the circumbinary disk and the disks of the binary holes.\\
   \begin{table}
   \caption{Model simulation results for the non-periodic synchrotron outburst
   peaking  at 2457380 and its comparison with that for the major periodic
   outburst in December/2015 (peaking at 2457360). $\rm{S_b}$=3.5\,mJy 
   (R-band). $t$\,=\,flare time\,=\,day--2457000. Explanation of the parameters
   and units as in Table 6.}
   \begin{flushleft}
   \centering
   \begin{tabular}{llllll}
   \hline
   $t$ & $\Gamma$ & ${\delta}_{max}$ & ratio & $\rm{S_{int}}$ & FWHM \\
   \hline
   360 & 9.5 & 18.88 & 4.11 & 1.33$\times{10^{-4}}$ & 5.9 \\
   380 & 13.5 & 26.77 & 7.29 & 1.67$\times{10^{-5}}$ & 2.2\\
    \hline
   \end{tabular}
   \end{flushleft} 
   \end{table}
   \section{Discussion}
   We have applied the precessing jet nozzle model previously proposed by
   Qian et al. (e.g., \cite{Qi91a}, \cite{Qi13}, \cite{Qi19})
   to investigate the optical variations observed in OJ287 and tried to
   clarify the nature of emission for the outbursts.\\
   It is found that the multi-wavelength variations (in NIR-optical-UV bands;
   Kushwaha et al. \cite{Ku18}) of the periodic major outburst in 
  December/2015 (peaking at JD2457360) are
  very similar to those of the non-periodic highly-polarized synchrotron 
  outburst in March/2016 (peaking at JD2457450).  The multi-wavelength 
   light curves of both the outbursts can be well simulated by symmetric 
   profiles and interpreted in terms of lighthouse effect due to the helical
    motion of one superluminal optical knot through two helical revolutions.
   This result seems important, indicating that the December/2015 outburst
   may like the March/2016 outburst and also originate from synchrotron
    process.
   \footnote{The December/2105 outburst was identified as 
   the ``impact (bremsstrahlung) flare'' according to the disk-impact model.} 
   Its association with the simultaneous $\gamma$-ray flare supports this 
   interpretation.\\
    The five periodic outbursts observed in 1983.00, 1984.10,
    1994.75, 2005.76 and 2007.70 (at V-band), and a few isolated 
    non-periodic flares
    have also been simulated. We find that all the periodic outbursts can be 
    decomposed into a number of subbursts (or ``elementary flares'').
    The light curves of all these elementary flares can be simulated by
    symmetric profiles with similar rising and decaying time-scales and and 
    interpreted in terms of the lighthouse model. The isolated non-periodic
    flares show their variability behavior similar to these elementary
    flares.\\
   In combination  with the simulation results for the December/2015 
    and March/2016, we tentatively suggest that the periodic optical
    outbursts observed during 1983\,--\,2015 may all originate from synchrotron
    process in the relativistic jet and they may be produced by lighthouse 
    effect due to the helical motion of superluminal optical knots 
    (blobs or shocks). This interpretation is consistent with 
    the requirement of ``single mechanism'', which is derived from the color 
    stability during the optical outbursts (Sillanp\"a\"a et al. \cite{Si96a},
    Gupta et al. \cite{Gu16}). The low polarization of the first flares of 
    the double-peaked outbursts can also be understood, because synchrotron
     flares can have a large range of polarization degree, as typically 
    observed in OJ287 (from  $<$2\% to $\sim$40\% 
    (Villforth et al. \cite{Vil10}, Kushwaha et al. \cite{Ku18}). The 
    close connection between the radio/mm and optical variations
    (e.g., observed in the 1995.84 outburst) can also be explained.\\
    We have shown that the entire optical variability in OJ287 could only be
    explained by invoking lighthouse effect due to the helical motion 
    of superluminal optical knots.  This result may have been expected, based
     on the magnetohydrodynamical (MHD) theories for jet formation in spinning
   black hole\,--\,accretion disk systems, in which relativistic jets are 
   produced  in the rotating magnetospheres with strong toroidal magnetic 
   fields and strong helical fields should be permeated  in the jets near the
   black holes (e.g., Blandford \& Znajek \cite{Bl77}, Blandford \& Payne 
   \cite{Bl82}, Camenzind \cite{Cam90}, Li et al. \cite{Lizy92}, Beskin 
   \cite{Be10}, Valhakis \& K\"onigl \cite{Vl04}, Meier \cite{Mei13}, 
    \cite{Mei01}). It would be a natural phenomenon that superluminal optical
   knots move along helical trajectories, producing optical outbursts through 
  lighthouse effect. Unfortunately, there seems only a few observational events
   revealing this phenomenon (e.g., Schramm et al. \cite{Sc93}, Dreissigacker
   \cite{Dr96a}, Dreissigacker \& Camenzind \cite{Dr96b}, Camenzind \&
   Kronkberger \cite{Cam92}, Wagner et al. \cite{Wa95}, Qian \cite{Qi15}). This
   work demonstrates that helical motion of superluminal optical components 
  may be a general phenomenon in blazar OJ287 and thus provide some 
   observational evidence for the existence of helical magnetic fields in the 
  inner jet regions of blazars.\\
   Under the binary black hole scenario, we have tentatively proposed  
   a unified and plausible relativistic jet
    model for fully explaining the optical activities in OJ287 (including its
    periodic and non-periodic outbursts), invoking lighthouse effect due
    to the helical motion of superluminal optical components. The chain of the
    physical processes in this model may be: a succession of discrete accretion
    events (including the double-stream accretion flows; e.g., 
    Tanaka \cite{Tan13} ) created by the  pericenter passages of the secondary
     hole (moving in an eccentric orbit) results in a succession of  ejection
     of superluminal optical components through the jet formation mechanism,
      producing a succession of the elementary optical flares which blend 
    together to form  major complex outbursts.\\
     The relativistic jet model tentatively suggested in this work 
     should be tested  by the future multi-wavelength (from radio to 
     $\gamma$-rays) observations. If this scenario is proved to be correct, the
    optical phenomena in OJ287 can be explained without needing to invoke the
    disk-impact mechanism, although this mechanism seems
    very attractive for testing the efffects of general relativity
   (Einstein \cite{Ei16}, \cite{Ei18}). However, the relativistic jet 
    scenario only concentrates on the
    explanation of the nature and characteristics of the optical activities
    (temporary and spectral variations of the outbursts), the quasi-periodicity
     of its optical variability and the double-structure of the periodic 
    outbursts remain to be interpreted. In principle, under the framework of
    binary black hole models, the quasi-periodicity can be related to the
    modulation of accretion rates via the pericenter passages of the 
    companion black hole in an eccentric orbit. As Sillanp\"a\"a et al.
    (\cite{Si96a}) originally suggested, the eccentric orbital  motion 
    of the secondary hole can cause quasi-periodically enhanced accretion 
    flows onto the primary hole, which consequently result in ejections 
    of superluminal optical knots via jet-formation mechanism(s), producing
    quasi-periodic optical outbursts. As regards the explanation of the 
    double-structure of periodic optical outbursts, cavity-accretion models
     as suggested by Tanaka (\cite{Tan13}) might be applicable. In the case
    of comparable-mass and eccentric binary systems, usually two gas streams
    are created per pericenter passage of the secondary hole 
    in the circumbinary disk and flow toward the binary black holes,
    producing a double-peaked outbursts. The cavity-accretion processes
    (dynamics and kinematics of the streams) might be quasi-regular, because
    the two streams would have to move across the Lagrange points of the binary
    system (e.g., Artymowicz \& Lubow \cite{Ar96}, Artymowicz \cite{Ar98},
    Tanaka \cite{Tan13}). However, in the case of cavity-accretion in binary 
    systems, complex processes are involved: e.g., eccentric
    motion of the binary around the mass-center, interaction between the
    binary and the circumbinary disk, creation of the pair of gas streams,
    jet formation, precession and  ejection of superluminal
    components in binary systems, etc. Thus it seems that cavity-accretion
     models can not 
    accurately predict the appearing times of the periodic optical outbursts, 
    because the timing of the periodic outbursts are not determined by the
    orbital motion only. Stochasticity in the circumbinary disk accretion 
    and in the dynamics and kinematics of the stream flows could result in
    some scattering of the appearing times of the double-peaked outbursts.
    In fact, even in the disk-impact model (Lehto \& Valtonen \cite{Le96},
    Dey et al. \cite{De18}) where the outburst timing is assumed to 
    mainly depend on the orbital motion, the strength of the double-peaked
    outbursts seems not closely related to the orbital phases 
    (or outburst timing). In Table 12
    the relation between the peak flux densities of the impact outbursts and
    their impact distances (and secondary hole velocities) are listed,
     which does not demonstrate any connection between these parameters:
    strong outbursts do not necessarily appear at small impact distances.
    This seems that some significant physical processes might have been missed
    for determining the strength of the periodic outbursts.\\
    In the relativistic jet model proposed in this work, the quasi-periodicity
    in optical variability and the double-peak structure can be ascribed to 
     the accretion processes in binary systems. In  the cavity-accretion models,
     the modulation of the accretion rate onto the binary holes and their 
     disks by the orbital period might be the most plausible mechanism for 
     explaining the quasi-periodicity of the optical variability observed in 
     OJ287. Moreover, the two streams of accretion flows created per
     pericenter passages of the secondary hole may be invoked to explain
     the double-peak structure of the periodic outbursts.  The timing
     mechanism for the periodic outbursts could be investigated along with
     the appearance of non-periodic outbursts. Detailed modeling
     based on HD/MHD simulations is imperatively required.
    \begin{table}
     \caption{Relation between the strength ($\rm{S_{v}}$, peak flux density
     at V-band) of the impact outbursts, the impact distance 
    ($\rm{R_{imp}}$) and the secondary velocity ${v_0}$/c. 
    t\,=\,flare time. The data are arranged in 
    sequence of impact distance. $\rm{S_{v,obs}}$\,=\,$\rm{S_v}$+$\rm{S_b}$.}
     \begin{flushleft}
     \centering
     \begin{tabular}{lllll}
     \hline
     $t$ & $R_{imp}$(AU) & $v_0$/c & $\rm{S_{v,obs}}$(mJy) & $\rm{S_v}$(mJy)\\
     \hline
     2007.70 & 3259 & 0.264 & 12.0 & 6.5 \\
     1995.84 & 3855 & 0.245 & 5.0 & 3.5 \\
     1983.00 & 4633 & 0.224 & 32.0 & 27.0 \\ 
     1984.10 & 5387 & 0.205 & 20.0 & 17.2 \\
     1994.75 & 7079 & 0.173 & 6.0  & 5.0 \\
     2005.76 & 12427 & 0.106 & 11.0 & 9.0 \\
     2015.87 & 17566 & 0.058 & 18.0 & 14.5 \\
     \hline
     \end{tabular}
     \end{flushleft}
     \end{table}
    \begin{acknowledgement}
   I wish to thank Prof. A. Sillanp\"a\"a for affording the
   V-band data on OJ287 during 1993.9\,--\,1995.3 (OJ-94 project),
    which was very helpful to this work.
   \end{acknowledgement}

  \end{document}